\documentclass[11pt]{article}

\usepackage[english]{babel}
\usepackage{amsmath,amssymb,graphicx}
\usepackage{hyperref}
\usepackage[nosort]{cite}
\usepackage{multicol,longtable}
\usepackage{tensor}
\usepackage{multirow}
\usepackage{nicefrac}
\usepackage{cases}
\usepackage[vcentermath]{youngtab}
\usepackage[top=2.5cm,bottom=2.5cm,inner=2.3cm,outer=2.3cm]{geometry}
\usepackage{tikz}

\usepackage{cite}
\usepackage{amsmath,amsfonts,amssymb} 
\allowdisplaybreaks[1] 
\usepackage[small,bf,labelsep=space]{caption}
\addto\captionsenglish{}
\usepackage{slashed}
\usepackage{subfig}
\usepackage{latexsym,epsfig}

\usepackage[final]{showkeys}
\usepackage{amsfonts}
\usepackage{mathrsfs}
\usepackage{dsfont}
\usepackage[Symbol]{upgreek}

\usepackage{tcolorbox}
\usepackage{empheq}

\DeclareMathAlphabet{\matcal}{OMS}{cmsy}{m}{n}

\newcommand{\be}{\begin{align}}
\newcommand{\ee}{\end{align}}

\makeatletter

\@addtoreset{equation}{section}
\makeatother

\newcommand{\bea}{\begin{eqnarray}}
\newcommand{\eea}{\end{eqnarray}}

\def\*{\partial}

\def\={\!=\!}

\usepackage{cite}
\usepackage{amsmath,amsfonts,amssymb}

\usepackage{color}
\definecolor{red}{rgb}{1,0,0}
\definecolor{lred}{rgb}{0.3,0,0}
\definecolor{green}{rgb}{0,0.6,0}
\definecolor{blue}{rgb}{0,0,1}
\definecolor{violet}{rgb}{0.8,0,0.8}

\definecolor{darkred}{rgb}{0.65,0.15,0}
\definecolor{darkgreen}{rgb}{.05,.5,.05}
\hypersetup{pdfborder={0 0 0},colorlinks=true,urlcolor=darkred,citecolor=darkred,linkcolor=darkred,linktocpage=true}

\def\SO{\ensuremath{\mathrm{SO}}\xspace}

\def\Odd{\ensuremath{\mathrm{O}(d,d,\mathbb{R})}\xspace}
\def\odd{\ensuremath{\mathfrak{o}(d,d,\mathbb{R})}\xspace}

\def\GL{\ensuremath{\mathrm{GL}}\xspace}

\newcommand{\F}{\ensuremath{\mathcal{F}}\xspace}
\renewcommand{\S}{\ensuremath{\mathcal{S}}\xspace}
\renewcommand{\H}{\ensuremath{\mathcal{H}}\xspace}

\def\d{\ensuremath{\mathrm{d}}\xspace}
\newcommand{\Tr}[1]{\ensuremath{\mathrm{Tr}\left(#1\right)}\xspace}

\newdimen\squaresize \squaresize=12pt
\newdimen\thickness \thickness=0.7pt

\def\square#1{\hbox{\vrule width \thickness
   \vbox to \squaresize{\hrule height \thickness\vss
      \hbox to \squaresize{\hss#1\hss}
   \vss\hrule height\thickness}
\unskip\vrule width \thickness} \kern-\thickness}

\def\cut#1{\hbox{\vrule width-1 \thickness
   \vbox to \squaresize{\hrule height-1 \thickness\vss
      \hbox to \squaresize{\hss#1\hss}
   \vss\hrule height-1\thickness}
\unskip\vrule width +4 \thickness} \kern-\thickness}

\def\vsquare#1{\vbox{\square{$#1$}}\kern-\thickness}

\def\young#1{
\vbox{\smallskip\offinterlineskip \halign{&\vsquare{##}\cr #1}}}

\newcommand{\tinyyoung}[1]{
\squaresize=10pt \thickness=0.4pt \mbox{\scriptsize\young{#1}}
\squaresize=15pt \thickness=0.7pt}

\usepackage{xspace}
\def\S{\ensuremath{\mathcal{S}}\xspace}
\def\F{\ensuremath{\mathcal{F}}\xspace}

\usepackage[affil-it]{authblk}

\def\preprint{}

\makeatletter
\def\@maketitle{%
  \newpage
  \null\hfill\texttt{\preprint}
  \vskip 4em%
  \begin{center}%
  \let \footnote \thanks
    {\LARGE\bfseries \@title \par}%
    \vskip 2.5em%
    {\large
      \lineskip .5em%
      \begin{center}
        \begin{minipage}{0.95\textwidth}
            \begin{tabular}[t]{c}%
            \@author
            \end{tabular}
        \end{minipage}    
      \end{center}\par}%
    \vskip 1em%
  \end{center}%
  \par
  \vskip 1.5em}
\makeatother

\title{Duality Invariance and Higher Derivatives}
\author[1]{Camille Eloy}
\author[2]{Olaf Hohm}
\author[1]{Henning Samtleben}
\affil[1]{Univ Lyon, Ens de Lyon, Univ Claude Bernard, CNRS,
Laboratoire de Physique, F-69342 Lyon, France}
\affil[2]{Institute for Physics, Humboldt University Berlin, Zum Gro{\ss}en Windkanal 6, D-12489 Berlin, Germany}

\begin{document}
\maketitle
\thispagestyle{empty}

\begin{abstract}

We dimensionally reduce the spacetime action of bosonic string theory, and that of the bosonic sector of heterotic string theory after truncating the Yang-Mills gauge fields, 
on a $d$-dimensional torus including all higher-derivative corrections to first order in $\alpha'$. 
A systematic procedure is developed that brings 
this  action into a minimal form in which all fields except the metric carry only first order derivatives. 
This action is shown to be invariant under ${\rm O}(d,d,\mathbb{R})$ transformations that acquire $\alpha'$-corrections through a Green-Schwarz type mechanism. 
We prove that, up to a global pre-factor,  the first order $\alpha'$-corrections are uniquely determined by ${\rm O}(d,d,\mathbb{R})$ invariance. 

\end{abstract}

\newpage
\setcounter{page}{1} 


\tableofcontents

\section{Introduction}

String theory features the T-duality property according to which there is a non-linear  group action of ${\rm O}(d,d,\mathbb{Z})$ on  $d$-dimensional toroidal backgrounds such that 
all backgrounds in one orbit are physically equivalent. When restricting to the massless fields for compactifications on tori, i.e., when performing dimensional reduction, 
this duality implies invariance under the continuous  symmetry group \Odd. 
For the two-derivative effective action this symmetry was first shown explicitly for the (cosmological) reduction to one dimension by Veneziano and Meissner in Refs.~\cite{Veneziano:1991ek,Meissner:1991zj} 
and later generalized to arbitrary $d$ by Maharana and Schwarz~\cite{Maharana:1992my}. 

It was proven by Sen, using closed string field theory, that the \Odd symmetry of dimensionally reduced theories is present to all order in $\alpha'$ \cite{Sen:1991zi},  
but it remains  as a highly non-trivial problem to actually display this symmetry when higher-derivative $\alpha'$-corrections are included. 
First significant progress was due to Meissner who investigated the dimensional reduction to one dimension including the four-derivative terms that appear in string theory to first order in $\alpha'$ \cite{Meissner:1996sa}
(for earlier work on the heterotic string see Ref.~\cite{Bergshoeff:1995cg}).
He uncovered the expected  \Odd symmetry, but this required a series of elaborate field redefinitions 
(that in particular cannot all originate from covariant field redefinitions before reduction). Subsequent work considered the reduction on a single circle~\cite{Kaloper:1997ux} 
and reductions on a general torus but truncating out all `off-diagonal' field components~\cite{Godazgar:2013bja}. 
In all these truncations there is a choice of field variables for which the \Odd transformations are undeformed, as is also suggested by 
string field theory \cite{Kugo:1992md}. In particular, this fact was used to classify all higher-derivative corrections in cosmology that, somewhat surprisingly, only require  
(higher powers of) first-order time derivatives \cite{Hohm:2015doa,Hohm:2019jgu}.

Recently, the higher-derivative $\alpha'$-corrections of string theory have been the focus of attention in the framework of double field theory. 
Double field theory is a formulation featuring a manifest \Odd invariance \textit{before} dimensional reduction by virtue of a generalized  spacetime
with doubled coordinates transforming covariantly under \Odd \cite{Siegel:1993th,Hull:2009mi,Hohm:2010jy,Hohm:2010pp}. 
While the two-derivative double field theory can be written naturally in terms of a `generalized metric' that encodes metric and $B$-field (\textit{c.f.}~Eq.~(\ref{HOdd}) below), 
there are obstacles when including higher derivatives that require  a deformation of the framework, see Refs.~\cite{Hohm:2013jaa,Hohm:2014eba,Hohm:2014xsa,Marques:2015vua,Lescano:2016grn,Naseer:2016izx,Baron:2017dvb,Baron:2018lve}. 
It was proven in Refs.~\cite{Hohm:2016lge,Hohm:2016yvc} that the general $\alpha'$-corrections of bosonic and heterotic string theory 
cannot be written in terms of the generalized metric, so that in particular the \Odd transformations of double field theory get $\alpha'$-deformed. 
Alternatively, one may set up a generalized frame formalism for which \Odd remains undeformed while the local frame transformations receive 
$\alpha'$-corrections \cite{Marques:2015vua,Hohm:2016lge,Hohm:2016yvc}.

In this paper we complete the existing literature by giving the complete dimensionally reduced action for bosonic string theory to first order in $\alpha'$, i.e., including all 
four-derivative terms,  
and prove its \Odd invariance, presenting results that have recently been announced in Ref.~\cite{Eloy:2019hnl}. 
In particular, we prove that the first order $\alpha'$-corrections are uniquely determined by \Odd invariance, up to an overall constant 
whose value depends on the string theory under consideration. 
While this \Odd invariance is also implied by the existence of $\alpha'$-deformed double field theory, whose dimensional reduction has already 
been explored  in Ref.~\cite{Baron:2017dvb}, until now it has not been systematically investigated whether some of the unexpected new features arising in double field theory 
also show up in the dimensional reduction of conventional (non-extended) theories, nor has the dimensionally reduced action been displayed in a 
sufficiently simplified form that allows for applications (and comparison with some of the earlier results cited above). 
To our surprise we find that there is no choice of field variables so that the full dimensionally reduced action can be written in terms of familiar \Odd covariant 
variables (like the generalized metric); rather, a generalized Green-Schwarz mechanism is required under which the (external) singlet $B$-field acquires 
non-trivial transformations under  \Odd, hence implying that the \Odd action gets $\alpha'$-deformed. 
This effect has been invisible in all truncations investigated so
far, but it does mimic the situation in double field theory before reduction. 
Intriguingly, the $\alpha'$-deformations needed in double field theory 
can thus not be blamed entirely on its novel geometric structure, but such deformations also emerge in completely conventional dimensional reductions.

On a technical level, the present investigation requires full control over all possible field redefinitions, both redefinitions that are covariant 
in the usual sense (i.e.~$\GL(d)$ covariant) and covariant with respect to \Odd. 
As one of the main technical results of this paper we present a fully systematic procedure to test \Odd invariance, generalizing that of Refs.~\cite{Hohm:2015doa,Hohm:2019jgu}
to higher dimensions. One first dimensionally reduces the action 
as usual and then uses $\GL(d)$ covariant field redefinitions to bring the action into a form in which all fields apart from the metric appear only with first-order 
derivatives. Next, one employs \Odd covariant redefinitions in order to find the minimal set of \Odd  invariant four-derivative terms, which then 
are decomposed under $\GL(d)$ with the aim to match with the dimensionally reduced terms. 
Our analysis applies to bosonic string theory but also to the bosonic sector of heterotic string theory after truncating out the Yang-Mills gauge fields, 
which still features a gravitational Chern-Simons form (due to the original Green-Schwarz mechanism).

The rest of the paper is organized as follows.
In Sec.~\ref{sec:two-derivative}, we review the dimensional reduction of the leading two-derivative action
of the bosonic string,
and its manifestly \Odd symmetric formulation revealed in Ref.~\cite{Maharana:1992my}.
In order to set up a systematic analysis of its higher-order corrections,
we outline how to organize and fix the ambiguities related to
partial integration and higher-order field redefinitions.
In Sec.~\ref{sec:basis}, we present a general counting of independent higher-derivative terms 
upon modding out these ambiguities. At order $\alpha'$, 
we construct an explicit 61-dimensional basis of independent \Odd invariant four-derivative terms,
which is algebraic in first order derivatives and the Riemann tensor.
Sec.~\ref{sec:reduc4d} presents the explicit torus reduction of the four-derivative action
of the bosonic string. In particular, we show how all second-order derivatives in the reduced action
can be eliminated by suitable field redefinitions.
Comparing the result to our explicit basis, 
we show in Sec.~\ref{sec:alpha'action} that apart from a single term
the entire reduced action can be rewritten in terms of manifestly \Odd invariant terms.
Restoring \Odd invariance of the full action then requires a Green-Schwarz type mechanism 
inducing a non-trivial \Odd transformation of order $\alpha'$ of the two-form $B_{\mu\nu}$\,.
In Sec.~\ref{sec:frame}, we embed this structure into a frame formalism in which the \Odd symmetry remains 
undeformed, while the local frame transformations acquire $\alpha'$ deformations.
Finally, in Sec.~\ref{sec:het}, we extend the analysis to the bosonic sector of heterotic supergravity
and present its dimensionally reduced action in manifestly \Odd invariant form.
The appendices collect a number of explicit technical results.

\section{Two-derivative action and systematics of field redefinitions}
\label{sec:two-derivative}

A main goal of this paper is to compute the dimensional reduction of the bosonic string
on a $d$-dimensional torus including the first order in $\alpha'$
and to make the resulting \Odd symmetry manifest. 
In this section, we review the reduction of the two-derivative action and its manifestly
\Odd symmetric formulation first exhibited in Ref.~\cite{Maharana:1992my}.
We then discuss its field equations and the systematics of non-linear field redefinitions
as a starting point for the subsequent systematic analysis of the higher order corrections.

\subsection[Reduction and \texorpdfstring{$\Odd$}{Odd} symmetry]{Reduction and \texorpdfstring{$\boldsymbol{\Odd}$}{Odd} symmetry}
Let us start from the two-derivative effective action for the bosonic string in $D+d$ dimensions, with metric $\hat{g}_{\hat\mu\hat\nu}$, antisymmetric Kalb-Ramond field $\hat B_{\hat\mu\hat\nu}$ and dilaton~$\hat\phi$:
\begin{equation}
\label{eq:I0D+d}
\widehat{I}_{0} = \int \d^{D+d}X\,\sqrt{-\hat{g}}\,e^{-\hat{\phi}}\,\left(\hat{R} + \partial_{\hat\mu}\hat\phi\,\partial^{\hat\mu}\hat\phi-\frac{1}{12}\,\hat H^{2}\right), 
\end{equation}
where indices $\hat\mu$ run over the $(D+d)$ dimensional space, and $\hat{H}^{2}= \hat H^{\hat\mu\hat\nu\hat\rho}\hat H_{\hat\mu\hat\nu\hat\rho}$ with the field strength $\hat H_{\hat\mu\hat\nu\hat\rho}=3\,\partial_{[\hat\mu}\hat B_{\hat\nu\hat\rho]}$. To compactify on the spatial torus ${\rm T}^{d}$, we use the index split $X^{\hat\mu}=(x^{\mu},y^{m}),\,$ with $\mu\in[\![1,D]\!], \;m\in[\![1,d]\!]$, for curved indices and $\{{\hat \alpha}\}=\{\alpha,a\}$, with $\alpha\in[\![1,D]\!],\;a\in[\![1,d]\!]$ for flat indices, and drop the dependence of all fields on the internal coordinates $y^{m}$. For the metric $\hat{g}_{\hat\mu\hat\nu}$, we use the vielbein formalism and consider the standard Kaluza-Klein ansatz
 \begin{equation}
 \hat{e}_{\hat{\mu}}^{\ \hat{\alpha}}= \begin{pmatrix}
                                      {e_{\mu}}^{\alpha} & A_{\mu}^{(1)\,n}\,{E_{n}}^{a} \\
                                      0 & {E_{m}}^{a}
                              \end{pmatrix},
 \end{equation}
in terms of the $D$-dimensional vielbein $e_{\mu}{}^{\alpha}$, Kaluza-Klein vector fields $A_{\mu}^{(1)m}$, and the internal vielbein ${E_{m}}^{a}$. The metric $\hat{g}_{\hat{\mu}\hat{\nu}}=\hat{e}_{\hat{\mu}}^{\ \hat{\alpha}}\eta_{\hat\alpha\hat\beta}\hat{e}_{\hat{\nu}}^{\ \hat{\beta}}$ then takes the form
\begin{equation}
\label{eq:ansatzg}
\hat{g}_{\hat{\mu}\hat{\nu}}= \begin{pmatrix}
                                      g_{\mu\nu}+A_{\mu}^{(1)p}G_{pq}A_{\nu}^{(1)q} & A_{\mu}^{(1)p}G_{pn} \\[.5ex]
                                      G_{mp}A_{\nu}^{(1)p} & G_{mn}
                              \end{pmatrix}
                              \;,
\end{equation}
where $g_{\mu\nu}={e_{\mu}}^{\alpha}\eta_{\alpha\beta}{e_{\nu}}^{\beta}$ and $G_{mn} = {E_{m}}^{a}\delta_{ab}{E_{n}}^{b}$ 
denote the $D$-dimensional metric and the internal metric, respectively.

Similarly, the 2-form $\hat{B}_{\hat\mu\hat\nu}$, is parametrized as~\cite{Maharana:1992my}
\begin{equation}
\label{eq:ansatzB}
\hat{B}_{\hat{\mu}\hat{\nu}}= \begin{pmatrix}
                                B_{\mu\nu}-A_{[\mu}^{(1)\,m}A_{\nu]\,m}^{(2)}+A_{\mu}^{(1)\,m}\,B_{mn}\,A_{\nu}^{(1)\,n} & A^{(2)}_{\mu\,n}-B_{np}\,A_{\mu}^{(1)\,p} \\
                                -A^{(2)}_{\nu\,m}+B_{mp}\,A_{\nu}^{(1)\,p} & B_{mn}
                              \end{pmatrix},
\end{equation}
in terms of $D$-dimensional scalars $B_{mn}=-B_{nm}$, vector fields $A_{\mu\,m}^{(2)}$, and a 2-form $B_{\mu\nu}$. The lower-dimensional components of $\hat{H}_{\hat\mu\hat\nu\hat\rho}$ are defined using the standard Kaluza-Klein procedure~\cite{Maharana:1992my}: first converting $\hat{H}$ to flat indices, block decomposing, and finally converting 
back to curved indices using the lower-dimensional blocks ${e_{\mu}}^{\alpha}$ and ${E_{m}}^{a}$. 
This amounts to converting a curved index $\hat\mu$ to a curved index $\mu$ using contraction 
with ${e_{\mu}}^{\alpha}\hat{e}_{\alpha}^{\ \,\hat\mu}$ and to $m$ contracting with ${E_{m}}^{a}\hat{e}_{a}^{\ \,\hat\mu}$,
such that the resulting fields transform covariantly under internal diffeomorphisms\footnote{
Note that it is not the procedure that is used on $\hat{B}$, as pointed out in Ref.~\cite{Hohm:2014sxa}.}. 
With Eq.~\eqref{eq:ansatzg}, this leads to
\begin{equation}
  \label{eq:Hreduced}
  \begin{cases}
  H_{\mu\nu\rho} = 3\partial_{[\mu}B_{\nu\rho]}-\dfrac{3}{2}\left(A^{(1)\,m}_{[\mu}F_{\nu\rho]\,m}^{(2)}+F_{[\mu\nu}^{(1)\,m}A_{\rho]\,m}^{(2)}\right)\;,\\
  H_{\mu\nu m} = F_{\mu\nu\,m}^{(2)} - B_{mn}F_{\mu\nu}^{(1)n}\;,\\
  H_{\mu mn} = \nabla_{\mu}B_{mn}\;,\\
  H_{mnp} =0\;,
  \end{cases}
\end{equation}
where we have defined the abelian field strengths
\begin{equation}
\begin{cases}
F_{\mu\nu}^{(1)m}=\partial_{\mu}A_{\nu}^{(1)\,m}-\partial_{\nu}A_{\mu}^{(1)\,m}\;, \\
F_{\mu\nu\,m}^{(2)}=\partial_{\mu}A_{\nu\,m}^{(2)}-\partial_{\nu}A_{\mu\,m}^{(2)}\;.
\end{cases}
\end{equation}

In terms of these objects, after dimensional reduction, the action (\ref{eq:I0D+d}) then takes the form~\cite{Maharana:1992my}
\begin{equation}
\begin{aligned}
I_{0}=&\int\d^{D}x\,\sqrt{-g}\,e^{-\Phi}\,\bigg(R+\partial_{\mu}\Phi\,\partial^{\mu}\Phi-\frac{1}{12}\,H_{\mu\nu\rho}H^{\mu\nu\rho} +\frac{1}{4}\,\Tr{\partial_{\mu}G\partial^{\mu}G^{-1}} \\
& +\frac{1}{4} \Tr{G^{-1}\partial_{\mu}BG^{-1}\partial^{\mu}B}-\frac{1}{4}\,F_{\mu\nu}^{(1)\,m}G_{mn}F^{(1)\,\mu\nu\,n}-\frac{1}{4} \,H_{\mu\nu m}G^{mn}{H^{\mu\nu}}_{n}\bigg),
\label{eq:I0Gld}
\end{aligned}
\end{equation}
with the rescaled dilaton $\Phi = \hat{\phi}-\dfrac{1}{2}\,\log(\det(G_{mn}))$. 
In this form, the action features an explicit $\GL(d)$ symmetry, as guaranteed by toroidal reduction. 
The symmetry enhancement to \Odd can be made manifest upon regrouping the vector 
fields $A^{(1)\,m}_{\mu}$ and $A^{(2)}_{\mu\,m}$ into a single \Odd vector
\begin{equation}
 {\cal A}_{\mu}{}^{M} = \begin{pmatrix}
                        A_{\mu}^{(1)\,m} \\
                        A_{\mu\,m}^{(2)}
                        \end{pmatrix}\;
                        \label{AOdd}\;,
\end{equation}
and the scalar fields $G_{mn}$, $B_{mn}$ into an  \Odd matrix ${\cal H}_{MN}$ as
\begin{equation}
  {\cal H}_{MN} = \begin{pmatrix}
  G_{mn}-B_{mp}G^{pq}B_{qn} & B_{mp}G^{pn}\\[.5ex]
                    -G^{mp}B_{pn} & G^{mn}                 
                    \end{pmatrix}\;.
                    \label{HOdd}
\end{equation}
Throughout, the fundamental \Odd indices are raised and lowered using the constant \Odd-invariant matrix
\begin{equation}
  \label{etamatrix} 
  \eta^{MN} = \begin{pmatrix}
                    0 & {\delta^{m}}_{n} \\
                    {\delta_{m}}^{n} & 0
                  \end{pmatrix}\;,
\end{equation}
so that $\H^{-1}$ is defined as $\H^{MN} = \eta^{MP}\H_{PQ}\eta^{QN}$.
In terms of the fields~(\ref{AOdd}), (\ref{HOdd}), the reduced action~(\ref{eq:I0Gld}) may be 
cast into the manifestly \Odd invariant form~\cite{Maharana:1992my}
\begin{align}
I_{0}=&\!\int\!\d^{D}x\,\sqrt{-g}\,e^{-\Phi}\Big(R+\partial_{\mu}\Phi\,\partial^{\mu}\Phi 
+\frac{1}{8}\,
\partial_{\mu}\H_{MN} \partial^{\mu}\H^{MN}
-\frac{1}{4}\F_{\mu\nu}{}^{M} \H_{M N} \F^{\mu\nu\,N}
-\frac{1}{12} H_{\mu\nu\rho}H^{\mu\nu\rho}
\Big)\;,
\label{eq:MaharanaSchwarz}
\end{align} 
where $\F_{\mu\nu}{}^{M}=2\partial_{[\mu}{\cal A}_{\nu]}{}^{M}$ is the abelian field-strength associated to 
the vectors (\ref{AOdd}). 
In terms of the covariant objects~\eqref{AOdd} and \eqref{HOdd},
the infinitesimal \Odd variations of the fields are given by
\begin{equation}
\label{eq:oddvariation}
\begin{cases}
  \delta_{\Gamma}g_{\mu\nu} = 0\;, \\
  \delta_{\Gamma}B_{\mu\nu} =0\;,
\end{cases} \qquad
\begin{cases}
  \delta_{\Gamma}\H_{MN} = \Gamma_{M}{}^{P}\H_{PN}+\Gamma_{N}{}^{P}\H_{MP}\;, \\
  \delta_{\Gamma}\F_{\mu\nu}{}^{M} = - \F_{\mu\nu}{}^{N}\,\Gamma_{N}{}^{M}\;,
\end{cases}
\end{equation}
for $\Gamma_{M}{}^{N}\in\odd$. The action~\eqref{eq:MaharanaSchwarz} is manifestly invariant 
under these transformations.
For later convenience, we also rewrite the action in terms of the matrix ${\S_{M}}^{N} = {\cal H}_{MP}\,\eta^{PN}$
\begin{equation}  
\label{eq:MaharanaSchwarzS}
I_{0}=\!\int\!\d^{D}x\,\sqrt{-g}\,e^{-\Phi}\!\left(\!R+\partial_{\mu}\Phi\,\partial^{\mu}\Phi+\frac{1}{8}\,\Tr{\partial_{\mu}\S\,\partial^{\mu}\S}-\frac{1}{4}\,\F_{\mu\nu}{}^{M}{\S_{M}}^{N}\F^{\mu\nu}{}_{N}
-\frac{1}{12}\,H_{\mu\nu\rho}H^{\mu\nu\rho}\right)
\;.
\end{equation}
Note that $\S\S=\mathds{1}$, so that $\S$ is a constrained field. 

\subsection[\texorpdfstring{${\rm GL}(d)$}{GL(d)} fields redefinitions]{\texorpdfstring{$\boldsymbol{{\rm GL}(d)}$}{GL(d)} fields redefinitions}
\label{sec:GLdredef}
Our aim is an extension of the previous construction to higher orders in $\alpha'$.
As usual, the study of higher-derivative terms requires to carefully handle the ambiguities due to
the possible non-linear field redefinitions.
In particular, the symmetry enhancement to \Odd will only be possible after identification
of the proper field redefinitions.
In this section, we describe the systematics of higher-order field redefinitions based on the two-derivative action~(\ref{eq:MaharanaSchwarz}), inspired by Refs.~\cite{Hohm:2019jgu,Hohm:2015doa}.

We consider the $\alpha'$ extension of Eq.~(\ref{eq:MaharanaSchwarz}) as a perturbation series
\bea
I=I_{0}+I_{1}+{\cal O}(\alpha'{}^2)
\;,
\eea
with the first order term $I_1 \sim {\cal O}(\alpha')$\,. In order to organize the possible ambiguities in $I_1$, we consider   
field redefinitions of the form
\begin{equation}
  \label{eq:fieldvariation}
 \varphi \rightarrow \varphi + \alpha'\,\delta\varphi,
\end{equation}
where $\varphi$ denotes a generic field. Under such redefinitions of its fields, the 
variation of $I$ to order $\alpha'$ arises exclusively from the variation of $I_0$ and takes the form
\begin{align}
\label{eq:deltaI0}
\delta I_{0} = \alpha'\int \d^{D}x\,\sqrt{-g}\,e^{-\Phi}\bigg[ & {\cal E}_{\Phi}\,\delta\Phi 
+ ({\cal E}_{g}){}_{\mu\nu}\,\delta g^{\mu\nu}+({\cal E}_{B}){}_{\mu\nu} \,\delta B^{\mu\nu}
+ ({\cal E}_{G}){}_{mn}\,\delta G^{mn} \nonumber \\
&\quad+
({\cal E}_{B}){}^{mn}\,\delta B_ {mn}+({{{\cal E}_{A^{(1)}}}){}^{\mu}}_{m}\,\delta A_{\mu}^{(1)\,m}
+({{\cal E}_{A^{(2)}}}){}^{\mu\,m}\,\delta A_{\mu\,m}^{(2)}\bigg]\;,
\end{align}
proportional to the field equations associated with the two-derivative action $I_0$
\begin{equation}
\begin{aligned}
\displaystyle {\cal E}_{\Phi} &= -2\,\Box\Phi - R + \nabla_{\mu}\Phi\,\nabla^{\mu}\Phi+\frac{1}{12}\,H^{2}-\frac{1}{8}\,\Tr{\nabla_{\mu}\S\,\nabla^{\mu}\S}+
\frac{1}{4}\,\F_{\mu\nu}{}^{M}\,{\S_{M}}^{N}\,\F^{\mu\nu}{}_{N}, \medskip\\
\displaystyle ({\cal E}_{g}){}_{\mu\nu} &= R_{\mu\nu}+\nabla_{\mu}\nabla_{\nu}\Phi - \frac{1}{4}H_{\mu\nu}^{2}+ \frac{1}{8}\,\Tr{\nabla_{\mu}\S\,\nabla_{\nu}\S}-\frac{1}{2}\,\F_{\mu\rho}^{M}\,{\S_{M}}^{N}\,{{\F_{\nu}}^{\rho}}_{N}+\frac{1}{2}g_{\mu\nu}\,{\cal E}_{\Phi},  \medskip\\
\displaystyle ({\cal E}_{B}){}_{\mu\nu} &= \frac{1}{2}\,\left(\nabla^{\rho}H_{\rho\mu\nu}-\nabla^{\rho}\Phi\,H_{\rho\mu\nu}\right),  \medskip\\
\displaystyle ({\cal E}_{G}){}_{mn} &= \frac{1}{2}\Big[-\Box G_{mn}+\nabla_{\mu}\Phi\,\nabla^{\mu}G_{mn}-\left(\nabla_{\mu}G\nabla^{\mu}G^{-1}G\right)_{mn}+\left(\nabla_{\mu}BG^{-1}\nabla^{\mu}B\right)_{mn} \medskip\\
& \qquad\quad {}
+\frac{1}{2}\,G_{mp}F_{\mu\nu}^{(1)\,p}F^{(1)\,\mu\nu\,q}G_{qn}-\frac{1}{2}\,H_{\mu\nu m}{H^{\mu\nu}}_{n}\Big]\,,  \medskip\\[1ex]
\displaystyle{({\cal E}_{B}})^{mn} &= \frac{1}{2}\Big[\left(G^{-1}\Box BG^{-1}\right)^{mn}-\nabla_{\mu}\Phi\left(G^{-1}\nabla^{\mu}BG^{-1}\right)^{mn}+\left(G^{-1}\nabla_{\mu}B\nabla^{\mu}G^{-1}\right)^{mn} \medskip \\
&  \qquad\quad{} +\left(\nabla_{\mu}G^{-1}\nabla^{\mu}BG^{-1}\right)^{mn}+\frac{1}{2}\,G^{mp}H_{\mu\nu p}F^{(1)\,\mu\nu\,n}-\frac{1}{2}\,F^{(1)\,\mu\nu\,m}G^{np}H_{\mu\nu p}\Big] \,,  \medskip\\[1ex]
\displaystyle{({{\cal E}_{A^{(1)}}})^{\nu}}_{n} &= \nabla_{\mu}F^{(1)\,\mu\nu\,m} G_{mn}-\nabla_{\mu}\Phi \,F^{(1)\mu\nu\,m} G_{mn} -\frac{1}{2}\,H^{\mu\nu\rho}H_{\mu\rho n} -{({\cal E}_{A^{(2)}}})^{\nu\,m}B_{mn}\medskip\\
&  \qquad {} +F^{(1)\,\mu\nu\,m}\nabla_{\mu}G_{mn}-{H^{\mu\nu}}_{m}{\left(G^{-1}\nabla_{\mu}B\right)^{m}}_{n}+ ({\cal E}_{B}){}^{\mu\nu}\left(A_{\mu\,n}^{(2)}-B_{nm}A_{\mu}^{(1)\,m}\right)\,,\medskip\\[1ex]
{({\cal E}_{A^{(2)}}})^{\nu\,m} &= \nabla_{\mu}{H^{\mu\nu}}_{n}G^{nm}-\nabla_{\mu}\Phi\,{H^{\mu\nu}}_{n}G^{nm} + {H^{\mu\nu}}_{n}\nabla_{\mu}G^{nm} +\frac{1}{2}\,H^{\mu\rho\nu}F_{\mu\rho}^{(1)\,m}  \medskip\\
&  \qquad {} + ({\cal E}_{B}){}^{\mu\nu}A_{\mu}^{(1)\,m} \,.
\end{aligned}
\label{eq:eomGLd}
\end{equation}
Here, $H^{2}_{\mu\nu}=H_{\mu\rho\sigma}H_{\nu}{}^{\rho\sigma}$, $\nabla_{\mu}$ denotes the covariant derivative with respect to $g_{\mu\nu}$ 
and accordingly $\Box=\nabla_{\mu}\nabla^{\mu}$.
At order $\alpha'$, the action thus is unique up to contributions proportional to the lowest order field equations. 
In the next section, we will show that by field redefinitions~(\ref{eq:fieldvariation}), the transformation~(\ref{eq:deltaI0}) 
together with partial integrations allows to map all terms at order $\alpha'$ to a basis which carries 
only first derivatives of all fields (except for the two-derivative terms within the Riemann tensor).

As an example, let us show how a term carrying the factor $\Box \Phi$ 
can be replaced by terms carrying only products of first derivatives. 
Consider a generic term of $I_{1}$ of the form
\begin{equation}
 Z = \alpha'\, \int\d^{D}x\,\sqrt{-g}\,e^{-\Phi}\,X\,\Box\Phi,
\end{equation}
where $X$ is a function of $\Phi$, $R_{\mu\nu\rho\sigma}$, $H_{\mu\nu\rho}$, $G_{mn}$, $B_{mn}$, $F_{\mu\nu}^{(1)\,m}$ and $H_{\mu\nu\,m}$ (and their derivatives) 
which carries exactly two derivatives. Redefining the dilaton and the metric as Eq.~(\ref{eq:fieldvariation})
with $\delta g_{\mu\nu}=\lambda\,g_{\mu\nu}$, 
Eq.~\eqref{eq:deltaI0} yields the transformation
\begin{equation}
 \begin{aligned}
 \delta I_{0}=\alpha'\,\int \d^{D}x\,\sqrt{-g}\,e^{-\Phi}\,  & \bigg[\,\Box\Phi\,\Big(-2\,\delta\Phi + \lambda \left(D+1\right)\Big) +  \frac{1}{2}\,R \,\Big(-2\,\delta\Phi +\lambda \left(D+2\right)\Big)  \\ 
 &\quad+ \nabla_{\mu}\Phi\,\nabla^{\mu}\Phi\, \Big(\delta\Phi - \frac{D}{2}\,\lambda\Big)+\frac{1}{24}\,H^{2} \,\Big(2\,\delta\Phi- \lambda\left(D+6\right)\Big) \\
 &\quad-\frac{1}{16}\,\Tr{\nabla_{\mu}\S\,\nabla^{\mu}\S}\,\Big(-2\,\delta\Phi+\lambda\left(D+2\right)\Big) \\
 &\quad+\frac{1}{8}\,\F_{\mu\nu}{}^{M}\,{\S_{M}}^{N}\,\F^{\mu\nu}{}_{N}\,\Big(2\,\delta\Phi-\lambda\left(D+4\right)\Big)\bigg]\;.
 \end{aligned}
 \label{eq:deltaI0ex}
\end{equation}
With the particular choice
\begin{equation}
\begin{cases}
 \delta\Phi = \dfrac{1}{2}\left(D+2\right) X\;,\\[1ex]
 \lambda = X\;,
\end{cases}
\end{equation}
the new terms (\ref{eq:deltaI0ex}) cancel the term $Z$ and replace it by
\begin{equation}
Z' = \alpha'\,\int\d^{D}x\,\sqrt{-g}\,e^{-\Phi}\,X\left(\nabla_{\mu}\Phi\,\nabla^{\mu}\Phi-\frac{1}{6}\,H^{2}-\frac{1}{4}\,\F_{\mu\nu}{}^{M}\,{\S_{M}}^{N}\,\F^{\mu\nu}{}_{N}\right),
\end{equation} 
which carries only products of first order derivatives.
In the same fashion, all the four-derivative terms carrying the leading two-derivative contributions from the field equations~\eqref{eq:eomGLd} can be transformed into terms carrying only products of first order derivatives. 
We may summarize the resulting replacement rules as
\begin{equation}
 \begin{aligned}
\label{eq:fieldredef0}
 \Box\Phi ~\longrightarrow~ {\cal Q}_{\Phi} &= \nabla_{\mu}\Phi\,\nabla^{\mu}\Phi-\frac{1}{6}\,H^{2}-\frac{1}{4}\,\F_{\mu\nu}^{M}\,{\S_{M}}^{N}\,\F^{\mu\nu}_{N}\;,\\[1ex]
 R_{\mu\nu} ~\longrightarrow~ {\cal Q}_{g\,\mu\nu} &= -\nabla_{((\mu}\nabla_{\nu))}\Phi+ \frac{1}{4}H_{\mu\nu}^{2} -\frac{1}{8}\,\Tr{\nabla_{\mu}\S\,\nabla_{\nu}\S} +\frac{1}{2}\,\F_{\mu\rho}^{M}\,{\S_{M}}^{N}\,{{\F_{\nu}}^{\rho}}_{N} \\
 &{} \quad\, -\frac{1}{D}\,g_{\mu\nu}\,\Big(\nabla_{\rho}\Phi\,\nabla^{\rho}\Phi-\frac{1}{6}\,H^{2}-\frac{1}{4}\,\F_{\rho\sigma}^{M}\,{\S_{M}}^{N}\,\F^{\rho\sigma}_{N}\Big)\;,\\[1ex]
  \nabla^{\mu}H_{\mu\rho\sigma} ~\longrightarrow~ {\cal Q}_{B\,\rho\sigma} &= \nabla^{\mu}\Phi\,H_{\mu\rho\sigma}\;,
  \\[1ex]
  \Box G_{mn}~\longrightarrow~{\cal Q}_{G\,mn} &= \nabla_{\mu}\Phi\,\nabla^{\mu} G_{mn} - \nabla_{\mu}G_{mp}\nabla^{\mu}G^{pq}G_{qn}-\frac{1}{2}\,H_{\mu\nu m}{H^{\mu\nu}}_{n} \\
 &{}\quad\, + \nabla_{\mu}B_{mp}G^{pq}\nabla^{\mu}B_{qn} + \frac{1}{2}\,F_{\mu\nu}^{(1)\,p}G_{pm} F^{(1)\,\mu\nu\,q}G_{qn} \;,\\[2ex]
 \Box B_{mn} ~\longrightarrow~ {\cal Q}_{B\,mn} &= \nabla_{\mu}\Phi\,\nabla^{\mu} B_{mn} - \nabla_{\mu}B_{mp}\nabla^{\mu}G^{pq}G_{qn} - G_{mp}\nabla^{\mu}G^{pq}\nabla_{\mu}B_{qn} \\
&{}\quad\, -\frac{1}{2}H_{\mu\nu m}F^{(1)\,\mu\nu\,p}G_{pn}+ \frac{1}{2} G_{mp}F^{(1)\,\mu\nu\,p}H_{\mu\nu n}\;,\\[1ex]
\nabla_{\mu} F^{(1)\,\mu\nu\,m} ~\longrightarrow~ {\cal Q}_{A^{(1)}}{}^{\nu\,m} &=\nabla_{\mu}\Phi\,F^{(1)\,\mu\nu\,m}+{H^{\mu\nu}}_{n}G^{np}\nabla_{\mu}B_{pq}G^{qm}  \\
&{}\quad\, +\frac{1}{2} H^{\mu\nu\rho}\,H_{\mu\rho\,n}G^{nm} - F^{(1)\,\mu\nu\,n}\nabla_{\mu}G_{np}G^{pm}\;,\\[1ex]
\nabla_{\mu} {H^{\mu\nu}}_{m} ~\longrightarrow~ {\cal Q}_{A^{(2)}}{}^{\nu}{}_{m} &= \nabla_{\mu}\Phi\,{H^{\mu\nu}}_{m}-{H^{\mu\nu}}_{n}\nabla_{\mu}G^{np}G_{pm}+\frac{1}{2} H^{\mu\nu\rho}\,F_{\mu\rho}^{(1)n}G_{nm}\;.
\end{aligned}
\end{equation}
Double parenthesis $((\dots))$ in the second line refer to traceless symmetrization.
The associated field redefinitions are collected in Tab.~\ref{tab:fieldredef}. As we will show in Sec.~\ref{sec:reduc4d}, 
all other four-derivative terms can be mapped into the terms listed in Tab.~\ref{tab:fieldredef} upon using partial 
integration and Bianchi identities.  

\begin{table}[t!]
\renewcommand{\arraystretch}{1.5}
\centering
 \begin{tabular}{c|c|c}
 Term in the action & Field redefinitions & Replacement \\\hline\hline
 \multirow{2}{*}{$\alpha'\,X\Box \Phi$} & \multirow{2}{*}{$\begin{cases}\delta\Phi = \dfrac{1}{2}\left(D+2\right)\,X \smallskip\\ \delta g_{\mu\nu} = g_{\mu\nu}\,X\end{cases}$} & \multirow{2}{*}{$\alpha'\,X\,{\cal Q}_{\Phi}$} \\
 &  & \\[1ex]
  \multirow{2}{*}{$\alpha'\,X^{\mu\nu}\,R_{\mu\nu}$}& \multirow{2}{*}{$\begin{cases} \delta g_{\mu\nu} = -X_{(\mu\nu)}-\dfrac{1}{D}\,g_{\mu\nu}\,X_{\rho}{}^{\rho}\\ \delta\Phi = -\dfrac{1}{D}\,X_{\mu}{}^{\mu}\end{cases}$}& \multirow{2}{*}{$\alpha'\,X^{\mu\nu}\,{\cal Q}_{g\,\mu\nu}$} \\
 &  & \\[1ex]
 $\alpha'\,X^{\mu\nu}\,\nabla^{\rho} H_{\rho\mu\nu}$ & $\delta B_{\mu\nu} = -2\,X_{[\mu\nu]}$ & $\alpha'\,X^{\mu\nu}\,{\cal Q}_{B\,\mu\nu}$ \\
 $\alpha'\,X^{mn}\,\Box G_{mn}$ & $\delta G^{mn} = 2\,X^{(mn)}$ & $\alpha'\,X^{mn}\,{\cal Q}_{G\,mn}$ \\
 $\alpha'\,X^{mn}\,\Box B_{mn}$ & $\delta B_{mn} = -2\,G_{mp}X^{[pq]}G_{qn}$ & $\alpha'\,X^{mn}\,{\cal Q}_{B\,mn}$ \\[1ex]
 \multirow{3}{*}{$\alpha'\,X_{\nu m}\,\nabla_{\mu}F^{(1)\mu\nu\,m}$}& \multirow{3}{*}{$\begin{cases} \delta A_{\mu}^{(1)\,m} = -X_{\mu n}G^{nm} \\ \delta A_{\mu m}^{(2)} = -X_{\mu n}G^{np}B_{mp} \\\delta B_{\mu\nu} = \left(A_{[\mu|m}^{(2)}X_{|\nu]n}^{\phantom{(1)}}G^{mn} - B_{mn}A_{[\mu}^{(1) n}X_{\nu]p}^{\phantom{(1)}}G^{mp}\right)\end{cases}$\hspace{-0.65cm} } & \multirow{3}{*}{$\alpha'\,X_{\nu m}\,{\cal Q}_{A^{(1)}}{}^{\nu\,m}$}\\
 &  & \\
 &  & \\[1ex]
 \multirow{2}{*}{$\alpha'\,X_{\nu}{}^{m}\,\nabla_{\mu}H^{\mu\nu}{}_{m}$}& \multirow{2}{*}{$\begin{cases} \delta A_{\mu m}^{(2)} = -X_{\mu}{}^{n}G_{nm} \\ \delta B_{\mu\nu} = A_{[\mu}^{(1) m}X_{\nu]}^{\ \,n}G_{mn}\end{cases}$}& \multirow{2}{*}{$\alpha'\,X_{\nu}{}^{m}\,{\cal Q}_{A^{(2)}}{}^{\nu}{}_{m}$} \\
 &  &
 \end{tabular}
 \caption{Replacement rules for the terms carrying the leading two-derivative contribution from the field equations descending from the two-derivative action~\eqref{eq:I0Gld} and associated field redefinitions. The explicit replacement rules are given in Eq.~\eqref{eq:fieldredef0}.}
 \label{tab:fieldredef}
\end{table}

\section{\texorpdfstring{$\boldsymbol{\Odd}$}{} invariant basis at order
 \texorpdfstring{$\boldsymbol{\alpha'}$}{alpha'}}
\label{sec:basis}

In this section, we present the construction of an explicit \Odd-invariant basis for the four-derivative terms in $D$ dimensions.
We discuss the general counting of independent terms for building an action upon modding out field redefinitions and
partial integrations. At order $\alpha'$ we find that the number of independent terms is 61 and coincides with the
number of terms that can be built from products of first order derivatives (and the Riemann tensor). 
We confirm the number by an explicit
construction of a 61-dimensional basis which we use subsequently in order to organize the result of the
explicit torus reduction.

\subsection{Counting independent terms}

Following the general discussion of field redefinition ambiguities of the last section, we first count the number of independent
terms modulo the two-derivative field equations (\ref{eq:fieldredef0}) and Bianchi identities. 
At this stage we do not yet restrict to Lorentz scalars, i.e.\ we keep all $D$-dimensional space-time indices uncontracted.
In a second step we will restrict to Lorentz scalars and account for the freedom of partial integration.
We start by defining the alphabet whose letters are the \Odd-invariant building blocks in the various matter sectors
(dilaton, scalars, vectors, 2-forms, metric) before identifying all possible symmetric words in these letters.
We only count manifestly \Odd and gauge invariant terms, i.e.\ neglect possible Chern-Simons and topological terms
which we will have to treat separately.

\paragraph{Dilaton}

The independent building blocks carrying the dilaton are given by powers of derivatives
\begin{equation}
{\cal B}_{{\rm dil}} = \left\{ \nabla_{((\mu_1} \dots \nabla_{\mu_n))} \Phi  
\;\big|\; n\in \mathbb{N}^*, \{\mu_1, \dots, \mu_{n}\}
\right\}
\;,
\label{dil}
\end{equation}
with the double parentheses $((\dots))$ indicating traceless symmetrization in order to divide out
field equations.
We may encode the set of letters (\ref{dil}) into a partition function
\begin{equation}
{\cal Z}_{{\rm dil}} = 
u\left(
\frac{1-q^2}{(1-q)^{{\rm\bf v}_D}} - 1\right)
\;,
\label{Zdil}
\end{equation}
such that upon expanding (\ref{Zdil}) into a series in $q$ every term represents a letter with
exponents counting the number of derivatives.
We have also added  a factor $u$ to keep track of the dilaton power when combining (\ref{Zdil})
with  the other building blocks of the theory.
We use the notation
\begin{align}
(1-q)^{{\rm\bf v}_D}  &= 1 - q\,{\rm\bf v}_D
+q^2\,({\rm\bf v}_D\otimes {\rm\bf v}_D)_{\rm alt} - \dots
\;,\nonumber\\
\frac{1}{(1-q)^{{\rm\bf v}_D}} &= 1 + q\,{\rm\bf v}_D
+q^2\,({\rm\bf v}_D\otimes {\rm\bf v}_D)_{\rm sym} + \dots
\;,
\label{vDseries}
\end{align}
with the ${\rm SO}(D)$ vectorial representation ${\rm\bf v}_D$,
in order to describe the tower of traceless symmetrized vectors. $({\bf v}_D\otimes {\bf v}_D)_{\rm alt}$ is the antisymmetric tensor product of two ${\rm SO}(D)$ vectors,

\paragraph{Coset scalars}

The scalar fields parametrize the ${\rm SO}(d,d)/({\rm SO}(d)\times {\rm SO}(d))$ matrix 
${\cal H}_{MN}$.
In order to directly implement all constraints deriving from the coset structure, it is convenient to turn to the vielbeins
\begin{equation}
{\cal H}_{MN} = E_M{}^A \,\delta_{AB}\,E_N{}^B
\quad\Longrightarrow\quad
\partial_\mu {\cal H} {\cal H}^{-1}
=
2\,E\,P_\mu\,E^{-1}
\;,
\label{vielbein}
\end{equation}
with the coset currents defined by
\begin{equation}
E^{-1}\partial_\mu E = Q_\mu + P_\mu \in \mathfrak{k} \oplus  \mathfrak{p} = \mathfrak{so}(d,d)
\;,
\end{equation}
where $\mathfrak{k}=\mathfrak{so}(d)\oplus \mathfrak{so}(d)$ and $\mathfrak{p}$ is its
(non-compact) orthogonal complement.
In terms of the currents $Q_\mu$ and $P_\mu$, global ${\rm SO}(d,d)$ invariance is ensured, 
and the counting problem reduces to identifying combinations that are invariant under local 
${\rm SO}(d)\times {\rm SO}(d)$ transformations, i.e.\ built from $P_\mu$'s and
covariant derivatives $D_\mu = \partial_\mu +{\rm ad}_{Q_\mu}$\,.
Moreover, we have integrability conditions
\begin{equation}
[D_\mu,D_\nu] = Q_{\mu\nu} \propto [P_\mu,P_\nu] \;,\qquad
D_{[\mu} P_{\nu]} = 0
\;,
\label{int}
\end{equation}
and field equations with leading second order term $D^\mu P_\mu$
which implies that a basis of on-shell independent combinations is given by
\begin{equation}
{\cal B}_P = \left\{ \nabla_{((\mu_1} \dots \nabla_{\mu_n} P_{\mu_{n+1}))}  \in \mathfrak{p}
\;\big|\; n\in \mathbb{N}, \{\mu_1, \dots, \mu_{n+1}\}
\right\}
\;,
\label{B}
\end{equation}
counted by the partition function 
\begin{equation}
{\cal Z}_{P} = p\left(
\frac{1-q^2}{(1-q)^{{\rm\bf v}_D}} - 1\right)
\;,
\label{alphaP}
\end{equation}
with the charge $p$ introduced to count the power of $P_\mu$'s.
It remains to count the independent ${\rm SO}(d)\times {\rm SO}(d)$ invariant single-trace combinations 
in the letters (\ref{B}). With $P_\mu$ transforming in the $(\boldsymbol{d},\boldsymbol{d})$ representation of ${\rm SO}(d)\times {\rm SO}(d)$,
this amounts to counting ordered monomials and dividing out transpositions and cyclic shifts of even length\footnote{
In this counting, we neglect all the identities induced by the finite size ($2d$) of the
${\rm SO}(d,d)$ matrices, i.e.\ formally we count for $d=\infty$\,.}.
The result then follows from Polya's counting theorem \cite{PolyaRead} as
\begin{equation}
{\cal Z}_{\rm sing. trace} 
 =
-\frac12\,\sum_{n} \frac{\varphi(n)}{n}\, 
{\rm log}\left(1-{\cal Z}_{{\cal P},n}^{2}\right) 
+\frac{{\cal Z}_{{\cal P},2}}{2\,(1- {\cal Z}_{{\cal P},2})}
\;,
\qquad
\label{cyclic}
\end{equation}
with Euler's totient function $\varphi(n)$
and ${\cal Z}_{{\cal P},n}={\cal Z}_{{\cal P}}(p^n,q^n)$\,.

\paragraph{Vectors}

The (manifestly) gauge invariant building blocks in terms of the vector field are obtained by
derivatives of its field strength subtracting Bianchi identities and contractions proportional to the field equations
\begin{equation}
{\cal B}_{\cal F} = \left\{ \nabla_{((\mu_1} \dots \nabla_{\mu_n))} {\cal F}_{\nu_1\nu_2}{}^M  
- \mbox{traces \& Bianchi}
\;\big|\; n\in \mathbb{N}
\right\}
\;,
\label{vec}
\end{equation}
counted by the partition function (see e.g. Ref.~\cite{Morales:2004xc})
\begin{equation}
{\cal Z}_{{\cal F}}= 
  \sum_{n=0}^\infty \bigg(\tikz[baseline=-0.9ex]{ \draw (0,0) node {$\tinyyoung{\nu_{1}&\mu_{1}& & &\mu_{n}\cr \nu_{2}\cr}$};}- {\rm traces}\bigg)
=
f\left(
\frac{1}{q}-\frac{1-{\rm\bf v}_D\, q\,(1-q^{2})-q^4}{q\,(1-q)^{{\rm\bf v}_D}}
\right)
\;,
\label{ZF}
\end{equation}
where $f$ is a charge for the powers of $\F_{\mu\nu}{}^{M}$\,. However, the letters (\ref{vec}) are not \Odd singlets but rather carry a fundamental vector index.
\Odd invariant combinations are built from bilinears of Eq.~(\ref{vec}) with the two \Odd vector indices contracted
by products of the \Odd invariant $\eta_{MN}$, the scalar matrix ${\cal H}_{MN}$, and its derivatives\,.
This is most conveniently counted by using the vielbeins (\ref{vielbein}) to convert the \Odd indices of Eq.~(\ref{vec}) 
into ${\rm SO}(d)\times{\rm SO}(d)$ indices, such that the flattened field strength ${\cal F}_{\mu\nu}{}^M E_M{}^A$ decomposes
into $(\boldsymbol{d},\boldsymbol{1})\boldsymbol{\oplus} (\boldsymbol{1},\boldsymbol{d})$ contributions which we denote by ${\cal F}_L$ and ${\cal F}_R$, respectively.
The flattened letters (\ref{vec}) are then contracted out by arbitrary chains of letters from Eq.~(\ref{B}). This gives rise
to three different types of terms
\begin{align}
& (\nabla\dots\nabla {\cal F}_L) \,(\mbox{even chain of } \nabla\dots\nabla P) \,(\nabla\dots\nabla {\cal F}_L)\;,\nonumber\\
& (\nabla\dots\nabla {\cal F}_R) \,(\mbox{even chain of } \nabla\dots\nabla P) \,(\nabla\dots\nabla {\cal F}_R)\;,\nonumber\\
& (\nabla\dots\nabla {\cal F}_L) \,(\mbox{odd chain of } \nabla\dots\nabla P)\,( \nabla\dots\nabla {\cal F}_R)\;.
\label{FF3}
\end{align}
Upon taking into account the reflection symmetries of the first two chains, the counting of
\Odd invariant building blocks in the vector sector yields
\begin{align}
{\cal Z}_{{\cal FF}} &=
\frac12\left(
\frac{{\cal Z}_{{\cal F},2}}{1-{\cal Z}_{{\cal P},2}}
+
\frac{{\cal Z}^2_{\cal F}}{1-{\cal Z}^2_{\cal P}}
\right)+
\frac12\left(
\frac{{\cal Z}_{{\cal F},2}}{1-{\cal Z}_{{\cal P},2}}
+
\frac{{\cal Z}^2_{\cal F}}{1-{\cal Z}^2_{\cal P}}
\right)
+
{\cal Z}_{{\cal F}} \frac{{\cal Z}_{{\cal P}}}{1-{\cal Z}_{{\cal P}}^2} {\cal Z}_{{\cal F}}
\nonumber\\[1ex]
&=
 \frac{{\cal Z}^2_{{\cal F}}}{1-{\cal Z}_{{\cal P}}}+\frac{{\cal Z}_{{\cal F},2}}{{1-{\cal Z}_{{\cal P},2}}}
 \;.
 \label{ZFF}
\end{align}

\paragraph{Two-form}

Similarly, the independent (manifestly gauge-invariant) building blocks carrying the 2-form $B_{\mu\nu}$ 
are counted by powers of derivatives on the field strength $H_{\mu\nu\rho}$ upon 
subtracting Bianchi identities and contractions proportional to the field equations
\begin{equation}
{\cal B}_H = \left\{ \nabla_{((\mu_1} \dots \nabla_{\mu_n))} H^{\nu_1\nu_2\nu_3}  - \mbox{traces \& Bianchi}
\;\big|\; n\in \mathbb{N}
\right\}
\;,
\label{two}
\end{equation}
giving rise to a partition function
\begin{align}
{\cal Z}_{H} &= 
  \sum_{n=0}^\infty \bigg(\tikz[baseline=-0.9ex]{ \draw (0,0) node {$\tinyyoung{\nu_{1}&\mu_{1}& & &\mu_{n}\cr \nu_{2}\cr \nu_{3} \cr}$};}-{\rm traces}\bigg)
\nonumber\\
&=
h\left(
 \frac{1-q^6-q\,(1-q^4)\,{\bf v}_D+q^2\,(1-q^2)\,({\bf v}_D\otimes {\bf v}_D)_{\rm alt}}{q^2\,(1-q)^{{\rm\bf v}_D}}
  -\frac{1}{q^2}
 \right)
\;,
\label{ZH}
\end{align}
where $h$ is a charge for the powers of $H_{\mu\nu\rho}$\,.

\paragraph{Metric}

For the external metric $g_{\mu\nu}$, we count derivatives of its Weyl tensor $C_{\nu_1\nu_2\nu_3\nu_4}$,
subtracting traces and Bianchi identities, giving rise to the letters
\begin{equation}
{\cal B}_C = \left\{ \nabla_{((\mu_1} \dots \nabla_{\mu_n))} C_{\nu_1\nu_2\nu_3\nu_4}  - \mbox{traces \& Bianchi}
\;\big|\; n\in \mathbb{N}
\right\}
\;,
\label{weyl}
\end{equation}
which are counted as
\begin{align}
{\cal Z}_{C}&= 
  \sum_{n=0}^\infty \bigg(\tikz[baseline=-0.9ex]{ \draw (0,0) node {$\tinyyoung{\nu_{1}& \nu_{2} & \mu_{1}& & &\mu_{n}\cr \nu_{3} & \nu_{4} \cr}$};}-{\rm traces}\bigg)
\nonumber\\
&=
c\left(
 \frac{q\,(1-q^2)\,({\bf v}_D\otimes{\bf v}_D)_{\rm sym}-(1-q^4)\,{\bf v}_D }{q\,(1-q)^{{\rm\bf v}_D}}
 +({\bf v}_D\otimes{\bf v}_D)_{\rm alt}+\frac{1}{q}\,{\bf v}_D
\right)
\;,
\label{ZC}
\end{align}
where $c$ is a charge for the powers of the Weyl tensor (or equivalently, the Riemann tensor).

\subsection{Space-time singlets and partial integration}

Putting everything together, we have identified the manifestly 
\Odd and gauge invariant building blocks in the various sectors,
\begin{equation}
{\cal Z}_{0} =
{\cal Z}_{\rm dil}
+{\cal Z}_{{H}}
+{\cal Z}_{{C}}
+{\cal Z}_{\rm sing. trace} 
+{\cal Z}_{{\cal FF}}
\;,
\label{Z0}
\end{equation}
with the different terms defined in Eqs.~(\ref{Zdil}), (\ref{ZH}), (\ref{ZC}), (\ref{cyclic}), and (\ref{ZFF}), respectively.
>From these objects, we can construct the most general \Odd and gauge invariant
terms as arbitrary polynomials in the letters of Eq.~(\ref{Z0}), counted as
\begin{equation}
{\cal Z}_{\rm inv}   = {\rm exp} \bigg[ \sum_k \frac1{k}\, {\cal Z}_{0,k}
\bigg]
\;.
\label{all_terms}
\end{equation}

So far, we have been counting combinations 
in all possible $\SO(D)$ representations,
without restricting to $\SO(D)$ Lorentz scalars.
In order to count the independent space-time actions, we
first project ${\cal Z}_{\rm inv}$ to Lorentz scalars.
Next, in order to subtract the ambiguities from partial integrations,
we extract from ${\cal Z}_{\rm inv}$ all possible $\SO(D)$ vectors ${\cal J}_\mu$
each of which gives rise to an ambiguity $\d \,\ast {\cal J}$ of the
space-time Lagrangian.
On the other hand, currents with (off-shell) vanishing divergence $\d \,\ast {\cal J}=0$
do not define ambiguities, these are of the form ${\cal J}=\ast \,\d \ast {\cal J}_2$
for a 2-form ${\cal J}_2$\,. Unless $\ast {\cal J}_2$ is of vanishing divergence thus
defined by a 3-form ${\cal J}_3$, etc.
To summarize, a basis of independent space-time Lagrangians,
after dividing out the freedom of partial integrations,
is given by
\begin{equation}
{\cal Z}_{\rm Lag} =
{\cal Z}_{\rm inv}\,(1-uq)^{{\rm\bf v}_D}  \Big|_{{\rm SO}(D) \;{\rm singlets}}
\;\;,
\label{all_Lag}
\end{equation}
in the notation of Eq.~(\ref{vDseries})\footnote{
Here, we have inserted a dilaton charge $u$, since 
all terms carry a global dilaton power $e^{-\Phi}$ such that partial integration
brings in an extra dilaton derivative.}.

\subsection{Some examples}

\paragraph{Evaluation in $\boldsymbol{D=10}$}
As a first test of the counting formula (\ref{all_Lag}), we may evaluate it
to order $\alpha'$ in $D=10$ dimensions, i.e.\ for $d=0$, upon truncating
out the vector and scalar sector which do not exist at $d=0$\,.
Then, in Eq.~(\ref{Z0}) only the contributions from metric, two-form and 
dilaton are taken into account.
Evaluating Eq.~(\ref{all_Lag}) gives rise to the following types of terms at the
four-derivative order
\begin{equation}
\Big\{R^2\; [1]
\,,\;
\nabla^2 H^2\; [1]
\,,\;
R \,H^2\;\; [1]
\,,\;
H^4\; [3]
\,,\;
H^2 \nabla^2 \Phi\; [1]
\,,\;
\nabla^2 \Phi\,\nabla^2 \Phi\; [1] \Big\}\;,
\label{q4D10}
\end{equation}
where the multiplicities $[n]$ indicate the number of independent terms of the same type. 
This precisely reproduces the counting from Ref.~\cite{Metsaev:1987zx} (c.f.\ their Eq.~(2.36)).
Let us recall that our counting only includes manifestly gauge invariant terms, 
so it does not account for the possible ten-dimensional gravitational Chern-Simons couplings.

\paragraph{Evaluation in $\boldsymbol{D=1}$} Upon reduction to only one dimension,
we can evaluate the counting formulas to all orders in closed form. In particular ${\cal Z}_H={\cal Z}_{{\cal F}}=0$,
while
\begin{equation}
{\cal Z}_{\rm dil} = u q\;,\quad
{\cal Z}_{P} = pq \;,\quad
{\cal Z}_{\rm sing. trace} = 
  \frac{p^2q^2}{1-p^2q^2}
\;,
\end{equation}
and
\begin{equation}
{\cal Z}_{C} = -c\,q^2~\longrightarrow~ -p^2 q^2
\;,
\end{equation}
reflecting the fact that in $D=1$ the Einstein equations pose a constraint on the energy-momentum tensor.
For Eqs.~(\ref{Z0}), (\ref{all_terms}), we thus find
\begin{equation}
{\cal Z}_0 = 
  \frac{p^4q^4}{1-p^2q^2}+ uq
  \qquad\Longrightarrow\qquad
  {\cal Z}_{\rm inv} =
\prod_{n>1} \frac1{1-p^{2n}q^{2n}}   \times \frac1{1-uq}
  \;,
\end{equation}
upon removing total derivatives (\ref{all_Lag}) thus
\begin{equation}
{\cal Z}_{\rm Lag} =
(1-qu)\,{\cal Z}_{\rm inv} = \prod_{n>1} \frac1{1-p^{2n}q^{2n}}
\;,
\end{equation}
which precisely reproduces the counting from Ref.~\cite{Hohm:2019jgu}.

\subsection{Basis at order \texorpdfstring{$\alpha'$}{alpha'}}
\label{subsec:basis}

Evaluating the counting formula (\ref{all_Lag}) in generic dimension $D$ we infer that
at order $\alpha'$ there are 61 independent manifestly \Odd invariant four-derivative terms.
While the general counting only determines the number of independent terms without selecting
a particular basis, it turns out that at order $\alpha'$ there is a distinguished explicit basis which
is built from polynomials in terms carrying only first order derivatives (and the Riemann tensor).
Indeed, truncating the partition functions (\ref{Zdil}), (\ref{ZH}), (\ref{ZC}), (\ref{alphaP}), (\ref{ZF})
to first order in derivatives, we may count from Eq.~(\ref{all_terms}) the number of independent terms
that carry first derivatives only, and find precisely 61 terms at order $\alpha'$.\footnote{
At order $\alpha'{}^2$ this pattern breaks down. The general counting (\ref{all_Lag}) reveals 1817 
independent terms at order $\alpha'{}^2$
whereas there are only 1212 independent polynomials that can be constructed in terms of first order derivatives.
This general case differs from the situation encountered in the reduction to $D=1$ dimensions
where one can always find a basis carrying no more than first-order time derivatives
\cite{Hohm:2015doa}.
}

The basis at order $\alpha'$ can thus be given in terms of polynomials in
$R_{\mu\nu\rho}{}^\sigma$, $H_{\mu\nu\rho}$, ${\cal F}_{\mu\nu}{}^M$, $\nabla_\mu {\S}_M{}^N$, and
$\nabla_\mu \Phi$.
Schematically, its elements take the form
\begin{align}
&
\Big\{
R^2\; [1]
\,,\;
H^4\; [3]
\,,\;
(\nabla\Phi)^4\; [1]
\,,\;
(\nabla\S)^4 \;[5]
\,,\;
\F^4 \;[12] 
\,,\;
R\,H^2 \;[1]
\,,\;
R\,\F^2 \;[2]
\,,\;
\nonumber\\
&{}\quad
H^2\,(\nabla\Phi)^2\; [2]
\,,\;
H^2\,(\nabla\S)^2 \;[2]
\,,\;
H^2\,\F^2 \;[8]
\,,\;
(\nabla\Phi)^2\,(\nabla\S)^2 \; [2]
\,,\;
(\nabla\Phi)^2\,\F^2 \; [4]
\,,\;
\nonumber\\
&{}\quad
(\nabla\S)^2\,\F^2 \; [10]
\,,\;
H\,\nabla\Phi\,\F^2 \;[2]
\,,\;
H\,\nabla\S\,\F^2 \;[3]
\,,\;
\nabla\Phi\,\nabla\S\,\F^2 \; [3]
\Big\}
\;.
\label{basisalpha1}
\end{align}
We give the explicit expressions for all the basis elements in App.~\ref{app:order1basis}.
In the following we will exhibit \Odd invariance of the dimensionally reduced action by expanding
the reduced action in the basis~(\ref{basisalpha1}).


\section{Compactification of the four-derivative action}
\label{sec:reduc4d}


The first order $\alpha'$ extension of the action of the bosonic string~\eqref{eq:I0D+d} 
has been known for some time~\cite{Metsaev:1987zx} 
and is given up to field redefinitions by
\begin{equation}
\begin{aligned}
\label{eq:alpha'D+d}
\widehat{I}_{1} = \frac{1}{4}\,\alpha'
\int \d^{D+d}X\,\sqrt{-\hat{g}}\,e^{-\hat{\phi}}\,\Big(&\,\hat{R}_{\hat\mu\hat\nu\hat\rho\hat\sigma}\hat{R}^{\hat\mu\hat\nu\hat\rho\hat\sigma} -\frac{1}{2}\, \hat{H}^{\hat\mu\hat\nu\hat\lambda}\hat{H}^{\hat\rho\hat\sigma}{}_{\hat\lambda}\, \hat{R}_{\hat\mu\hat\nu\hat\rho\hat\sigma}  -\frac{1}{8}\,\hat{H}^{2}_{\hat\mu \hat\nu}\hat{H}^{2\,\hat\mu \hat\nu} \\
& +\frac{1}{24}\,\hat{H}_{\hat\mu \hat\nu \hat\rho}\hat{H}^{\hat\mu}{}_{\hat\sigma}{}^{\hat\lambda}\hat{H}^{\hat\nu}{}_{\hat\lambda}{}^{\hat\tau}\hat{H}^{\hat\rho}{}_{\hat\tau}{}^{\hat\sigma}\Big).
\end{aligned}
\end{equation}
In this section, we compactify separately all of its terms on a $d$-torus, 
using the ans\"atze~\eqref{eq:ansatzg} and~\eqref{eq:ansatzB}. 
We fix the freedom of partial integration and possible field redefinitions,
by converting all terms into polynomials of first order derivatives (and the Riemann tensor).
To do so, we systematically use partial integration and Bianchi identities to bring all terms carrying 
second order derivatives into a form corresponding to the first column of Tab.~\ref{tab:fieldredef},
which can then be converted to the desired form by means of field redefinitions
as discussed in Sec.~\ref{sec:GLdredef}.
In the next section, we then compare the result to the 
\Odd basis of Sec.~\ref{subsec:basis}.

The reduction of the three-form field strength $\hat{H}_{\hat\mu\hat\nu\hat\rho}$ is given in Eq.~\eqref{eq:Hreduced}. 
For the reduction of the Riemann tensor, we follow the results of Ref.~\cite{Bao:2007fx}, 
and give the lower-dimensional components in flat indices as
\begin{align}
\label{eq:Rreduced}  
  \hat{R}_{\alpha\beta\gamma\delta} &= R_{\alpha\beta\gamma\delta} -\dfrac{1}{2}\left[-G_{mn}F^{(1)\,m}_{\alpha[\gamma}F^{(1)\,n}_{\delta]\beta}+G_{mn}F^{(1)\,m}_{\alpha\beta}F^{(1)\,n}_{\gamma\delta}\right]
  \;,\nonumber\\
  \hat{R}_{\alpha\beta\gamma d} &= \left[\nabla_{[\alpha}F^{(1)\,p}_{\beta]\gamma}-\dfrac{1}{2}\left(G_{mn}\nabla_{[\alpha}G^{np}F^{(1)\,m}_{\beta]\gamma}-F^{(1)\,m}_{\alpha\beta}G_{mn}\nabla_{\gamma}G^{np}\right)\right]\,E_{p\,d}
  \;,\nonumber\\
  \hat{R}_{\alpha\beta cd} &= \dfrac{1}{2}\left[F^{(1)\,\gamma\,m}_{\alpha}F^{(1)\,q}_{\gamma\beta}-\nabla_{\alpha}G^{mn}G_{np}\nabla_{\beta}G^{pq}\right]\,E_{m\,[c}E_{\vert q\vert\,d]}
  \;,\nonumber\\
  \hat{R}_{\alpha b\gamma d} &= \dfrac{1}{4}\left[2\,\nabla_{\alpha}\nabla_{\gamma}G^{mq}-2\,\nabla_{\alpha}G^{mn}G_{np}\nabla_{\gamma}G^{pq}-\nabla_{\gamma}G^{mn}G_{np}\nabla_{\alpha}G^{pq}+F^{(1)\,\,m}_{\gamma\varepsilon}F^{(1)\,\varepsilon\,q}_{\alpha}\right]\,E_{m\,b}E_{q\,d}
  \;,\nonumber\\
  \hat{R}_{a b\gamma d} &= -\dfrac{1}{2}F^{(1)\,\,m}_{\gamma\varepsilon}\nabla^{\varepsilon}G^{np}\,E_{m\,[a}E_{\vert n\vert\,b]}E_{p\,d}
  \;,\nonumber\\
  \hat{R}_{abcd} &= -\dfrac{1}{2}\nabla_{\varepsilon}G^{mn}\nabla^{\varepsilon}G^{pq}E_{m\,a}E_{p\,b}E_{n\,[c}E_{\vert q\vert\,d]}
  \;.
\end{align}

\subsection{Reduction of the various terms}

We reduce the action (\ref{eq:alpha'D+d}) term by term.

\paragraph{Reduction of \texorpdfstring{$\boldsymbol{\hat{H}_{\hat\mu\hat\nu}^{2}\hat{H}^{2\,\hat\mu\hat\nu}}$}{H2H2}}
Upon compactification, we obtain
\begin{equation}
 \begin{aligned}
\hat{H}_{\hat\mu\hat\nu}^{2}\hat{H}^{2\,\hat\mu\hat\nu} =~& H^{2}_{\mu\nu}H^{2\,\mu\nu} + 
4 \,H^{2}_{\mu\nu}{H^{\mu}}_{\rho m}H^{\nu\rho m} - 2\,H^{2\,\mu\nu}H_{\mu\nu m}{H_{\nu}}^{nm} + 4\, H_{\mu\nu{m}}H^{\mu\rho m}H^{\nu\sigma n}H_{\rho\sigma n} \\
& -4\, H^{\mu\rho p}{H^{\nu}}_{\rho p}H_{\mu mn}{H_{\nu}}^{nm} + H_{\mu mn}{H_{\nu}}^{mn}{H^{\mu}}_{pq}H^{\nu qp} +2\, H^{\mu\nu\lambda}{H^{\rho\sigma}}_{\lambda}H_{\mu\nu m}{H_{\rho\sigma}}^{m} \\
& +8\,H^{\mu\nu\rho}H_{\mu\nu m}{H_{\sigma}}^{mn}{H^{\sigma}}_{\rho n} - 8\, H_{\mu\rho m} H^{\mu mn}{H^{\nu}}_{np}{H_{\nu}}^{\rho p} + H_{\mu\nu m}{H_{\rho\sigma}}^{m} H^{\mu\nu n}{H^{\rho\sigma}}_{n}\\
&-4\, H_{\mu\nu m}{H_{\rho}}^{mn}{H^{\rho}}_{np}H^{\mu\nu p} + 4\, H_{\mu mn}H^{\nu np} H_{\nu pq}H^{\mu qm} + 4\,H_{mnp}H_{\mu\nu\rho}H^{mn\rho}H^{\mu\nu p} \\
& +8\, H_{mnp}H_{\mu\nu q}H^{\mu pq}H^{\nu mn}+2\,H_{mnp}H_{\mu qr}H^{\mu mn}H^{pqr}+2\,H_{mnp}H^{mnq} H_{\mu\nu}{}^{p}H^{\mu\nu q} \\
& +4\,H_{mnp}H^{mnq}H^{\mu pr}H_{\mu qr} + H_{mnr}H^{mns}H_{pqs}H^{pqr} .                    
\end{aligned}
\end{equation}
Using Eq.~\eqref{eq:Hreduced}, this takes the form
\begin{equation}
\begin{aligned}
\label{eq:H2H2reduced}
\hat{H}_{\hat\mu\hat\nu}^{2}\hat{H}^{2\,\hat\mu\hat\nu}& = H^{2}_{\mu\nu}H^{2\,\mu\nu} +4\,\Tr{\nabla_{\mu}BG^{-1}\nabla^{\mu}BG^{-1}\nabla_{\nu}BG^{-1}\nabla^{\nu}BG^{-1}}   \\
&+ \Tr{\nabla_{\mu}BG^{-1}\nabla_{\nu}BG^{-1}}\Tr{\nabla^{\mu}BG^{-1}\nabla^{\nu}BG^{-1}} + H_{\mu\nu m}G^{mn}H_{\rho\sigma n} {H^{\mu\nu}}_{p}G^{pq}{H^{\rho\sigma}}_{q}  \\
& +4\, H_{\mu\nu m}G^{mn}{H^{\mu\rho}}_{n} H_{\rho\sigma p}G^{pq}{H^{\nu\sigma}}_{q} -2\,H^{2\,\mu\nu}\Tr{\nabla_{\mu}BG^{-1}\nabla_{\nu}BG^{-1}} \\
& +4 \,H^{2\,\mu\nu}H_{\mu\rho m} G^{mn}{{H_{\nu}}^{\rho}}_{n} -4\,\Tr{\nabla^{\mu}BG^{-1}\nabla^{\nu}BG^{-1}} H_{\mu\rho m} G^{mn}{{H_{\nu}}^{\rho}}_{n} \\
& +2\,H^{\mu\nu\lambda}{H^{\rho\sigma}}_{\lambda} H_{\mu\nu m}G^{mn}H_{\rho\sigma n}-8\,H_{\mu\rho m}\left(G^{-1}\nabla^{\mu}BG^{-1}\nabla_{\nu}BG^{-1}\right)^{mn}{H^{\nu\rho}}_{n} \\
&-8\,H^{\mu\nu\rho}H_{\mu\nu m}\left(G^{-1}\nabla^{\sigma}BG^{-1}\right)^{mn}H_{\rho\sigma n}-4\,H_{\mu\nu m}\left(G^{-1}\nabla_{\rho}BG^{-1}\nabla^{\rho}BG^{-1}\right)^{mn}{H^{\mu\nu}}_{n}\;,
\end{aligned}
\end{equation}
where all terms carry first order derivatives only, i.e.\ are already of the desired form.

\paragraph{Reduction of \texorpdfstring{$\boldsymbol{\hat{H}_{\hat\mu \hat\nu \hat\rho}\hat{H}^{\hat\mu}{}_{\hat\sigma}{}^{\hat\lambda}\hat{H}^{\hat\nu}{}_{\hat\lambda}{}^{\hat\tau}\hat{H}^{\hat\rho}{}_{\hat\tau}{}^{\hat\sigma}}$}{HHHH}}
Upon compactification, we obtain
\begin{equation}
\begin{aligned}
\hat{H}_{\hat\mu \hat\nu \hat\rho}\hat{H}^{\hat\mu}{}_{\hat\sigma}{}^{\hat\lambda}&\hat{H}^{\hat\nu}{}_{\hat\lambda}{}^{\hat\tau}\hat{H}^{\hat\rho}{}_{\hat\tau}{}^{\hat\sigma}= H_{\mu\nu\rho}H^{\mu}{}_{\sigma}{}^{\lambda}H^{\nu}{}_{\lambda}{}^{\tau}H^{\rho}{}_{\tau}{}^{\sigma} + 6\,H^{\mu\nu\lambda}{H^{\rho\sigma}}_{\lambda}H_{\mu\rho m}{H_{\nu\sigma}}^{m} \\
& -12\, H^{\mu\nu\rho}H_{\mu\sigma m}{H_{\nu}}^{mn}{{H_{\rho}}^{\sigma}}_{n} +4\, H^{\mu\nu\rho}H_{\mu mn}{H_{\nu}}^{np}{H_{\rho p}}^{m} + 3\,H_{\mu\nu m}{H_{\rho\sigma}}^{m}{H^{\mu\rho}}_{n}H^{\nu\sigma n}  \\
&-12\, H_{\mu\nu m}{H_{\rho}}^{mn}{H^{\nu}}_{np}H^{\mu\rho p} +3\,H_{\mu mn}{H_{\nu}}^{np}{H^{\mu}}_{pq}H^{\nu pm} +4\,H_{mnp}H^{m}{}_{\mu\nu}H_{\rho}{}^{\mu n}H^{\rho\nu p}\\
& +12\, H_{mnp}H^{\mu mq}H^{\nu n}{}_{q}H_{\mu\nu}{}^{p} +6\,H_{mnr}H_{pq}{}^{r}H^{mp}{}_{\mu}H^{nq\mu}+H_{mnp}H^{m}{}_{q}{}^{r}H^{n}{}_{r}{}^{s}H^{p}{}_{s}{}^{q}  \;.
\end{aligned}
\end{equation}
Using Eq.~\eqref{eq:Hreduced}, this takes the form
\begin{equation}
\begin{aligned}
\label{eq:HHHHreduced}
\hat{H}_{\hat\mu \hat\nu \hat\rho}&\hat{H}^{\hat\mu}{}_{\hat\sigma}{}^{\hat\lambda}\hat{H}^{\hat\nu}{}_{\hat\lambda}{}^{\hat\tau}\hat{H}^{\hat\rho}{}_{\hat\tau}{}^{\hat\sigma}= H_{\mu\nu\rho}H^{\mu}{}_{\sigma}{}^{\lambda}H^{\nu}{}_{\lambda}{}^{\tau}H^{\rho}{}_{\tau}{}^{\sigma} +3\,H_{\mu\nu m}G^{mn}H_{\rho\sigma n} {H^{\mu\rho}}_{p}G^{pq}{H^{\nu\sigma}}_{q} \\
&  + 6\,H^{\mu\nu\lambda}{H^{\rho\sigma}}_{\lambda}H_{\mu\rho m}G^{mn}H_{\nu\sigma n} +3\,\Tr{\nabla_{\mu}BG^{-1}\nabla_{\nu}BG^{-1}\nabla^{\mu}BG^{-1}\nabla^{\nu}BG^{-1}}\\
& -12\,H^{\mu\nu\rho}H_{\mu\sigma m}\left(G^{-1}\nabla_{\nu}BG^{-1}\right)^{mn}{{H_{\rho}}^{\sigma}}_{n} -12\, H_{\mu\nu m}\left(G^{-1}\nabla_{\rho}BG^{-1}\nabla^{\nu}BG^{-1}\right)^{mn}{H^{\mu\rho}}_{n}\\
&+4\,H^{\mu\nu\rho}\Tr{\nabla_{\mu}BG^{-1}\nabla_{\nu}BG^{-1}\nabla_{\rho}BG^{-1}}\;,
\end{aligned}
\end{equation}
where again all terms carry first order derivatives only, i.e.\ are already of the desired form.

\paragraph{Reduction of \texorpdfstring{$\boldsymbol{\hat{R}_{\hat\mu\hat\nu\hat\rho\hat\sigma}\hat{R}^{\hat\mu\hat\nu\hat\rho\hat\sigma}}$}{RR}}
\label{sec;reducRR}
Splitting the $D+d$ indices $\hat\mu$ as $\hat\mu\rightarrow\{\mu, m\}$, we obtain
\begin{equation}
\begin{aligned}
\hat{R}_{\hat\mu\hat\nu\hat\rho\hat\sigma}\hat{R}^{\hat\mu\hat\nu\hat\rho\hat\sigma} =~& \hat{R}_{\mu\nu\rho\sigma}\hat{R}^{\mu\nu\rho\sigma}+4\,\hat{R}_{\mu\nu\rho m}\hat{R}^{\mu\nu\rho m} + 2 \,\hat{R}_{\mu\nu mn}\hat{R}^{\mu\nu mn}  \\
&+4\,\hat{R}_{\mu m\nu n}\hat{R}^{\mu m\nu n}+4\,\hat{R}_{mn\mu p}\hat{R}^{mn\mu p}+\hat{R}_{mnpq}\hat{R}^{mnpq}\;.
\end{aligned}
\end{equation}
Upon using Eq.~\eqref{eq:Rreduced}, the reduction of the first term of the action~\eqref{eq:alpha'D+d} then yields
\begin{align}
\label{eq:RRreducedorder2}
&\!\!\!\!\!\!
\frac{\alpha'}{4} \int\d^{D+d}X\,\sqrt{-\hat{g}}\,e^{-\hat{\phi}} \,\hat{R}_{\hat\mu\hat\nu\hat\rho\hat\sigma}\hat{R}^{\hat\mu\hat\nu\hat\rho\hat\sigma}  ~\longrightarrow
\nonumber \\
&\frac{\alpha'}{4}
\int\d^{D}x\,\sqrt{-g}\,e^{-\Phi} \,\Big[
R_{\mu\nu\rho\sigma}R^{\mu\nu\rho\sigma} -\frac{3}{2}\,R^{\mu\nu\rho\sigma}F_{\mu\nu}^{(1)\,m}G_{mn}F_{\rho\sigma}^{(1)\,n}+ \frac{3}{2}\,\Tr{\nabla_{\mu}G^{-1}\nabla^{\mu}G\nabla_{\nu}G^{-1}\nabla^{\nu}G}\nonumber \\
&+\frac{5}{8}\,\Tr{\nabla_{\mu}G^{-1}\nabla_{\nu}G\nabla^{\mu}G^{-1}\nabla^{\nu}G}+\frac{1}{8}\,\Tr{\nabla_{\mu}G^{-1}\nabla_{\nu}G}\Tr{\nabla^{\mu}G^{-1}\nabla^{\nu}G}\nonumber \\
&+\frac{3}{8}\,F_{\mu\nu}^{(1)\,m}G_{mn}F_{\rho\sigma}^{(1)\,n} F^{(1)\mu\nu\,p}G_{pq}F^{(1)\rho\sigma\,q} +\frac{1}{8}\,F_{\mu\nu}^{(1)\,m}G_{mn}F_{\rho\sigma}^{(1)\,n} F^{(1)\mu\rho\,p}G_{pq}F^{(1)\nu\sigma\,q}\nonumber \\
&+\frac{1}{2}\,F_{\mu\nu}^{(1)\,m}G_{mn}F^{(1)\mu\rho\,n} F_{\rho\sigma}^{(1)\,p}G_{pq}F^{(1)\nu\sigma\,q}- \frac{1}{2}\,\Tr{\nabla^{\mu}G^{-1}\nabla_{\nu}G}F_{\mu\rho}^{(1)\,m}G_{mn}F^{(1)\,\nu\rho\,n}\\
&-\frac{3}{2}\,F_{\mu\nu}^{(1)\,m}\left(\nabla_{\rho}G\nabla^{\rho}G^{-1}G\right)_{mn} F^{(1)\mu\nu\,n}+\frac{1}{2}\,F_{\mu\nu}^{(1)\,m}\left(\nabla_{\rho}G\nabla^{\mu}G^{-1}G\right)_{mn} F^{(1)\rho\nu\,n}\nonumber\\
&+\Tr{\nabla_{\mu}\nabla_{\nu}G^{-1}G\nabla^{\mu}\nabla^{\nu}G^{-1}G}+3\,\Tr{\nabla_{\mu}\nabla_{\nu}G^{-1}\nabla^{\mu}G\nabla^{\nu}G^{-1}G}-6\,\nabla_{\rho}F_{\mu\nu}^{(1)\,m}\nabla^{\mu}G_{mn}F^{(1)\nu\rho\,n}\nonumber\\
&+ F_{\mu\nu}^{(1)\,m}\left(G\nabla^{\mu}\nabla_{\rho}G^{-1}G\right)_{mn} F^{(1)\rho\nu\,n}-2\,\nabla_{\rho}F_{\mu\nu}^{(1)\,m}G_{mn}\nabla^{\mu}F^{(1)\nu\rho\,n} \Big]\;. \nonumber
\end{align}
Apart from the Riemann tensor,
only the five last terms contain second order derivatives. Using partial integration and Bianchi identities, it is possible to transform those terms so that all second order derivatives appear as the leading two-derivative contribution from the field Eqs.~\eqref{eq:eomGLd},
i.e.\ appear within the first column of Tab.~\ref{tab:fieldredef}. Details are given in App.~\ref{app:IPP}.
Specifically, the remaining second order derivative terms combine into
\begin{equation}
\begin{aligned}
\label{eq:RRreducedeom}
&\frac{\alpha'}{4} 
\int \d^{D}x\,\sqrt{-g}\,e^{-\Phi}\,\Big[\Tr{\Box G^{-1}G\Box G^{-1} G}-2\,\nabla_{\mu}\Phi\,\Tr{\Box G^{-1}G\nabla^{\mu}G^{-1}} \\
& +2\,\Tr{\Box G^{-1}G\nabla_{\nu}G^{-1}\nabla^{\nu}G}+\frac{1}{2}\, \Tr{\Box GG^{-1}\nabla_{\nu}G\nabla^{\nu}G^{-1}} -\frac{5}{4}\,F_{\mu\nu}^{(1)\,m}\Box G_{mn}F^{(1)\mu\nu\,n}\\
& +\left(R_{\mu\nu}+\nabla_{\mu}\nabla_{\nu}\Phi\right)\left(\Tr{\nabla^{\mu}G^{-1}\nabla^{\nu}G}-2\,F_ {\mu\rho}^{(1)\,m}G_{mn}F_{\ \nu}^{(1)\,\rho\,n}\right) \\
&+2\,\nabla^{\mu}F_{\mu\nu}^{(1)\,m}G_{mn}\left(\nabla_{\rho}F^{(1)\rho\nu\,n}-\nabla_{\rho}\Phi F^{(1)\rho\nu\,n}\right) \\
&+ \left(-2\,\nabla^{\mu}\Phi F_{\mu\nu}^{(1)\,m}G_{mn}+3\,F_{\mu\nu}^{(1)\,m}\nabla^{\mu}G_{mn}\right)\nabla_{\rho}F^{(1)\rho\nu\,n}\Big]\;,
\end{aligned}
\end{equation}
and can be eliminated by field redefinitions according to the rules defined in Tab.~\ref{tab:fieldredef}.
The explicit induced field redefinitions  are collected in Eq.~\eqref{eq:fieldrefRR}.
The final result of the reduction (\ref{eq:RRreducedorder2}) then takes the form
\begin{align}
\label{eq:RRreduced1order}
&\!\!\!\!\!\!
\frac{\alpha'}{4} \int\d^{D+d}X\,\sqrt{-\hat{g}}\,e^{-\hat{\phi}} \,\hat{R}_{\hat\mu\hat\nu\hat\rho\hat\sigma}\hat{R}^{\hat\mu\hat\nu\hat\rho\hat\sigma}  ~\longrightarrow
\nonumber \\
& \frac{\alpha'}{4} \int\d^{D}x\,\sqrt{-g}\,e^{-\Phi}\,\Bigg[ R_{\mu\nu\rho\sigma}R^{\mu\nu\rho\sigma} -\frac{1}{2}\,R^{\mu\nu\rho\sigma}F_{\mu\nu}^{(1)\,m}G_{mn}F_{\rho\sigma}^{(1)\,n} + \frac{1}{2}\,\Tr{\nabla_{\mu}G\nabla^{\mu}G^{-1}\nabla_{\nu}BG^{-1}\nabla^{\nu}BG^{-1}}\nonumber \\
&\qquad
+ \Tr{\nabla_{\mu}BG^{-1}\nabla^{\mu}BG^{-1}\nabla_{\nu}BG^{-1}\nabla^{\nu}BG^{-1}}+\frac{1}{8}\,\Tr{\nabla_{\mu}G^{-1}\nabla_{\nu}G\nabla^{\mu}G^{-1}\nabla^{\nu}G}\nonumber \\
&\qquad
-\frac{1}{8}\,\Tr{\nabla_{\mu}G^{-1}\nabla_{\nu}G}\Tr{\nabla^{\mu}G^{-1}\nabla^{\nu}G}-\frac{1}{4}\,\Tr{\nabla_{\mu}BG^{-1}\nabla_{\nu}BG^{-1}}\Tr{\nabla^{\mu}G^{-1}\nabla^{\nu}G}\nonumber \\
&\qquad
+\frac{1}{4}\,H_{\mu\nu m}G^{mn}H_{\rho\sigma n}{H^{\mu\nu}}_{p}G^{pq}{H^{\rho\sigma}}_{q} +\frac{1}{8}\,F_{\mu\nu}^{(1)\,m}G_{mn}F_{\rho\sigma}^{(1)\,n} F^{(1)\mu\rho\,p}G_{pq}F^{(1)\nu\sigma\,q}\nonumber \\
&\qquad
-\frac{1}{2}\,F_{\mu\nu}^{(1)\,m}G_{mn}F^{(1)\mu\rho\,n} F_{\rho\sigma}^{(1)\,p}G_{pq}F^{(1)\nu\sigma\,q} - H_{\mu\nu m}G^{mn}{H^{\mu\rho}}_{n} F_{\rho\sigma}^{(1)\,p}G_{pq}F^{(1)\nu\sigma\,q}\nonumber \\
&\qquad
+\frac{1}{8}\,F_{\mu\nu}^{(1)\,m}H_{\rho\sigma m} F^{(1)\mu\nu\,n} {H^{\rho\sigma}}_{n}+\frac{1}{4}\,H^{2\,\mu\nu}\Tr{\nabla_{\mu}G^{-1}\nabla_{\nu}G}-\frac{1}{2}\,H^{2}_{\mu\nu} F^{(1)\,\mu\rho\,m}G_{mn}F^{(1)\,\nu\ \,n}_{\quad\ \ \rho}\nonumber\\
&\qquad
 +\frac{1}{2}\,\Tr{\nabla_{\mu}BG^{-1}\nabla_{\nu}BG^{-1}}F^{(1)\,\mu\rho\,m}G_{mn}F^{(1)\,\nu\ \,n}_{\quad\ \ \rho}+\frac{1}{2}\,\Tr{\nabla_{\mu}G^{-1}\nabla_{\nu}G}{H^{\mu\rho}}_{m}G^{mn}{H^{\nu}}_{\rho\, n}\nonumber \\
&\qquad
+\frac{1}{2}\,\Tr{\nabla_{\mu}G^{-1}\nabla_{\nu}G}F^{(1)\,\mu\rho\,m}G_{mn}F^{(1)\,\nu\ \,n}_{\quad\ \ \rho} +\frac{1}{2}\,H^{\mu\nu\lambda}{H^{\rho\sigma}}_{\lambda} H_{\mu\nu m}G^{mn}H_{\rho\sigma n}  \nonumber \\
&\qquad
-2\,H^{\mu\nu\rho}H_{\mu\nu m}\left(G^{-1}\nabla^{\sigma}BG^{-1}\right)^{mn}H_{\rho\sigma n} -\frac{1}{2}\,H^{\mu\nu\rho}H_{\mu\nu m}{\left(G^{-1}\nabla^{\sigma}G\right)^{m}}_{n}F_{\rho\sigma}^{(1)\,n} \nonumber \\
&\qquad
-\frac{1}{4}\,F_{\mu\nu}^{(1)\,m}\left(\nabla_{\rho}BG^{-1}\nabla^{\rho}B\right)_{mn} F^{(1)\mu\nu\,n} -\frac{1}{4}\,H_{\mu\nu m}\left(\nabla_{\rho}G^{-1}\nabla^{\rho}GG^{-1}\right)^{mn}{H^{\mu\nu}}_{n} \nonumber \\
&\qquad
-H_{\mu\nu m}\left(G^{-1}\nabla_{\rho}BG^{-1}\nabla^{\rho}BG^{-1}\right)^{mn}{H^{\mu\nu}}_{n}-\frac{1}{2}\,F_{\mu\nu}^{(1)\,m}\left(\nabla_{\rho}G\nabla^{\nu}G^{-1}G\right)_{mn} F^{(1)\mu\rho\,n} \nonumber \\
&\qquad
-2\,H_{\mu\rho m}\left(G^{-1}\nabla^{\mu}BG^{-1}\nabla_{\nu}BG^{-1}\right)^{mn}{H^{\nu\rho}}_{n} -H_{\mu\rho m}{\left(G^{-1}\nabla^{\mu}BG^{-1}\nabla_{\nu}G\right)^{m}}_{n}F^{(1)\,\nu\rho\,n}\Bigg]\;.
\end{align}

\paragraph{Reduction of \texorpdfstring{$\boldsymbol{\hat{R}_{\hat\mu\hat\nu\hat\rho\hat\sigma}\hat{H}^{\hat\mu\hat\nu\hat\lambda}\hat{H}^{\hat\rho\hat\sigma}{}_{\hat\lambda}}$}{RHH}}
Let us finally consider the reduction of the term $RHH$. The index split gives
\begin{equation}
\begin{aligned}
\hat{R}_{\hat\mu\hat\nu\hat\rho\hat\sigma}\hat{H}^{\hat\mu\hat\nu\hat\lambda}\hat{H}^{\hat\rho\hat\sigma}{}_{\hat\lambda} =~&\hat{R}_{\mu\nu\rho\sigma}\hat{H}^{\mu\nu\lambda}\hat{H}^{\rho\sigma}{}_{\lambda} + \hat{R}_{\mu\nu\rho\sigma}\hat{H}^{\mu\nu}{}_{m}\hat{H}^{\rho\sigma m} -4\,\hat{R}_{\mu\nu\rho m}\hat{H}^{\mu\nu\lambda}\hat{H}^{\rho\ \,m}_{\ \,\lambda}  \\
&-4\,\hat{R}_{\mu\nu\rho m}\hat{H}^{\rho nm}\hat{H}^{\mu\nu}{}_{n} +2\,\hat{R}_{\mu\nu mn}\hat{H}^{\mu\nu\rho}\hat{H}_{\rho}{}^{mn}+4\,\hat{R}_{\mu m\nu n}\hat{H}^{\mu\rho m}\hat{H}^{\nu\ \,n}_{\ \,\rho} \\
&+4\,\hat{R}_{\mu m\nu n}\hat{H}^{\mu mp}\hat{H}^{\nu n}{}_{p}-4\,\hat{R}_{\mu mnp}\hat{H}^{\mu\nu m}\hat{H}_{\nu}{}^{np}+\hat{R}_{mnpq}\hat{H}^{\mu m n}\hat{H}_{\mu}{}^{pq} \\ 
& +2\,\hat{R}_{\mu\nu mn}\hat{H}^{\mu\nu}{}_{p}\hat{H}^{mnp}+4\,\hat{R}_{\mu mnp}\hat{H}^{\mu m}{}_{q}\hat{H}^{npq}+\hat{R}_{mnpq}\hat{H}^{mn}{}_{r}\hat{H}^{pqr} .
\end{aligned}
\end{equation}
Then, using Eqs.~\eqref{eq:Hreduced} and \eqref{eq:Rreduced}, the reduction of the corresponding term in the action~\eqref{eq:alpha'D+d} gives
\begin{align}
\label{eq:RHHreducedorder2}
&\!\!\!\!\!\!\!\!-\frac{1}{8}\,\alpha' \int\d^{D+d}X\,\sqrt{-\hat{g}}\,e^{-\hat{\phi}} \,\hat{R}_{\hat\mu\hat\nu\hat\rho\hat\sigma}\hat{H}^{\hat\mu\hat\nu\hat\lambda}\hat{H}^{\hat\rho\hat\sigma}{}_{\hat\lambda} ~\longrightarrow
\nonumber \\
&\frac{\alpha'}{4} \int\d^{D}x\,\sqrt{-g}\,e^{-\Phi} \Big[-\frac{1}{2}\,R_{\mu\nu\rho\sigma}H^{\mu\nu\lambda}H^{\rho\sigma}{}_{\lambda} -\frac{1}{2}\, R_{\mu\nu\rho\sigma}{H^{\mu\nu}}_{m}G^{mn}{H^{\rho\sigma}}_{n}  \nonumber \\
& -\frac{1}{4}\,\Tr{\nabla_{\mu}G^{-1}\nabla_{\nu}B\nabla^{\mu}G^{-1}\nabla^{\nu}B}-\Tr{\nabla_{\mu}B\nabla^{\mu}G^{-1}G\nabla_{\nu}G^{-1}\nabla^{\nu}BG^{-1}}\nonumber \\
& -\frac{1}{2}\,\Tr{\nabla_{\mu}G^{-1}G\nabla_{\nu}G^{-1}\nabla^{\mu}BG^{-1}\nabla^{\nu}B} +\frac{1}{4}\,F_{\mu\nu}^{(1)\,m}G_{mn}F_{\rho\sigma}^{(1)\,n}{H^{\mu\nu}}_{p}G^{pq}{H^{\rho\sigma}}_{q}\nonumber \\
& +\frac{1}{4}\,F_{\mu\nu}^{(1)\,m}G_{mn}F_{\rho\sigma}^{(1)\,n}{H^{\mu\rho}}_{p}G^{pq}{H^{\nu\sigma}}_{q} -\frac{1}{2}\,F_{\mu\nu}^{(1)\,m}H_{\rho\sigma m} F^{(1)\rho\nu\,n} {H^{\mu\sigma}}_{n}+\frac{1}{4}\,H^{\mu\nu\lambda}{H^{\rho\sigma}}_{\lambda} F_{\mu\rho}^{(1)\,m}G_{mn}F_{\nu\sigma}^{(1)\,n}\nonumber \\
&+\frac{1}{4}\,H^{\mu\nu\lambda}{H^{\rho\sigma}}_{\lambda}F_{\mu\nu}^{(1)\,m}G_{mn}F_{\rho\sigma}^{(1)\,n}-H^{\mu\nu\rho}F_{\mu\nu}^{(1)\,m}{\left(G\nabla^{\sigma}G^{-1}\right)_{m}}^{n}H_{\rho\sigma n} -H^{\mu\nu\rho}F_{\mu\sigma}^{(1)\,m}{\left(G\nabla_{\nu}G^{-1}\right)_{m}}^{n}H_{\rho\ \, n}^{\ \,\sigma}\nonumber \\
&-\frac{1}{2}\, H^{\mu\nu\rho}F_{\mu\sigma}^{(1)\,m}\nabla_{\nu}B_{mn} F_{\ \,\rho}^{(1)\,\sigma\,n} \nonumber -F_{\mu\nu}^{(1)\,m}{\left(\nabla_{\rho}B\nabla^{\nu}G^{-1}\right)_{m}}^{n}{H^{\mu\rho}}_{n} +\frac{1}{2}\,F_{\mu\nu}^{(1)\,m}\left(\nabla_{\rho}BG^{-1}\nabla^{\nu}B\right)_{mn} F^{(1)\mu\rho\,n}\\
&-F_{\mu\nu}^{(1)\,m}{\left(G\nabla_{\rho}G^{-1}\nabla^{\rho}BG^{-1}\right)_{m}}^{n}{H^{\mu\nu}}_{n}-F_{\mu\nu}^{(1)\,m}{\left(G\nabla_{\rho}G^{-1}\nabla^{\nu}BG^{-1}\right)_{m}}^{n}{H^{\mu\rho}}_{n} \nonumber \\
&-\frac{1}{2}\,H_{\mu\nu m}\left(\nabla_{\rho}G^{-1}\nabla^{\nu}GG^{-1}\right)^{mn}{H^{\mu\rho}}_{n}-H_{\mu\rho m}\left(\nabla^{\mu}G^{-1}\nabla_{\nu}GG^{-1}\right)^{mn}{H^{\nu\rho}}_{n}\nonumber \\
& -\frac{1}{2}\,H^{\mu\nu\rho}\Tr{\nabla_{\mu}G^{-1}G\nabla_{\nu}G^{-1}\nabla_{\rho}B} +\Tr{\nabla_{\mu}\nabla_{\nu}G^{-1}\nabla^{\mu}BG^{-1}\nabla^{\nu}B}-{H^{\mu\rho}}_{m}\nabla_{\mu}\nabla_{\nu}G^{mn}{H^{\nu}}_{\rho n}\nonumber \\
&-2\,\nabla_{\mu}F_{\nu\rho}^{(1)\, m}{\left(\nabla^{\rho}BG^{-1}\right)_{m}}^{n}{H^{\mu\nu}}_{n} +2\,H^{\mu\nu\lambda}\nabla_{\mu}F_{\nu\rho}^{(1)\,m}{H^{\rho}}_{\lambda m}\Big]\;.
\end{align}
Apart from the Riemann tensor, the four last terms contain second order derivatives. Just as for the Riemann squared term (\ref{eq:RRreducedorder2}), 
upon partial integration, one can transform these terms such that all second order derivatives appear as the leading two-derivative contribution from the field Eqs.~\eqref{eq:eomGLd}. Details are given in App.~\ref{app:IPP}. Specifically, the remaining second order derivative terms combine into
\begin{equation}
\begin{aligned}
\label{eq:RHHreducedeom}
&\int\d^{D}x\,\sqrt{-g}\,e^{-\Phi}\,\frac{\alpha'}{4}\Bigg[\frac{1}{2}\,\Tr{\Box G^{-1}\nabla_{\nu}BG^{-1}\nabla^{\nu}B}-\Tr{\Box BG^{-1}\nabla_{\nu}B\nabla^{\nu}G^{-1}} \\
&-\frac{1}{2}\,F_{\mu\nu}^{(1)\,m}\left(\Box BG^{-1}\right)_{m}{}^{n}{H^{\mu\nu}}_{n} -\frac{1}{4}\,{H^{\mu\nu}}_{m}\Box G^{mn}H_{\mu\nu n}-\nabla_{\mu}H^{\mu\nu\rho}F_{\nu\sigma}^{(1)\,m}{H^{\sigma}}_{\rho m}\\
&-\frac{1}{2}\,\nabla_{\mu}{H^{\mu\nu}}_{m}\left(2\,\nabla^{\rho}G^{mn}H_{\nu\rho n}-H_{\nu\rho\sigma}F^{(1)\rho\sigma\,m}+2\,{\left(G^{-1}\nabla^{\rho}B\right)^{m}}_{n}F_{\nu\rho}^{(1)\,n}\right) \\
&-\frac{1}{2}\, \nabla_{\mu}F^{(1)\mu\nu\,m}\Big(2{\left(\nabla^{\rho}BG^{-1}\right)_{m}}^{n}H_{\nu\rho n}-H_{\nu\rho\sigma}{H^{\rho\sigma}}_{m}\Big)\Bigg],
\end{aligned}
\end{equation}
and can be eliminated by field redefinitions according to the rules defined in Tab.~\ref{tab:fieldredef}.
The explicit induced field redefinitions  are collected in Eq.~\eqref{eq:fieldrefRHH}.
The final result of the reduction (\ref{eq:RHHreducedorder2}) then takes the form
\begin{align}
\label{eq:RHHreduced1order}
&\!\!\!\!\!\!\!\!\!\!\!\!\!
-\frac{1}{8}\,\alpha' \int \d^{D+d}X\,\sqrt{-\hat{g}}\,e^{-\hat{\phi}} \,\hat{R}_{\hat\mu\hat\nu\hat\rho\hat\sigma}\hat{H}^{\hat\mu\hat\nu\hat\lambda}\hat{H}^{\hat\rho\hat\sigma}_{\quad\hat\lambda} 
~\longrightarrow
\nonumber \\
& \frac{\alpha'}{4}
\int\d^{D}x\,\sqrt{-g}\,e^{-\Phi}\,\Big[-\frac{1}{2}\,R_{\mu\nu\rho\sigma}H^{\mu\nu\lambda}H^{\rho\sigma}_{\quad\lambda} -\frac{1}{2}\,R_{\mu\nu\rho\sigma}{H^{\mu\nu}}_{m}G^{mn}{H^{\rho\sigma}}_{n}\nonumber \\
&\qquad
+\frac{1}{4}\,\Tr{\nabla_{\mu}G^{-1}\nabla_{\nu}B\nabla^{\mu}G^{-1}\nabla^{\nu}B} -\frac{1}{2}\,\Tr{\nabla_{\mu}G\nabla^{\mu}G^{-1}\nabla_{\nu}BG^{-1}\nabla^{\nu}BG^{-1}}\nonumber \\
&\qquad
-\frac{1}{2}\, \Tr{\nabla_{\mu}BG^{-1}\nabla^{\mu}BG^{-1}\nabla_{\nu}BG^{-1}\nabla^{\nu}BG^{-1}}-\frac{1}{2}\,\Tr{\nabla_{\mu}G^{-1}G\nabla_{\nu}G^{-1}\nabla^{\mu}BG^{-1}\nabla^{\nu}B}\nonumber \\
&\qquad
 - \frac{1}{8}\,H_{\mu\nu m}G^{mn}H_{\rho\sigma n}{H^{\mu\nu}}_{p}G^{pq}{H^{\rho\sigma}}_{q} +\frac{1}{4}\,F_{\mu\nu}^{(1)\,m}G_{mn}F_{\rho\sigma}^{(1)\,n}{H^{\mu\rho}}_{p}G^{pq}{H^{\nu\sigma}}_{q} \nonumber \\
&\qquad
+\frac{1}{2}\,F_{\mu\nu}^{(1)\,m}H_{\rho\sigma m} F^{(1)\rho\nu\,n} {H^{\mu\sigma}}_{n} -\frac{1}{8}\,F_{\mu\nu}^{(1)\,m}H_{\rho\sigma m} F^{(1)\mu\nu\,n} {H^{\rho\sigma}}_{n}\nonumber \\
&\qquad
+\frac{1}{4}\,H^{\mu\nu\lambda}{H^{\rho\sigma}}_{\lambda} F_{\mu\rho}^{(1)\,m}G_{mn}F_{\nu\sigma}^{(1)\,n}-\frac{1}{4}\,H^{\mu\nu\lambda}{H^{\rho\sigma}}_{\lambda} H_{\mu\nu m}G^{mn}H_{\rho\sigma n} \nonumber \\
&\qquad
+H^{\mu\nu\rho}H_{\mu\nu m}\left(G^{-1}\nabla^{\sigma}BG^{-1}\right)^{mn}H_{\rho\sigma n} +\frac{1}{2}\,H^{\mu\nu\rho}H_{\mu\nu m}{\left(G^{-1}\nabla^{\sigma}G\right)^{m}}_{n}F_{\rho\sigma}^{(1)\,m} \nonumber \\
&\qquad
-H^{\mu\nu\rho}F_{\mu\sigma}^{(1)\,m}{\left(G\nabla_{\nu}G^{-1}\right)_{m}}^{n}H_{\rho\ \, n}^{\ \,\sigma}+\frac{1}{2}\,H^{\mu\nu\rho}F_{\mu\sigma}^{(1)\,m}\nabla_{\nu}B_{mn} F_{\ \,\rho}^{(1)\,\sigma\,n}\nonumber \\
&\qquad
 +\frac{1}{4}\,F_{\mu\nu}^{(1)\,m}\left(\nabla_{\rho}BG^{-1}\nabla^{\rho}B\right)_{mn} F^{(1)\mu\nu\,n}+\frac{1}{2}\,H_{\mu\nu m}\left(G^{-1}\nabla_{\rho}BG^{-1}\nabla^{\rho}BG^{-1}\right)^{mn}{H^{\mu\nu}}_{n} \nonumber\\
&\qquad
+\frac{1}{4}\,H_{\mu\nu m}\left(G^{-1}\nabla_{\rho}G\nabla^{\rho}G^{-1}\right)^{mn}{H^{\mu\nu}}_{n}-F_{\mu\nu}^{(1)\,m}{\left(G\nabla_{\rho}G^{-1}\nabla^{\nu}BG^{-1}\right)_{m}}^{n}{H^{\mu\rho}}_{n} \nonumber \\
&\qquad
+F_{\mu\nu}^{(1)\,m}{\left(\nabla_{\rho}B\nabla^{\nu}G^{-1}\right)_{m}}^{n}{H^{\mu\rho}}_{n} -\frac{1}{2}\,F_{\mu\nu}^{(1)\,m}\left(\nabla_{\rho}BG^{-1}\nabla^{\nu}B\right)_{mn} F^{(1)\mu\rho\,n}\nonumber \\
&\qquad
-\frac{1}{2}\,H_{\mu\nu m}\left(\nabla_{\rho}G^{-1}\nabla^{\nu}GG^{-1}\right)^{mn}{H^{\mu\rho}}_{n}+H_{\mu\rho m}\left(G^{-1}\nabla^{\mu}BG^{-1}\nabla_{\nu}BG^{-1}\right)^{mn}{H^{\nu\rho}}_{n}\nonumber \\
&\qquad
 -F_{\mu\rho}^{(1)\,m}{\left(\nabla^{\mu}GG^{-1}\nabla_{\nu}BG^{-1}\right)_{m}}^{n}{H^{\nu\rho}}_{n} -\frac{1}{2}\,H^{\mu\nu\rho}\Tr{\nabla_{\mu}G^{-1}G\nabla_{\nu}G^{-1}\nabla_{\rho}B} \Big]\;.
\end{align}

In the next section, we will match the result of the explicit reduction against the basis 
(\ref{basisalpha1}) in order to establish \Odd invariance of the reduced action.

\subsection{Field redefinitions}

By partial integration and suitable field redefinitions, we have thus cast the reduced action at order $\alpha'$
into a form which is polynomial in first order derivatives and the Riemann tensor. 
As an illustration and for potential applications requiring the dictionary between the lower-dimensional
fields and the fields featuring in the original action (\ref{eq:I0D+d}), let us list the full set of induced
field redefinitions, put together from Eqs.~\eqref{eq:fieldrefRR} and \eqref{eq:fieldrefRHH}:
\begin{align}
    \delta \Phi &=  \dfrac{1}{4}\Big[-\,F_{\mu\nu}^{(1)\, m}G_{mn}F^{(1)\mu\nu\,n}+\dfrac{1}{2}\,\Tr{\nabla_{\mu}G^{-1}\nabla^{\mu}G}\Big]\;,\nonumber\\
    \delta g_{\mu\nu} &= \dfrac{1}{4}\Big[2\,F_{\mu\rho}^{(1)\, m}G_{mn}F_{\ \nu}^{(1)\rho\,n}-\Tr{\nabla_{(\mu}G^{-1}\nabla_{\nu)}G}\Big]\;,\nonumber\\
    \delta B_{\mu\nu} &=\dfrac{1}{8}\Big[\Big(-2\,\nabla^{\rho}F_{\rho\mu}^{(1)\,m}+2\,\nabla^{\rho}\Phi F_{\rho\mu}^{(1)\,m} +\dfrac{1}{2}\,H_{\mu\rho\sigma}{H^{\rho\sigma}}_{p}G^{pm} \nonumber\\
    &\qquad{}+F_{\mu\rho}^{(1)\,p}{\left(\nabla^{\rho}GG^{-1}\right)_{p}}^{m}+H_{\mu\rho p}\left(G^{-1}\nabla^{\rho}BG^{-1}\right)^{pm}\Big)\Big(A_{\nu\,m}^{(2)}-B_{mn}A_{\nu}^{(1)n}\Big) \nonumber\\
     &\qquad{} -A_{\mu}^{(1)m}{\left(G\nabla_{\rho}G^{-1}\right)_{m}}^{n}H_{\nu\rho n} -A_{\mu}^{(1)m} \nabla^{\rho}B_{mn}F^{(1)n}_{\nu\rho} +2\,F_{\mu\rho}^{(1)m}{H^{\rho}}_{\nu m} \nonumber\\
     &\qquad{}+\dfrac{1}{2}\,A_{\mu}^{(1)m}H_{\nu\rho\sigma}G_{mn}F^{(1)\rho\sigma\,n} \Big] - \Big(\mu\leftrightarrow\nu\Big)\;,\nonumber\\
    \delta G^{mn} &=\dfrac{1}{4}\Big[-2\,\Box G^{mn}+2\,\nabla_{\mu}\Phi\nabla^{\mu}G^{mn}-\dfrac{1}{2}\,G^{mp}H_{\mu\nu p}G^{nq}{H^{\mu\nu}}_{q} - \dfrac{3}{2}\,F_{\mu\nu}^{(1)\,m}F^{(1)\mu\nu\,n}\nonumber\\
     &\qquad{}-\left(G^{-1}\nabla_{\mu}G\nabla^{\mu}G^{-1}\right)^{mn}+\,\left(G^{-1}\nabla_{\mu}BG^{-1}\nabla^{\mu}BG^{-1}\right)^{mn}\Big]\;,\nonumber\\
    \delta B_{mn} &=\dfrac{1}{4}\Big[\left(\nabla_{\mu}B\nabla^{\mu}G^{-1}G\right)_{mn}+\left(G\nabla_{\mu}G^{-1}\nabla^{\mu}B\right)_{mn} \nonumber\\
     &\qquad{}- \dfrac{1}{2}\, H_{\mu\nu m}F^{(1)\mu\nu\,p}G_{pn} + \dfrac{1}{2}\,G_{mp}F^{(1)\mu\nu\,p}H_{\mu\nu n}\Big]\;,\nonumber\\
    \delta A_{\mu}^{(1)m} &=\dfrac{1}{4}\Big[-2\,\nabla^{\nu}F_{\nu\mu}^{(1)\,m}+2\,\nabla^{\nu}\Phi F_{\nu\mu}^{(1)\,m} +\dfrac{1}{2}\,H_{\mu\nu\rho}{H^{\nu\rho}}_{n}G^{nm} \nonumber\\
     &\qquad{}+F_{\mu\nu}^{(1)\,n}{\left(\nabla^{\nu}GG^{-1}\right)_{n}}^{m}+H_{\mu\nu n}\left(G^{-1}\nabla^{\nu}BG^{-1}\right)^{nm}\Big]\;,\nonumber\\
    \delta A_{\mu\,m}^{(2)} &=\dfrac{1}{4}\Big[2\,\nabla^{\nu}F_{\nu\mu}^{(1)\,n}B_{nm}-2\,\nabla^{\nu}\Phi F_{\nu\mu}^{(1)\,n}B_{nm} -\dfrac{1}{2}\,H_{\mu\nu\rho}{H^{\nu\rho}}_{n}{\left(G^{-1}B\right)^{n}}_{m} \nonumber\\
     &\qquad{}-F_{\mu\nu}^{(1)\,n}\left(\nabla^{\nu}GG^{-1}B\right)_{nm}-H_{\mu\nu n}{\left(G^{-1}\nabla^{\nu}BG^{-1}B\right)^{n}}_{m}+H_{\mu\nu n}{\left(\nabla^{\nu}G^{-1}G\right)^{n}}_{m}\nonumber\\
     &\qquad{}-F^{(1)n}_{\mu\nu}\nabla^{\nu}B_{nm}-\dfrac{1}{2}\,H_{\mu\nu\rho}F^{(1)\nu\rho\,n}G_{nm}\Big]
     \;,
     \label{eq:fieldredef}
\end{align}
where we used the convention of Eq.~\eqref{eq:fieldvariation}.

\section{\texorpdfstring{$\boldsymbol{\Odd}$}{} invariance and a Green-Schwarz type mechanism}
\label{sec:alpha'action}

We have now set up all the elements allowing to systematically exhibit the \Odd invariance of the dimensionally reduced
theory at order $\alpha'$.
Having brought the reduced action into a form that is polynomial in first derivatives (and the Riemann tensor),
we have fully fixed the ambiguities due to field redefinitions and partial integration.
We can then compare the result to the distinguished manifestly \Odd invariant basis
constructed in Sec.~\ref{subsec:basis}, after breaking up the latter under ${\rm GL}(d)$\footnote{
See App.~\ref{app:OddtoGLd} for the $\GL(d)$ expressions of the relevant \Odd terms.}.
Different terms of the \Odd basis (\ref{basisalpha1}) do not share common terms in the decomposition under ${\rm GL}(d)$, i.e.\ every ${\rm GL}(d)$ invariant term we have obtained in the reduction in the previous section has a unique ancestor within the 
\Odd basis (\ref{basisalpha1}). It becomes thus a straightforward -- albeit lengthy -- task to recombine (if possible)
any collection of ${\rm GL}(d)$ terms into \Odd invariant  expressions.

The dimensionally reduced action is given by the sum of Eqs.~\eqref{eq:H2H2reduced}, \eqref{eq:HHHHreduced}, \eqref{eq:RRreduced1order}, and \eqref{eq:RHHreduced1order}. 
Upon combining these terms into the \Odd invariant expressions of the basis (\ref{basisalpha1}), we
can bring it into the form
\begin{equation}
{I}_{1} =
\underline{I_{1}} + O_1
\;,
\end{equation}
where $\underline{I_{1}}$ is the part of $I_1$ that can be 
organized into a linear combination of 
manifestly \Odd invariant basis elements as
\begin{align}
\label{eq:alpha'D}
\underline{I_{1}}~=~ & 
\frac{1}{4}\,\alpha' \int\d^{D}x\,\sqrt{-g}\,e^{-\Phi}\,\Bigg[
\nonumber \\
&\;\;\;\;R_{\mu\nu\rho\sigma}R^{\mu\nu\rho\sigma}-\frac{1}{2} R_{\mu\nu\rho\sigma}H^{\mu\nu\lambda}{H^{\rho\sigma}}_{\lambda}-\frac{1}{8}\,H^{2}_{\mu\nu}H^{2\,\mu\nu} + \frac{1}{24} H_{\mu\nu\rho}{{H^{\mu}}_{\sigma}}^{\lambda}{{H^{\nu}}_{\lambda}}^{\tau}{{H^{\rho}}_{\tau}}^{\sigma} \nonumber \\
&-\frac{1}{2} R_{\mu\nu\rho\sigma} \F^{\mu\nu\,M}{\S_{M}}^{N}{\F^{\rho\sigma}}_{N}+\frac{1}{16}\Tr{\nabla_{\mu}\S\nabla_{\nu}\S\nabla^{\mu}\S\nabla^{\nu}\S} - \frac{1}{32} \Tr{\nabla_{\mu}\S\nabla_{\nu}\S} \Tr{\nabla^{\mu}\S\nabla^{\nu}\S} \nonumber \\
&+\frac{1}{8}{\F_{\mu\nu}}^{M}{\F_{\rho\sigma}}_{M}\F^{\mu\rho\,N}{\F^{\nu\sigma}}_{N} -\frac{1}{2}{\F_{\mu\nu}}^{M}{\S_{M}}^{N}{\F^{\mu\rho}}_{N}\F^{\nu\sigma\,P}{\S_{P}}^{Q}\F_{\rho\sigma\,Q} \nonumber \\
&+\frac{1}{8} {\F_{\mu\nu}}^{M}{\S_{M}}^{N}\F_{\rho\sigma\,N}\F^{\mu\rho\,P}{\S_{P}}^{Q}{\F^{\nu\sigma}}_{Q}+\frac{1}{8}\,H^{2}_{\mu\nu}\Tr{\nabla^{\mu}\S\nabla^{\nu}\S}-\frac{1}{2}\,H^{2}_{\mu\nu}{{\F^{\mu}}_{\rho}}^{M}{\S_{M}}^{N}{\F^{\nu\rho}}_{N} \nonumber \\
& + \frac{1}{4}\F^{\mu\rho\,M}{\S_{M}}^{N}{\F^{\nu}}_{\rho\,N}\Tr{\nabla_{\mu}\S\nabla_{\nu}\S}+\frac{1}{4} H^{\mu\nu\lambda}{H^{\rho\sigma}}_{\lambda}{\F_{\mu\rho}}^{M}{\S_{M}}^{N}\F_{\nu\sigma\,N}  \nonumber \\
& -\frac{1}{2}\,H^{\mu\nu\rho}{\F_{\mu\sigma}}^{M}{\left(\S\nabla_{\nu}\S\right)_{M}}^{N}{{\F_{\rho}}^{\sigma}}_{N}-\frac{1}{2}{\F_{\mu\nu}}^{M}{\left(\S\nabla_{\rho}\S\nabla^{\nu}\S\right)_{M}}^{N} {\F^{\mu\rho}}_{N}  \Bigg]\;,
\end{align}
whereas the remaining part of the action $O_1$ is not manifestly \Odd invariant, but given by
\begin{equation}
O_1 =
-\frac{1}{8}\,\alpha'
\!\!\int\! \!\d^{D}x\,\sqrt{-g}\,e^{-\Phi} H^{\mu\nu\rho}\,{\rm Tr} \Big[  \nabla_{\mu}G^{-1}G\nabla_{\nu}G^{-1}\nabla_{\rho}B
-\frac{1}{3} \, \nabla_{\mu}BG^{-1}\nabla_{\nu}BG^{-1}\nabla_{\rho}BG^{-1} \Big]
\;.\;\;\;\;\;\;
\end{equation}
This suggests the definition
\begin{equation}
\label{Omegafirst}
\Omega_{\mu\nu\rho} = -\frac{3}{4}\,{\rm Tr}\big(\partial_{[\mu}G^{-1} G \partial_{\nu}G^{-1} \partial_{\rho]} B\big)
  +\frac{1}{4}\,{\rm Tr}\big(\partial_{[\mu}B G^{-1} \partial_{\nu}B G^{-1}\partial_{\rho]}B G^{-1}\big)\;,
\end{equation}
such that $O_1$ takes the form
\begin{equation}
O_1 = \frac16\,\alpha' \int\d^{D}x\,\sqrt{-g}\,e^{-\Phi}\,H_{\mu\nu\rho}\,\Omega^{\mu\nu\rho}
\;.
\label{eq:O1}
\end{equation}
The 3-form (\ref{Omegafirst}) descends from the 
non-vanishing cohomology $H^4$ of $\Odd/({\rm O}(d)\times {\rm O}(d))$~\cite{Ferretti:1992fg,DHoker:1995mfi},
although it is not \Odd invariant, its exterior derivative is\footnote{See App.~\ref{app:OddtoGLd} for the $\GL(d)$ expression.}
\begin{equation}
\label{Sinvariant}
4\,\partial_{[\mu}\Omega_{\nu\rho\sigma]}=\frac{3}{8}\,\Tr{\S\,\partial_{[\mu}\S\,\partial_{\nu}\S\,\partial_{\rho}\S\,\partial_{\sigma]}\S}
\,.
\end{equation}
For $\Gamma \in \odd$ this implies that $\d\delta_{\Gamma} \Omega = \delta_{\Gamma}\d  \Omega= 0$, i.e.\ 
the \Odd variation of $\Omega_{\mu\nu\rho}$ is closed and can locally be integrated to a 2-form $X_{\mu\nu}$ such that
\begin{equation}
\delta_{\Gamma}\Omega_{\mu\nu\rho} = 3\,\partial_{[\mu}X_{\nu\rho]}\;. 
\label{eq:CStransform}
\end{equation}
This observation together with the particular form of (\ref{eq:O1}) suggests a Green-Schwarz type mechanism in oder to restore \Odd invariance of the $D$-dimensional action. Specifically, the term (\ref{eq:O1}) can be absorbed into a deformation of the
two-derivative action (\ref{eq:MaharanaSchwarz}) upon redefining
\begin{equation}
  \label{eq:newHH}
  \widetilde{H}_{\mu\nu\rho}\equiv H_{\mu\nu\rho} -\alpha'\,\Omega_{\mu\nu\rho}\;, 
\end{equation}
such that the kinetic term now produces 
\begin{equation}
  -\frac{1}{12}\widetilde{H}^{\mu\nu\rho}\widetilde{H}_{\mu\nu\rho} =  -\frac{1}{12}{H}^{\mu\nu\rho}{H}_{\mu\nu\rho}
  +\frac{\alpha'}{6} H^{\mu\nu\rho} \Omega_{\mu\nu\rho} + {\cal O}(\alpha'{}^2)
  \;.
\end{equation}
In view of Eq.~(\ref{eq:CStransform}), the deformed field strength (\ref{eq:newHH}) remains \Odd invariant,
if we impose on $B_{\mu\nu}$ a non-trivial \Odd transformation for $\Gamma\in\odd$ as
\begin{equation}
\label{eq:deltaB}
\delta_{\Gamma} B_{\mu\nu} = \alpha'\,X_{\mu\nu}
\quad\Longrightarrow\quad
\delta_{\Gamma} \widetilde{H}_{\mu\nu\rho} = 0
\;.
\end{equation}
The resulting theory is then fully \Odd-invariant to first order in $\alpha'$.
In order to compute an explicit expression for $X_{\mu\nu}$, 
we start from a general $\odd$ matrix parametrized as
\begin{equation}
\label{eq:gammaodd}
{\Gamma_{M}}^{N} = \begin{pmatrix}
                      {\mathfrak{a}_{m}}^{n} & \mathfrak{b}_{m n} \\
                      \mathfrak{c}^{m n} & -{\mathfrak{a}_{n}}^{m}
                    \end{pmatrix} ,
\end{equation}
with $\mathfrak{c}^{mn}$ and $\mathfrak{b}_{mn}$ antisymmetric. Further defining the \odd matrices
\begin{equation}
\mathfrak{A}(\mathfrak{a})_{M}{}^{N} = \begin{pmatrix}
                      {\mathfrak{a}_{m}}^{n} & 0 \\
                      0 & -{\mathfrak{a}_{n}}^{m}
                    \end{pmatrix} ,\ 
\mathfrak{B}(\mathfrak{b})_{M}{}^{N} = \begin{pmatrix}
                      0 & \mathfrak{b}_{m n} \\
                      0 & 0
                    \end{pmatrix} ,\ 
\mathfrak{C}(\mathfrak{c})_{M}{}^{N} = \begin{pmatrix}
                      0 & 0 \\
                      \mathfrak{c}^{m n} & 0
                    \end{pmatrix},
\end{equation}
the \odd algebra takes the form
\begin{equation}
  \label{eq:oddalgebra}
 \begin{cases}
  \left[\mathfrak{A}(\mathfrak{a}_{1}),\mathfrak{A}(\mathfrak{a}_{2})\right] = \mathfrak{A}\left(\left[\mathfrak{a}_{1},\mathfrak{a}_{2}\right]\right), \\
  \left[\mathfrak{A}(\mathfrak{a}),\mathfrak{B}(\mathfrak{b})\right] = \mathfrak{B}\left(\mathfrak{a}\mathfrak{b}+\mathfrak{b}\mathfrak{a}^{\rm t}\right), \\
  \left[\mathfrak{A}(\mathfrak{a}),\mathfrak{C}(\mathfrak{c})\right] = -\mathfrak{C}\left(\mathfrak{c}\mathfrak{a}+\mathfrak{a}^{\rm t}\mathfrak{c}\right),
 \end{cases}
 \qquad
 \begin{cases}
  \left[\mathfrak{B}(\mathfrak{b}_{1}),\mathfrak{B}(\mathfrak{b}_{2})\right] = 0\;, \\
  \left[\mathfrak{B}(\mathfrak{b}),\mathfrak{C}(\mathfrak{c})\right] = \mathfrak{A}\left(\mathfrak{b}\mathfrak{c}\right)\;, \\
  \left[\mathfrak{C}(\mathfrak{c}_{1}),\mathfrak{C}(\mathfrak{c}_{2})\right] = 0\;.
 \end{cases}
\end{equation}
The action of these generators on $G_{mn}$ and $B_{mn}$ is obtained from Eq.~\eqref{HOdd} as
\begin{equation}
 \begin{cases}
  \delta_{\Gamma}G = \mathfrak{a}\,G +G\,\mathfrak{a}^{\rm t}-G\,\mathfrak{c}\,B-B\,\mathfrak{c}\,G\;, \\
  \delta_{\Gamma}B = \mathfrak{a}\,B +B\,\mathfrak{a}^{\rm t}-B\,\mathfrak{c}\,B-G\,\mathfrak{c}\,G+\mathfrak{b}\;,
 \end{cases}
\end{equation}
which, together with Eq.~\eqref{Omegafirst}, yields the general \Odd variation of $\Omega_{\mu\nu\rho}$
\begin{equation}
 \delta_{\Gamma}\Omega_{\mu\nu\rho} = -\frac{3}{2}\Big[\Tr{\mathfrak{c}\,\partial_{[\mu}G\partial_{\nu}G^{-1}\partial_{\rho]}G}+\Tr{\mathfrak{c}\,\partial_{[\mu}B\partial_{\nu}G^{-1}\partial_{\rho]}B}\Big].
\end{equation}
Pulling out one derivative, we extract the explicit form of $X_{\mu\nu}$ from Eq.~(\ref{eq:CStransform}):
\begin{equation}
X_{\mu\nu}= \frac{1}{2}\Tr{\mathfrak{c}\,\partial_{[\mu}(G+B)\,G^{-1}\partial_{\nu]}(G+B)}\,.
\label{eq:trafX}
\end{equation}
According to Eq.~(\ref{eq:deltaB}), the 2-form thus acquires new transformations only along 
the nilpotent $\odd$ generators $\mathfrak{c}^{mn}$.
This is consistent with the fact that all the other $\odd$ generators have a geometric origin and by construction
represent manifest symmetries of the dimensionally reduced action.
Moreover, with the expression~(\ref{eq:trafX}), one can verify that
the algebra of \odd transformations~\eqref{eq:oddalgebra} closes on $B_{\mu\nu}$. 
Crucially, the deformed \odd action~\eqref{eq:deltaB} cannot be absorbed into a redefinition of the fields
but represents a genuine deformation of the \Odd transformation rules.

We may also consider the behavior of Eq.~(\ref{eq:O1}) under the $\mathbb{Z}_2$ invariance 
of bosonic string theory that sends $\hat{B}\rightarrow -\hat{B}$. 
On the \Odd matrix (\ref{HOdd}) this symmetry acts as~\cite{Hohm:2010pp}
 \begin{equation}\label{Z2metric}
  {\cal H} \rightarrow Z^T {\cal H} Z\,,\qquad Z\equiv 
  \begin{pmatrix}
                    1 & 0 \\
                    0 & -1
                  \end{pmatrix}\;. 
 \end{equation}
 The matrix $Z$ is not \Odd-valued since the metric (\ref{etamatrix}) transforms as 
 \begin{equation}
  \eta\rightarrow Z\eta Z^T = -\eta
  \qquad\Longrightarrow\qquad
    {\cal S}\rightarrow -Z{\cal S}Z\;. 
 \end{equation}
Thus, the \Odd invariant defined by the r.h.s.\ of Eq.~(\ref{Sinvariant}) is $\mathbb{Z}_2$ odd. 
This ensures $\mathbb{Z}_2$ invariance of the action (\ref{eq:O1}) 
since $B_{\mu\nu}$ and its field strength $H_{\mu\nu\rho}$ are also $\mathbb{Z}_2$ odd.

\paragraph*{}
Let us summarize the previous discussion. The bosonic string effective action, including its first order $\alpha'$-corrections, 
upon compactification on a $d$-torus exhibits a global \Odd symmetry, provided the \Odd transformations of the
two-derivative action acquire $\alpha'$-corrections according to Eq.~\eqref{eq:deltaB}. 
The full $\alpha'$-corrected transformations are given by
  \begin{equation}
  \label{eq:oddvariationalpha'}
  \begin{cases}
    \delta_{\Gamma}g_{\mu\nu} = 0\;, \\
    \delta_{\Gamma}B_{\mu\nu} =\dfrac{\alpha'}{2}\,\Tr{\mathfrak{c}\,\partial_{[\mu}(G+B)G^{-1}\partial_{\nu]}(G+B)},
  \end{cases} 
  \begin{cases}
    \delta_{\Gamma}\H_{MN} = \Gamma_{M}{}^{P}\H_{PN}+\Gamma_{N}{}^{P}\H_{MP}\;, \\
    \delta_{\Gamma}\F_{\mu\nu}{}^{M} =- \F_{\mu\nu}{}^{N}\,\Gamma_{N}{}^{M}\;,
  \end{cases}
  \end{equation}
for $\Gamma_{M}{}^{N}\in\odd$ parametrized as Eq.~\eqref{eq:gammaodd}. To order $\alpha'$, the \Odd 
invariant action is given by
\begin{align}
 \label{eq:fullaction}
 I=&\int\d^{D}x\,\sqrt{-g}\,e^{-\Phi}\Bigg[R+\partial_{\mu}\Phi\,\partial^{\mu}\Phi-\frac{1}{12}\,\widetilde{H}_{\mu\nu\rho}\widetilde{H}^{\mu\nu\rho}+\frac{1}{8}\,\Tr{\partial_{\mu}\S\,\partial^{\mu}\S}-\frac{1}{4}\,\F_{\mu\nu}^{M}\,{\S_{M}}^{N}\,\F^{\mu\nu}_{N} \nonumber \\
&+\frac{1}{4}\,\alpha' \,\Big(R_{\mu\nu\rho\sigma}R^{\mu\nu\rho\sigma}-\frac{1}{2}\, R_{\mu\nu\rho\sigma}{H}^{\mu\nu\lambda}{H}^{\rho\sigma}_{\ \ \,\lambda}   + \frac{1}{24}\, H_{\mu\nu\rho}{H}^{\mu\ \, \lambda}_{\ \,\sigma}{H}^{\nu\ \,\tau}_{\ \,\lambda}{H}^{\rho\ \,\sigma}_{\ \,\tau}
  \nonumber \\
  & -\frac{1}{8}\,{H}^{2}_{\mu\nu}{H}^{2\,\mu\nu}
 +\frac{1}{16}\,\Tr{\nabla_{\mu}\S\nabla_{\nu}\S\nabla^{\mu}\S\nabla^{\nu}\S} - \frac{1}{32}\, \Tr{\nabla_{\mu}\S\nabla_{\nu}\S} \Tr{\nabla^{\mu}\S\nabla^{\nu}\S} \nonumber \\
  &+\frac{1}{8}\, {\F_{\mu\nu}}^{M}{\S_{M}}^{N}\F_{\rho\sigma\,N}\F^{\mu\rho\,P}{\S_{P}}^{Q}{\F^{\nu\sigma}}_{Q} -\frac{1}{2}\,{\F_{\mu\nu}}^{M}{\S_{M}}^{N}{\F^{\mu\rho}}_{N}\F^{\nu\sigma\,P}{\S_{P}}^{Q}\F_{\rho\sigma\,Q} \nonumber \\
  &+\frac{1}{8}\,{\F_{\mu\nu}}^{M}{\F_{\rho\sigma}}_{M}\F^{\mu\rho\,N}{\F^{\nu\sigma}}_{N}  -\frac{1}{2}\, R_{\mu\nu\rho\sigma} \F^{\mu\nu\,M}{\S_{M}}^{N}{\F^{\rho\sigma}}_{N} \nonumber \\
  &+\frac{1}{8}\,H^{2}_{\mu\nu}\Tr{\nabla^{\mu}\S\nabla^{\nu}\S}-\frac{1}{2}\,H^{2}_{\mu\nu}{{\F^{\mu}}_{\rho}}^{M}{\S_{M}}^{N}{\F^{\nu\rho}}_{N} +\frac{1}{4}\, {H}^{\mu\nu\lambda}{H}^{\rho\sigma}_{\ \ \,\lambda}{\F_{\mu\rho}}^{M}{\S_{M}}^{N}\F_{\nu\sigma\,N} \nonumber\\
  &-\frac{1}{2}\,{\F_{\mu\nu}}^{M}{\left(\S\nabla_{\rho}\S\nabla^{\nu}\S\right)_{M}}^{N} {\F^{\mu\rho}}_{N} + \frac{1}{4}\,\F^{\mu\rho\,M}{\S_{M}}^{N}{\F^{\nu}}_{\rho\,N}\Tr{\nabla_{\mu}\S\nabla_{\nu}\S}\nonumber \\
  &-\frac{1}{2}\,H^{\mu\nu\rho}{\F_{\mu\sigma}}^{M}{\left(\S\nabla_{\nu}\S\right)_{M}}^{N}{{\F_{\rho}}^{\sigma}}_{N} \Big)\Bigg]
  ~~+~{\cal O}(\alpha'{}^2)\;,
  \end{align}
with the deformed field-strength $\widetilde{H}_{\mu\nu\rho}$ defined in Eq.~\eqref{eq:newHH}. This constitutes the main result of this paper.

Let us comment on the relation to Ref.~\cite{Godazgar:2013bja}, where
a similar analysis of the first order $\alpha'$-corrections is performed, 
however restricted
to the scalar sector, i.e.\ setting $A_{\mu}^{(1)\,m} =A_{\mu\,m}^{(2)}=B_{\mu\nu}=0$, $g_{\mu\nu}=\eta_{\mu\nu}$\,.
Their result is given in their Eq.~(74):
\begin{equation}
\begin{aligned}
I_{1} = &\frac{1}{8}\,\alpha'   \int\d^{D}x\,e^{-\Phi}\,\Bigg[-\Tr{\nabla_{\mu}\nabla_{\nu}\S\nabla^{\mu}\nabla^{\nu}\S} + \frac{1}{16} \Tr{\nabla_{\mu}\S\nabla_{\nu}\S} \Tr{\nabla^{\mu}\S\nabla^{\nu}\S} \\
&\qquad\qquad\qquad\qquad
+\Tr{\nabla_{\mu}\S\nabla^{\mu}\S\nabla_{\nu}\S\nabla^{\nu}\S}+\frac{1}{8}\Tr{\nabla_{\mu}\S\nabla_{\nu}\S\nabla^{\mu}\S\nabla^{\nu}\S}\Bigg]\;.
\end{aligned}
\end{equation}
Upon partial integration, this can be rewritten as
\begin{equation}
\begin{aligned}
I_{1} ~=~ &\frac{1}{8}\,\alpha'\int\d^{D}x\,\sqrt{-g}\,e^{-\Phi}\,\Bigg[-{\rm Tr}\,\Big(\left(\Box \S-\nabla_{\mu}\Phi\nabla^{\mu}\S\right)\left(\Box \S-\nabla_{\nu}\Phi\nabla^{\nu}\S\right)\Big)  \\
&+\left(R_{\mu\nu}+\nabla_{\mu}\nabla_{\nu}\Phi\right)\Tr{\nabla^{\mu}\S\nabla^{\nu}\S}  + \frac{1}{16} \Tr{\nabla_{\mu}\S\nabla_{\nu}\S} \Tr{\nabla^{\mu}\S\nabla^{\nu}\S}  \\
&+\Tr{\nabla_{\mu}\S\nabla^{\mu}\S\nabla_{\nu}\S\nabla^{\nu}\S}+\frac{1}{8}\Tr{\nabla_{\mu}\S\nabla_{\nu}\S\nabla^{\mu}\S\nabla^{\nu}\S}\Bigg]\;.
\end{aligned}
\end{equation}
As discussed in Sec.~\ref{sec:GLdredef},
we can then remove the second order derivative terms by performing the (\Odd covariant) field redefinitions
\begin{equation}
\begin{cases}
\delta\Phi = \dfrac{1}{16} \,\Tr{\nabla_{\mu}\S\nabla^{\mu}\S},\\
\delta g_{\mu\nu} = -\dfrac{1}{8}\,\Tr{\nabla_{\mu}\S\nabla_{\nu}\S}, \\
\delta\S =-\dfrac{1}{2}\,\left(\Box\S-\nabla_{\mu}\Phi\nabla^{\mu}\S\right) + \dfrac{1}{2}\, \S\nabla_{\mu}\S\nabla^{\mu}\S,
\end{cases}
\end{equation}
in the convention of Eq.~\eqref{eq:fieldvariation}, to bring the result into the equivalent form
\begin{align}
I_{1} ~= ~&\frac{1}{4}\,\alpha' \int\!\d^{D}x\,\sqrt{-g}\,e^{-\Phi}\,\Bigg[\frac{1}{16}\Tr{\nabla_{\mu}\S\nabla_{\nu}\S\nabla^{\mu}\S\nabla^{\nu}\S}\!-\! \frac{1}{32} \Tr{\nabla_{\mu}\S\nabla_{\nu}\S} \Tr{\nabla^{\mu}\S\nabla^{\nu}\S} \Bigg]\;.
\end{align}
This precisely coincides with the truncation of Eq.~\eqref{eq:alpha'D} to the scalar fields. 
Our result reproduces also the first order $\alpha'$ expressions of Refs.~\cite{Meissner:1996sa,Hohm:2015doa} for the reduction to $D=1$ dimensions.

\paragraph*{}
Let us finally point out that considering the most generic manifestly diffeomorphism invariant four-derivative 
action~\cite{Metsaev:1987zx}\footnote{
As in Eq.~(\ref{q4D10}) above, we restrict to manifestly diffeomorphism invariant terms. 
The potential gravitational Chern-Simons coupling which appears for the heterotic string is discussed in detail in Sec.~\ref{sec:het} below.}
\begin{align}
\label{eq:mostgeneralaction}
I_{1} &=\alpha' \int \d^{D+d}X\,\sqrt{-\hat{g}}\,e^{-\hat{\phi}}\,\Big(\gamma_{1}\,\hat{R}_{\hat\mu\hat\nu\hat\rho\hat\sigma}\hat{R}^{\hat\mu\hat\nu\hat\rho\hat\sigma}  + \gamma_{2}\, \hat{H}^{\hat\mu\hat\nu\hat\lambda}{\hat{H}^{\hat\rho\hat\sigma}}_{\ \ \,\hat\lambda}\, \hat{R}_{\hat\mu\hat\nu\hat\rho\hat\sigma} +\gamma_{3}\,\hat{H}_{\hat\mu \hat\nu \hat\rho}\hat{H}^{\hat\mu\ \hat\lambda}_{\ \,\hat\sigma}\hat{H}^{\hat\nu\ \hat\tau}_{\ \,\hat\lambda}\hat{H}^{\hat\rho\ \hat\sigma}_{\ \,\hat\tau} \nonumber \\
&+\gamma_{4}\,\hat{H}^{2}_{\hat\mu \hat\nu}\hat{H}^{2\,\hat\mu \hat\nu}+ \gamma_{5}\,(\hat{H}^{2})^{2} +\gamma_{6}\,\hat{H}^{2}_{\hat\mu\hat\nu}\partial^{\hat\mu}\hat\phi\partial^{\hat\nu}\hat\phi+\gamma_{7}\,\hat{H}^{2}\partial_{\hat\mu}\hat\phi\partial^{\hat\mu}\hat\phi+\gamma_{8}\,\partial_{\hat\mu}\hat\phi\partial^{\hat\mu}\hat\phi\partial_{\hat\nu}\hat\phi\partial^{\hat\nu}\hat\phi   \Big),
\end{align}
the only choice of coefficients that give rise to an \Odd invariant action after reduction on a generic 
$d$-dimensional torus is
\begin{equation}
\label{eq:gammacoeff}
\gamma_{2}=-\frac{\gamma_{1}}{2}\;,\ \gamma_{3} = \frac{\gamma_{1}}{24}\;,\ \gamma_{4}=-\frac{\gamma_{1}}{8}\;,\ \gamma_{5}=0\;,\ \gamma_{6}=0\;,\ \gamma_{7}=0\;,\ \gamma_{8}=0\;,
\end{equation}
corresponding to the action~\eqref{eq:alpha'D+d}. Indeed, as the definition of $\Phi$ imposes
\begin{equation}
 \partial_{\mu}\hat{\phi} = \partial_{\mu}\Phi+\frac{1}{2}\,\Tr{G^{-1}\partial_{\mu}G},
\end{equation}
the terms proportional to $\gamma_{6}$, $\gamma_{7}$ and $\gamma_{8}$ respectively in Eq.~\eqref{eq:mostgeneralaction} produce terms carrying a factor $\Tr{G^{-1}\partial_{\mu}G}$. However, there is no \Odd-invariant term in the basis (\ref{basisalpha1}) that contains such a factor, as shown in App.~\ref{app:OddtoGLd}. Moreover, these terms cannot cancel each other, as they come with different contraction structures. This imposes $\gamma_{6}=\gamma_{7}=\gamma_{8}=0$. The computations detailed in Secs.~\ref{sec:two-derivative} and \ref{sec:reduc4d} finally implies the remaining coefficients of Eq.~\eqref{eq:gammacoeff}. Only with this choice do the ${\rm GL}(d)$ terms combine into
the \Odd invariant terms of the basis (\ref{basisalpha1}).
Up to field redefinition, the action~\eqref{eq:alpha'D+d} thus is the unique four-derivative correction
exhibiting \Odd invariance upon dimensional reduction.

\section{Frame formulation}
\label{sec:frame}

In the previous section we have shown that invariance under rigid \Odd transformations 
requires an $\alpha'$-deformation of the transformation rules that resembles a Green-Schwarz  mechanism. 
We will now make this analogy more precise by introducing a frame formalism for which the \Odd symmetry remains 
undeformed, while the  local frame transformations acquire $\alpha'$-deformations.
This formulation uses the standard Green-Schwarz mechanism, albeit with composite gauge fields.

We introduce a frame field $E\equiv (E_{M}{}^{A})$ with inverse $E^{-1}\equiv (E_{A}{}^{M})$ from which the scalar matrix~(\ref{HOdd})
encoding $G$ and $B$ can be reconstructed via 
 \bea
  {\cal H}_{MN} = E_{M}{}^{A} E_{N}{}^{B}\kappa_{AB}\;, 
 \eea
where flat indices are split as $A=(a,\bar{a})$, and 
$\kappa_{AB}$ is a block-diagonal matrix with components $\kappa_{ab}$ and $\kappa_{\bar{a}\bar{b}}$. 
Furthermore, we constrain the frame field by demanding that the `flattened' \Odd metric is also block-diagonal according to 
 \bea\label{newCOnstr}
  \eta_{AB} \equiv E_{A}{}^{M} E_{B}{}^{N}\eta_{MN} =   \begin{pmatrix}  \kappa_{ab} & 0 \\[0.7ex]
 0 & - \kappa_{\bar{a}\bar{b}}
\end{pmatrix}\;, 
 \eea
with a relative sign in the space of barred indices reflecting the signature of the \Odd metric.   
In this formalism $\kappa_{ab}$ and $\kappa_{\bar{a}\bar{b}}$ need not be Kronecker deltas, and in particular can be spacetime dependent,  
 and so there is a local $\GL(d)\times \GL(d)$ frame invariance, 
with transformation rules 
 \bea\label{GLdGLd}
  \delta_{\Lambda}E_{A}{}^{M} =\Lambda_{A}{}^{B} E_{B}{}^{M}\,, \quad \Lambda_{A}{}^{B}= \begin{pmatrix}  \Lambda_{a}{}^{b} & 0 \\[0.7ex]
 0 & \bar\Lambda_{\bar{a}}{}^{\bar{b}}
\end{pmatrix}. 
 \eea
We could partially gauge fix  $\kappa_{AB}=\delta_{AB}$, which reduces the frame transformations to  $\SO(d)\times \SO(d)$, 
but in the following another gauge fixing is  convenient: 
we identify the components of $\kappa$ with the metric $G$ according to 
 \bea
 \kappa = \begin{pmatrix}  2G & 0 \\[0.7ex]
 0 & 2G
\end{pmatrix}\;, 
 \eea
where we used matrix notation.  A frame field satisfying the constraint (\ref{newCOnstr}) and leading to the familiar form of ${\cal H}_{MN}$ 
is then given by 
 \begin{equation}
 \label{sspecialgauge}
  E \equiv  (E_{M}{}^{A})\equiv  \frac{1}{2} \begin{pmatrix}
                  1+BG^{-1} & 1-BG^{-1} \\[0.7ex]
                  G^{-1} & -G^{-1}
                  \end{pmatrix}\, .
 \end{equation}

In order to derive composite connections from the frame field we define the  Maurer-Cartan forms 
 \bea\label{MC}
  (E^{-1}\partial_{\mu}E)_A{}^{B} \equiv \begin{pmatrix}  Q_{\mu a}{}^{b} & P_{\mu a}{}^{\bar{b}} \\[0.7ex]
 \bar{P}_{\mu \bar{a}}{}^{b} & \bar{Q}_{\mu \bar{a}}{}^{\bar{b}}
\end{pmatrix}\;. 
 \eea
>From this definition one finds that under $\GL(d)\times \GL(d)$ transformations (\ref{GLdGLd}) the 
$P_{\mu}$ transform as tensors, and  the $Q_{\mu}$ transform as connections: 
 \bea\label{Qgauge}
  \delta_{\Lambda}Q_{\mu a}{}^{b} = -D_{\mu}\Lambda_{a}{}^{b}\;, \qquad  \delta_{\Lambda}\bar{Q}_{\mu \bar{a}}{}^{\bar{b}} = -D_{\mu}\bar{\Lambda}_{\bar{a}}{}^{\bar{b}}\;, 
 \eea
with $D_{\mu}\Lambda_a{}^{b}=\partial_{\mu}\Lambda_a{}^{b}+[Q_{\mu},\Lambda]_a{}^{b}$ and a similar formula  for the barred expression. 
We can evaluate these connections for the gauge choice (\ref{sspecialgauge}), 
  \begin{equation}
   \label{explicitQ}
  \begin{cases}
  Q_{\mu} = -\dfrac{1}{2}\partial_{\mu}(G-B)G^{-1}, \smallskip\\  
  \bar{Q}_{\mu} = -\dfrac{1}{2}\partial_{\mu}(G+B)G^{-1}\;, 
 \end{cases} 
  \end{equation}
using again   matrix notation. 
  
Having constructed composite gauge fields from the frame field we can consider the familiar   
Chern-Simons three-forms built from them: 
 \bea
  {\rm CS}_{\mu\nu\rho}(Q) \equiv {\rm Tr}\Big(Q_{[\mu}\partial_{\nu}Q_{\rho]}+\frac{2}{3}Q_{[\mu}Q_{\nu}Q_{\rho]}\Big)\;. 
 \eea
These Chern-Simons forms transform under Eq.~\eqref{Qgauge} as 
 \bea\label{CStransform}
  \delta_{\Lambda}{\rm CS}_{\mu\nu\rho}(Q)=\partial_{[\mu}{\rm Tr}\big(\partial_{\nu}\Lambda\, Q_{\rho]}\big)\;, 
 \eea
with the barred formulas being analogous. 
Evaluating  the Chern-Simons-form with Eq.~\eqref{explicitQ} one recovers  precisely the expression \eqref{Omegafirst} encountered in the previous section, up to a global factor 3.  
Therefore, we can define a 3-form curvature with Chern-Simons modification: 
\begin{equation}
   \label{newHH}
  \widetilde{H}_{\mu\nu\rho}\equiv H_{\mu\nu\rho} -\frac{3}{2}\, \alpha'\left({\rm CS}_{\mu\nu\rho}(Q)-  {\rm CS}_{\mu\nu\rho}(\bar{Q})\right)\;, 
\end{equation}
which then reproduces the term proportional to $\Omega H$ encountered in the ${\cal O}(\alpha')$ action. 

We have thus succeeded to find a formulation for which the \Odd invariance is manifestly realized without deformation. Rather, the $\GL(d)\times \GL(d)$ gauge symmetry is deformed by having a  2-form transforming according to the Green-Schwarz mechanism,  
 \begin{equation}
  \delta B_{\mu\nu} = \frac{1}{2}\,\alpha'\,{\rm Tr}\big(\partial_{[\mu}\Lambda\, Q_{\nu]}\big)-\frac{1}{2}\,\alpha'\,{\rm Tr}\big(\partial_{[\mu}\bar{\Lambda}\, \bar{Q}_{\nu]}\big) \,. 
\end{equation}
Performing a  partial gauge fixing to ${\rm SO}(d)\times {\rm SO}(d)$, together with appropriate  field redefinitions, this Green-Schwarz mechanism relates to the reduction
of $\alpha'$-deformed double field theory \cite{Baron:2017dvb}. This formulation is related to the one of the previous section as follows: 
if one fully gauge fixes $\GL(d)\times \GL(d)$ the \Odd transformations acquire deformations through compensating frame transformations 
and hence the singlet $B_{\mu\nu}$ starts transforming non-trivially under \Odd.

Let us close this section by discussing how the $\mathbb{Z}_{2}$ invariance (\ref{Z2metric}) of bosonic string theory is realized in this frame formulation. 
The $\mathbb{Z}_{2}$ acts on the frame field as 
 \begin{equation}\label{ZonFrame}
  E\rightarrow Z^T E \tilde{Z}\;, \quad \tilde{Z}\equiv \begin{pmatrix}
                    0 & 1 \\
                    1 & 0
                  \end{pmatrix}. 
 \end{equation}
The matrix $\tilde{Z}$ exchanges  the two GL$(d)$ factors and hence exchanges the role of unbarred and barred indices. 
Indeed, under the transformation (\ref{ZonFrame}) the Maurer-Cartan forms (\ref{MC}) transform as $P_{\mu}\leftrightarrow \bar{P}_{\mu}$ 
and $Q_{\mu}\leftrightarrow \bar{Q}_{\mu}$, as one may verify by a quick computation and as is suggested by the explicit form~(\ref{explicitQ}). 
Thus,  the relative sign in Eq.~(\ref{newHH}) implies that the total Chern-Simons form is $\mathbb{Z}_{2}$ odd, which together with $B_{\mu\nu}\rightarrow -B_{\mu\nu}$
implies $\mathbb{Z}_{2}$ invariance of the action.


\section{Gravitational Chern-Simons form of the heterotic supergravity}
\label{sec:het}


In this section, we repeat the above analysis of the first order $\alpha'$-corrections
for the case of the heterotic string. In absence of the Yang-Mills field in ten dimensions, 
the bosonic part of the four-derivative effective action of the heterotic string
takes the form~\cite{Metsaev:1987zx}
\begin{equation}
\begin{aligned}
\label{eq:alpha'D+dhet}
\widehat{I}_{1} =\frac14\,\alpha' 
\int \d^{D+d}X\,\sqrt{-\hat{g}}\,e^{-\hat{\phi}}\,&\bigg[-\hat{H}^{\hat\mu\hat\nu\hat\rho}\, \hat{\Omega}^{(\hat\omega)}_{\hat\mu\hat\nu\hat\rho} + \frac{1}{2}\Big(\,\hat{R}_{\hat\mu\hat\nu\hat\rho\hat\sigma}\hat{R}^{\hat\mu\hat\nu\hat\rho\hat\sigma} -\frac{1}{2}\, \hat{H}^{\hat\mu\hat\nu\hat\lambda}{\hat{H}^{\hat\rho\hat\sigma}}_{\ \ \,\hat\lambda}\, \hat{R}_{\hat\mu\hat\nu\hat\rho\hat\sigma} \\
& \qquad
-\frac{1}{8}\,\hat{H}^{2}_{\hat\mu \hat\nu}\hat{H}^{2\,\hat\mu \hat\nu} +\frac{1}{24}\,\hat{H}_{\hat\mu \hat\nu \hat\rho}\hat{H}^{\hat\mu\ \hat\lambda}_{\ \,\hat\sigma}\hat{H}^{\hat\nu\ \hat\tau}_{\ \,\hat\lambda}\hat{H}^{\hat\rho\ \hat\sigma}_{\ \,\hat\tau}\Big) \bigg]\;,
\end{aligned}
\end{equation}
where $D+d=10$.
Apart from terms proportional to the $\alpha'$ corrections of the bosonic string~\eqref{eq:alpha'D+d}, the action features 
the gravitational Chern-Simons form 
$\hat{\Omega}^{(\hat\omega)}_{\hat\mu\hat\nu\hat\rho}$, defined as
\begin{equation}
\hat{\Omega}^{(\hat\omega)}_{\hat\mu\hat\nu\hat\rho} = \Tr{\hat\omega_{[\hat\mu}\partial_{\hat\nu}\,\hat\omega_{\hat\rho]}}+\frac{2}{3}\,\Tr{\hat\omega_{[\hat\mu}\,\hat\omega_{\hat\nu}\,\hat\omega_{\hat\rho]}}
\;,
\label{eq:gravCS}
\end{equation}
in terms of the spin connection
\begin{equation}
\label{eq:defomega}
\hat\omega_{\hat\mu\,\hat\alpha}{}^{\hat\beta} = \nabla_{\hat\mu}\hat{e}_{\hat{\nu}}{}^{\hat{\beta}} \hat{e}_{\hat\alpha}{}^{\hat\nu}\;.
\end{equation}
With the  \Odd invariant form of the bosonic string discussed in 
Secs.~\ref{sec:reduc4d} and~\ref{sec:alpha'action} above, it thus remains to 
reduce the first term of Eq.~(\ref{eq:alpha'D+dhet}).
We follow the same systematics outlined above.

In the flat basis, after dimensional reduction, 
the non-vanishing components of the spin connection are given by
\begin{align}
  \hat{\omega}_{\alpha,\,\beta\gamma} &= \omega_{\alpha,\,\beta\gamma}\;,\nonumber\\
  \hat{\omega}_{\alpha,\,\beta a} &= \dfrac{1}{2} e_{\alpha}{}^{\mu}e_{\beta}{}^{\nu}\eta_{ab}E_{m}{}^{b}\,F_{\mu\nu}^{(1)\,m} \;,\nonumber\\
  \hat{\omega}_{\alpha,\,a b} &= e_{\alpha}{}^{\mu}\,\widetilde{Q}_{\mu\,a}{}^{c}\eta_{cb}\;,\nonumber\\
  \hat{\omega}_{a,\,\alpha\beta} &= - \dfrac{1}{2} e_{\alpha}{}^{\mu}e_{\beta}{}^{\nu}\eta_{ab}E_{m}{}^{b}\,F_{\mu\nu}^{(1)\,m}\;,\nonumber\\
  \hat{\omega}_{a,\,b \alpha} &= e_{\alpha}{}^{\mu}\,\widetilde{P}_{\mu\,a}{}^{c}\eta_{cb}\;.
\end{align}
Here, $\widetilde{P}_{\mu\,a}{}^{b}$ and $\widetilde{Q}_{\mu\,a}{}^{b}$ are, respectively, the symmetric and antisymmetric parts of the ${\rm GL}(d)$ Maurer-Cartan form $\widetilde{J}_{\mu\,a}{}^{b}=E_{a}{}^{m}\,\partial_{\mu}E_{m}{}^{b}=\widetilde{P}_{\mu\,a}{}^{b}+\widetilde{Q}_{\mu\,a}{}^{b}$ and verify the integrability relations
\begin{equation}
 \label{eq:integrabilityPQ}
   \partial_{[\mu}\widetilde{J}_{\nu]\,a}{}^{b} = - (\widetilde{J}_{[\mu}\widetilde{J}_{\nu]})_{a}{}^{b}
   \qquad\Longleftrightarrow\qquad
 \begin{cases}
  \partial_{[\mu}\widetilde{P}_{\nu]\,a}{}^{b} = - (\widetilde{P}_{[\mu}\widetilde{Q}_{\nu]})_{\,a}{}^{b}- (\widetilde{Q}_{[\mu}\widetilde{P}_{\nu]})_{\,a}{}^{b}, \\
  \partial_{[\mu}\widetilde{Q}_{\nu]\,a}{}^{b} = - (\widetilde{Q}_{[\mu}\widetilde{Q}_{\nu]})_{\,a}{}^{b}- (\widetilde{P}_{[\mu}\widetilde{P}_{\nu]})_{\,a}{}^{b}.
 \end{cases}
\end{equation}
Defining the low-dimensional components of $\hat\Omega^{(\hat\omega)}$ in the same way as 
we did for $\hat H$ in Eq.~\eqref{eq:Hreduced}, we obtain
\begin{align}
\label{eq:Omegareduced}
  \Omega_{\mu\nu\rho}^{(\hat\omega)} &= \Omega_{\mu\nu\rho}^{(\omega)} -\dfrac{1}{3}\,\Tr{\widetilde{J}_{[\mu}\widetilde{J}_{\nu}\widetilde{J}_{\rho]}} -\dfrac{1}{2}\,F_{[\mu|\sigma}^{(1)\,m}G_{mn}\nabla_{\vert\nu}^{}F_{\rho]}^{(1)\,\sigma\,n} 
   +\dfrac{1}{4} \,F_{[\mu|\sigma}^{(1)\,m} \nabla^{\sigma}G_{mn}F_{\vert\nu\rho]}^{(1)\,n} 
     \nonumber\\
  &{}
- \dfrac{1}{4}\,e_{\alpha}{}^{\sigma} \nabla_{[\mu\vert}^{}e_{\tau}{}^{\alpha} F_{\vert\nu\rho]}^{(1)\,m}G_{mn}F_{\sigma}^{(1)\,\tau\,n}
   \;,\nonumber\\[1ex]
  \Omega_{\mu\nu m}^{(\hat\omega)} &= \dfrac{1}{6}\,R_{\mu\nu\rho\sigma}F^{(1)\rho\sigma\,n}G_{nm} -\dfrac{1}{12}\,F_{\mu\nu}^{(1)\,n}\left(\nabla_{\rho}GG^{-1}\nabla^{\rho}G\right)_{nm}-\dfrac{1}{6}\,F_{[\mu}^{(1)\,\rho\,n}\left(\nabla_{\nu]}GG^{-1}\nabla_{\rho}G\right)_{nm}\nonumber\\
 &{}- \dfrac{1}{24}\,G_{mn} F^{(1)\rho\sigma\,n}\, F_{\mu\nu}^{(1)\,p}G_{pq}F_{\rho\sigma}^{(1)\,q} - \dfrac{1}{12}\,G_{mn}F^{(1)\rho\sigma\,n} \, F_{[\mu\vert\rho}^{(1)\,p}G_{pq}F_{\vert\nu]\sigma}^{(1)\,q}\,-\dfrac{1}{6}\,\nabla_{[\mu\vert}\nabla^{\rho}G_{mn}F_{\vert\nu]\rho}^{(1)\,n}\nonumber\\
 &{}+\dfrac{1}{6}\,\nabla^{\rho}G_{mn}\nabla_{[\mu}^{}F_{\nu]\rho}^{(1)\,n} +\dfrac{1}{6}\,\nabla_{[\mu}\left(\nabla_{\nu]}e_{\sigma}{}^{\alpha}e_{\alpha}{}^{\rho} F_{\rho}^{(1)\,\sigma\,n}G_{nm}\right) \;,\nonumber\\[1ex]
  \Omega_{\mu mn}^{(\hat\omega)} &= \dfrac{1}{12}\,F_{\rho\sigma}^{(1)\,p}G_{p[m\vert}\,\nabla_{\mu}\left(F^{(1)\rho\sigma\,q}G_{q\vert n]}\right) - \dfrac{1}{6}\,F_{\mu\nu}^{(1)\,p}\nabla_{\rho}G_{p[m}G_{n]q}F^{(1)\rho\nu\,q}
   \nonumber\\
 &{}
 +\dfrac{1}{6}\left(\nabla_{\nu}GG^{-1}\nabla_{\mu}\nabla^{\nu}G\right)_{[mn]}\;.
\end{align}

We can now focus on the reduction of the action. 
Splitting the ten-dimensional indices $\hat\mu$ into $(\mu,m)$, we obtain
\begin{equation}
\hat{H}^{\hat\mu\hat\nu\hat\rho}\, \hat{\Omega}^{(\hat\omega)}_{\hat\mu\hat\nu\hat\rho} = H^{\mu\nu\rho}\Omega_{\mu\nu\rho}^{(\hat\omega)} + 3\,H^{\mu\nu m}\Omega_{\mu\nu m}^{(\hat\omega)} +3\,H^{\mu mn}\Omega_{\mu mn}^{(\hat\omega)} + H^{mnp}\Omega_{mnp}^{(\hat\omega)}\;.
\end{equation}
Using the explicit expressions 
of Eqs.~\eqref{eq:Hreduced} and \eqref{eq:Omegareduced}, the reduced Chern-Simons form then takes the form
\begin{equation}
  \label{eq:hetsecondorder}
 \begin{aligned}
  & \!\!\!\!\!\!\!\!\!\!\!\!\!\!\!\!\!\!\!\!\!\!\!\!
  -\frac{1}{4}\,\alpha' \int \d^{D+d}X\,\sqrt{-\hat{g}}\,e^{-\hat{\phi}} \,\hat{H}^{\hat\mu\hat\nu\hat\rho}\, \hat{\Omega}^{(\hat\omega)}_{\hat\mu\hat\nu\hat\rho} ~\longrightarrow \\
  &\frac{\alpha'}{4}  \int\d^{D}x\,\sqrt{-g}\,e^{-\Phi}\,\Big[ -H^{\mu\nu\rho}\Omega_{\mu\nu\rho}^{(\omega)}  + \frac{1}{3}\, H^{\mu\nu\rho}\,\Tr{\widetilde{J}_{\mu}\widetilde{J}_{\nu}\widetilde{J}_{\rho}}
  - \frac{1}{2}\,R^{\mu\nu\rho\sigma}F_{\mu\nu}^{(1)\,m}H_{\rho\sigma\,m} 
   \\
  &\qquad -\frac{1}{4}\,H^{\mu\nu\rho}F_{\mu\sigma}^{(1)\,m} \nabla^{\sigma}G_{mn}F_{\nu\rho}^{(1)\,n} +\frac{1}{8}\,F_{\mu\nu}^{(1)\,m}G_{mn}F_{\rho\sigma}^{(1)\,n} \,F^{(1)\rho\sigma\,p}H^{\mu\nu}{}_{p} \\
  &\qquad +\frac{1}{4}\,F_{\mu\nu}^{(1)\,m}H_{\rho\sigma\,m} \, F^{(1)\mu\rho\,p}G_{pq}F^{(1)\nu\sigma\,q} - \frac{1}{4}\,H_{\mu\nu\,m}\left(\nabla_{\rho}G^{-1}\nabla^{\rho}G\right)^{m}{}_{n}F^{(1)\mu\nu\,n} \\
  &\qquad- \frac{1}{4}\,F_{\mu\nu}^{(1)\,m}\left(\nabla_{\rho}BG^{-1}\nabla^{\rho}G\right)_{mn}F^{(1)\mu\nu\,n} + \frac{1}{2}\,H_{\mu\nu\,m}\left(\nabla_{\rho}G^{-1}\nabla^{\mu}G\right)^{m}{}_{n}F^{(1)\nu\rho\,n} \\
  &\qquad+ \frac{1}{2}\,F_{\mu\nu}{}^{(1)\,m}\left(\nabla_{\rho}GG^{-1}\nabla^{\mu}B\right)_{mn}F^{(1)\rho\nu\,n} + \frac{1}{4}\,H^{\mu\nu\rho}e_{\alpha}{}^{\sigma} \nabla_{\mu}^{}e_{\tau}{}^{\alpha} F_{\nu\rho}^{(1)\,m}G_{mn}F_{\sigma}^{(1)\,\tau\,n}  \\
  &\qquad  - \frac{1}{2}\,H^{\mu\nu}{}_{m}G^{mn}\nabla_{\mu}\left(\nabla_{\nu}e_{\sigma}{}^{\alpha}e_{\alpha}{}^{\rho} F_{\rho}^{(1)\,\sigma\,p}G_{np}\right)+ \frac{1}{2}\,\Tr{\nabla_{\mu}\nabla_{\nu}G\nabla^{\mu}G^{-1}\nabla^{\nu}BG^{-1}} \\
  &\qquad - \frac{1}{2}\,H_{\mu\nu\,m}\left(G^{-1}\nabla_{\rho}G\right)^{m}{}_{n}\nabla^{\mu}F^{(1)\nu\rho\,n}+ \frac{1}{2}\,H_{\mu\nu\,m}\left(G^{-1}\nabla^{\mu}\nabla_{\rho}G\right)^{m}{}_{n}F^{(1)\nu\rho\,n} \\
  &\qquad +\frac{1}{2}\,H^{\mu\nu\rho}F_{\mu\sigma}^{(1)\,m}G_{mn}\nabla_{\nu}^{}F_{\rho}^{(1)\,\sigma\,n} - \frac{1}{4}\,F_{\mu\nu}^{(1)\,m}\nabla_{\rho}B_{mn}\nabla^{\rho}F^{(1)\mu\nu\,n} \Big]\;.
 \end{aligned}
\end{equation}
Only the six last terms carry second order derivatives. Following the systematics of Sec.~\ref{sec:reduc4d},
these terms can be transformed by means of partial integration and Bianchi identities
such that all second order derivatives appear as the leading two-derivative contribution from the field Eqs.~\eqref{eq:eomGLd},
i.e.\ appear within the first column of Tab.~\ref{tab:fieldredef}. Details are given in App.~\ref{app:heterotic}.
Specifically, the remaining second order derivative terms combine into
\begin{align}
  &\frac{\alpha'}{8} \int\d^{D}x\,\sqrt{-g}\,e^{-\Phi}\,\Big[ \nabla_{\mu}H^{\mu\nu}{}_{m}F_{\rho}^{(1)\,\sigma\,m}\nabla_{\nu}e_{\sigma}{}^{\alpha}e_{\alpha}{}^{\rho} -\Tr{\Box BG^{-1}\nabla_{\nu}G\nabla^{\nu}G^{-1}
  -\Box G\nabla_{\nu}G^{-1}\nabla^{\nu}BG^{-1}
  }\nonumber\\
  &\qquad
  -\nabla_{\mu}F^{(1)\mu\nu\,m}\left(\nabla^{\rho}GG^{-1}\right)_{m}{}^{n}H_{\nu\rho\,n}-\frac{1}{2}\,H_{\mu\nu\,m}\left(G^{-1}\Box G\right)^{m}{}_{n}F^{(1)\mu\nu\,n} \nonumber\\
  &\qquad
   -\frac{1}{2}\,H_{\mu\nu\rho}\nabla_{\sigma}F^{(1)\sigma\mu\,m}G_{mn}F^{(1)\nu\rho\,n} -\nabla_{\mu}F^{(1)\mu\nu\,m}\nabla^{\rho}B_{mn}F^{(1)\,n}_{\nu\rho}\Big]\;,
\end{align}
and can be eliminated by field redefinitions according to the rules defined in Tab.~\ref{tab:fieldredef}.
These take the explicit form (in the convention of Eq.~\eqref{eq:fieldvariation})
\begin{align}
\label{eq:fieldredefhet}
    \delta B_{\mu\nu}&=
    \dfrac{1}{8}\,A_{[\mu}^{(1)\,m}\nabla_{\nu]}^{}e_{\sigma}{}^{\alpha}e_{\alpha}{}^{\rho}G_{mn}\,F_{\rho}^{(1)\,\sigma\,n}
    +\dfrac{1}{8}\,A_{[\mu\vert\,m}^{(2)}\nabla^{\rho}G^{mn}H_{\vert\nu]\rho\,n}
    -\dfrac{1}{8}\,B_{mn}A_{[\mu}^{(1)\,n}\nabla^{\rho}G^{mp}H_{\nu]\rho\,p}\nonumber\\
    &{} -\dfrac{1}{16}\,A_{[\mu\vert\,m}^{(2)} H_{\vert\nu]\rho\sigma}\,F^{(1)\rho\sigma\,m}
    +\dfrac{1}{16}\,B_{mn}A_{[\mu}^{(1)\,n} H_{\nu]\rho\sigma}\,F^{(1)\rho\sigma\,m}
    -\dfrac{1}{8}\,A_{[\mu\vert\,m}^{(2)}\left(G^{-1}\nabla^{\rho}B\right)^{m}{}_{n}\,F^{(1)\,n}_{\vert\nu]\rho}
    \nonumber\\
    &{} +\dfrac{1}{8}\,B_{mn}A_{[\mu\vert}^{(1)\,n}\left(G^{-1}\nabla^{\rho}B\right)^{m}{}_{p}\,F^{(1)\,p}_{\vert\nu]\rho}\;,
    \nonumber\\[1ex]
    \delta G^{mn}&=\dfrac{1}{4}\,\left(\nabla_{\mu}G^{-1}\nabla^{\mu}BG^{-1}\right)^{(mn)} -\dfrac{1}{8}\,F^{(1)\mu\nu\,(m}G^{n)p}H_{\mu\nu\,p}\;,\nonumber\\
    \delta B_{mn}&=\dfrac{1}{4}\,\left(\nabla_{\mu}G\nabla^{\mu}G^{-1}G\right)_{[mn]} \;,\nonumber\\
    \delta A_{\mu}^{(1)m}&=-\dfrac{1}{8}\,\nabla^{\nu}G^{mn}H_{\mu\nu\,n} +\dfrac{1}{16}\,H_{\mu\nu\rho}\,F^{(1)\nu\rho\,m} +\dfrac{1}{8}\,\left(G^{-1}\nabla^{\nu}B\right)^{m}{}_{n}F^{(1)\,n}_{\mu\nu} \;,\nonumber\\[1ex]
    \delta A_{\mu\,m}^{(2)}&=-\dfrac{1}{8}\,\nabla_{\mu}e_{\sigma}{}^{\alpha}e_{\alpha}{}^{\rho} G_{mn}\,F_{\rho}^{(1)\,\sigma\,n} -\dfrac{1}{8}\,B_{mn}\nabla^{\nu}G^{np}H_{\mu\nu\,p} +\dfrac{1}{16}\,H_{\mu\nu\rho}B_{mn}\,F^{(1)\nu\rho\,n}\nonumber\\
    &{} +\dfrac{1}{8}\,\left(BG^{-1}\nabla^{\nu}B\right)_{mn}\,F^{(1)\,n}_{\mu\nu}\;.  
\end{align}
After applying these field redefinitions,
the resulting reduced action does no longer carry any second order derivative (except within the Riemann tensor), 
and turns into
 \begin{align}
  &\!\!\!\!\!\!\!\!
  -\frac{1}{4}\,\alpha'
  \int \d^{D+d}X\,\sqrt{-\hat{g}}\,e^{-\hat{\phi}} \,\hat{H}^{\hat\mu\hat\nu\hat\rho}\, \hat{\Omega}^{(\hat\omega)}_{\hat\mu\hat\nu\hat\rho}
  ~\longrightarrow \nonumber \\
  & \frac{\alpha'}{4}
  \int\d^{D}x\,\sqrt{-g}\,e^{-\Phi}\,\Big[ -H^{\mu\nu\rho}\,\Omega_{\mu\nu\rho}^{(\omega)}  + \frac{1}{3}\, H^{\mu\nu\rho}\,\Tr{\widetilde{J}_{\mu}\widetilde{J}_{\nu}\widetilde{J}_{\rho}} \nonumber\\
  &\qquad - \frac{1}{2}\,R^{\mu\nu\rho\sigma}F_{\mu\nu}^{(1)\,m}H_{\rho\sigma\,m} + \frac{1}{8}\,H^{\mu\nu\lambda}H^{\rho\sigma}{}_{\lambda}F_{\mu\nu}^{(1)\,m}H_{\rho\sigma\,m} -\frac{1}{4}\,H^{\mu\nu\rho}F_{\mu\sigma}^{(1)\,m} \nabla^{\sigma}G_{mn}F_{\nu\rho}^{(1)\,n} \nonumber\\
  &\qquad -\frac{1}{4}\,H^{\mu\nu\rho}H_{\mu\sigma\,m}\nabla^{\sigma}G^{mn}H_{\nu\rho\,n} -\frac{1}{4}\,H^{\mu\nu\rho}F_{\mu\sigma}^{(1)\,m} \left(\nabla^{\sigma}BG^{-1}\right)_{m}{}^{n} H_{\nu\rho\,n} \nonumber\\
  &\qquad +\frac{1}{4}\,H^{\mu\nu\rho}H_{\mu\sigma\,m} \left(G^{-1}\nabla^{\sigma}B\right)^{m}{}_{n}F_{\nu\rho}^{(1)\,n} +\frac{1}{8}\,F_{\mu\nu}^{(1)\,m}G_{mn}F_{\rho\sigma}^{(1)\,n} \,F^{(1)\rho\sigma\,p}H^{\mu\nu}{}_{p} \nonumber\\
  &\qquad +\frac{1}{8}\,H_{\mu\nu\,m}G^{mn}H_{\rho\sigma\,n} \,F^{(1)\rho\sigma\,p}H^{\mu\nu}{}_{p}- \frac{1}{4}\,H_{\mu\nu\,m}\left(\nabla_{\rho}G^{-1}\nabla^{\rho}G\right)^{m}{}_{n}F^{(1)\mu\nu\,n} \nonumber\\
  &\qquad - \frac{1}{4}\,F_{\mu\nu}^{(1)\,m}\left(\nabla_{\rho}BG^{-1}\nabla^{\rho}G\right)_{mn}F^{(1)\mu\nu\,n} - \frac{1}{4}\,H_{\mu\nu\,m}\left(\nabla_{\rho}G^{-1}\nabla^{\rho}BG^{-1}\right)^{mn}H^{\mu\nu}{}_{n} \nonumber\\
  &\qquad - \frac{1}{4}\,H_{\mu\nu\,m}\left(G^{-1}\nabla_{\rho}BG^{-1}\nabla^{\rho}B\right)^{m}{}_{n}F^{(1)\mu\nu\,n} -\frac{1}{2}\,H_{\mu\nu\,m}\left(G^{-1}\nabla^{\mu}BG^{-1}\nabla_{\rho}B\right)^{m}{}_{n}F^{(1)\nu\rho\,n}\nonumber\\
  &\qquad+\frac{1}{2}\,F_{\mu\nu}^{(1)\,m}\left(\nabla^{\mu}GG^{-1}\nabla_{\rho}B\right)_{mn}F^{(1)\nu\rho\,n}- \frac{1}{2}\,F_{\mu\nu}^{(1)\,m}\left(\nabla^{\mu}G\nabla_{\rho}G^{-1}\right)_{m}{}^{n}H^{\nu\rho}{}_{n}\nonumber\\
  &\qquad+ \frac{1}{2}\,H_{\mu\nu\,m}\left(G^{-1}\nabla^{\mu}B\nabla_{\rho}G^{-1}\right)^{mn}H^{\nu\rho}{}_{n} -\frac{1}{2}\,\Tr{\nabla_{\mu}B\nabla^{\mu}G^{-1}\nabla_{\nu}G\nabla^{\nu}G^{-1}}\nonumber\\
  &\qquad+\frac{1}{2}\,\Tr{\nabla_{\mu}BG^{-1}\nabla^{\mu}B\nabla_{\nu}G^{-1}\nabla^{\nu}BG^{-1}}\Big]\;.
 \end{align}
The terms appearing in this expression
can finally be compared to the $\GL(d)$ decompositions of the \Odd basis 
as collected in App.~\ref{app:OddtoGLd}. 
This allows to recast the result into the form
\begin{equation}
\label{eq:CSgravreduced}
 \begin{aligned}
  &\!\!\!\!\!\!\!\!
  -\frac{1}{4}\,\alpha'
  \int \d^{D+d}X\,\sqrt{-\hat{g}}\,e^{-\hat{\phi}} \,\hat{H}^{\hat\mu\hat\nu\hat\rho}\, \hat{\Omega}^{(\hat\omega)}_{\hat\mu\hat\nu\hat\rho}
  ~\longrightarrow \\
  &\qquad\frac{\alpha'}{4}  \int\d^{D}x\,\sqrt{-g}\,e^{-\Phi}\,\Big[ -H^{\mu\nu\rho}\,\Omega_{\mu\nu\rho}^{(\omega)} +\frac{1}{16}\,\Tr{\S\nabla_{\mu}\S\nabla^{\mu}\S\nabla_{\nu}\S\nabla^{\nu}\S} \\
  &\qquad  + \frac{1}{16}\,\F_{\mu\nu}{}^{M}\F_{\rho\sigma\,M}\,\F^{\mu\nu\,P}\S_{P}{}^{Q}\F^{\rho\sigma}{}_{Q} - \frac{1}{4}\,R^{\mu\nu\rho\sigma}\F_{\mu\nu}^{M}\F_{\rho\sigma\,M} + \frac{1}{16}\,H^{\mu\nu\lambda}H^{\rho\sigma}{}_{\lambda}\F_{\mu\nu}^{M}\F_{\rho\sigma\,M} \\
  &\qquad  -\frac{1}{8}\,\F_{\mu\nu}{}^{M}\left(\nabla_{\rho}\S\nabla^{\rho}\S\right)_{M}{}^{N}\F^{\mu\nu}{}_{N} -\frac{1}{4}\,\F_{\mu\nu}{}^{M}\left(\nabla^{\mu}\S\nabla_{\rho}\S\right)_{M}{}^{N}\F^{\nu\rho}{}_{N}\\
  &\qquad-\frac{1}{4}\,H^{\mu\nu\rho}\F_{\mu\sigma}{}^{M}\nabla^{\sigma}\S_{M}{}^{N}\F_{\nu\rho\,N}+ \frac{1}{3}\, H^{\mu\nu\rho}\,\Tr{\widetilde{J}_{\mu}\widetilde{J}_{\nu}\widetilde{J}_{\rho}} \Big]\;,
 \end{aligned}
\end{equation}
manifestly \Odd invariant, except for the last term which carries the $\GL(d)$ Chern-Simons form
\begin{equation}
\Omega_{\mu\nu\rho}^{(\widetilde{J})}=\Tr{\widetilde{J}_{[\mu}\widetilde{J}_{\nu}\widetilde{J}_{\rho]}}
\;.
\end{equation}
This form is closed by virtue of the integrability relations~\eqref{eq:integrabilityPQ}. 
It can thus locally be integrated into a 2-form
\begin{equation}
\Omega_{\mu\nu\rho}^{(\widetilde{J})} = 3\,\partial_{[\mu}\theta_{\nu\rho]}^{\rm WZW}
\;,
\end{equation}
such that
the last term in Eq.~\eqref{eq:CSgravreduced} can be absorbed into a field 
redefinition 
\begin{equation}
\delta B_{\mu\nu} = \frac12\,\theta_{\mu\nu}^{\rm WZW}
\;.
\label{eq:delBCS}
\end{equation}
As $\Omega_{\mu\nu\rho}^{(\widetilde{J})}$ is \Odd-invariant, this does not affect 
 the behaviour of $B_{\mu\nu}$ under \Odd transformations. 
 Putting everything together, the reduced action for the bosonic part of heterotic supergravity 
 (in absence of the ten-dimensional Yang-Mills field) 
 is obtained by combining Eqs.~\eqref{eq:alpha'D} and \eqref{eq:CSgravreduced} into
\begin{align}
 \label{eq:fullactionhet}
 I=&\int\d^{D}x\,\sqrt{-g}\,e^{-\Phi}\Bigg[R+\partial_{\mu}\Phi\,\partial^{\mu}\Phi-\frac{1}{12}\,\widetilde{H}_{\mu\nu\rho}\widetilde{H}^{\mu\nu\rho}+\frac{1}{8}\,\Tr{\partial_{\mu}\S\,\partial^{\mu}\S}-\frac{1}{4}\,\F_{\mu\nu}^{M}\,{\S_{M}}^{N}\,\F^{\mu\nu}_{N} \nonumber \\
 &\quad-\frac{1}{4}\,\alpha'\,\Big(H^{\mu\nu\rho}\Omega_{\mu\nu\rho}^{(\omega)} -\frac{1}{16}\,\Tr{\S\nabla_{\mu}\S\nabla^{\mu}\S\nabla_{\nu}\S\nabla^{\nu}\S} - \frac{1}{16}\,\F_{\mu\nu}{}^{M}\F_{\rho\sigma\,M}\,\F^{\mu\nu\,P}\S_{P}{}^{Q}\F^{\rho\sigma}{}_{Q} \nonumber\\
  &\quad   + \frac{1}{4}\,R^{\mu\nu\rho\sigma}\F_{\mu\nu}^{M}\F_{\rho\sigma\,M} - \frac{1}{16}\,H^{\mu\nu\lambda}H^{\rho\sigma}{}_{\lambda}\F_{\mu\nu}^{M}\F_{\rho\sigma\,M} +\frac{1}{8}\,\F_{\mu\nu}{}^{M}\left(\nabla_{\rho}\S\nabla^{\rho}\S\right)_{M}{}^{N}\F^{\mu\nu}{}_{N} \nonumber \\
  &\quad +\frac{1}{4}\,\F_{\mu\nu}{}^{M}\left(\nabla^{\mu}\S\nabla_{\rho}\S\right)_{M}{}^{N}\F^{\nu\rho}{}_{N}+\frac{1}{4}\,H^{\mu\nu\rho}\F_{\mu\sigma}{}^{M}\nabla^{\sigma}\S_{M}{}^{N}\F_{\nu\rho\,N} \Big)\nonumber \\
&\quad+\frac{1}{8}\,\alpha' \,\Big(R_{\mu\nu\rho\sigma}R^{\mu\nu\rho\sigma}-\frac{1}{2}\, R_{\mu\nu\rho\sigma}{H}^{\mu\nu\lambda}{H}^{\rho\sigma}_{\ \ \,\lambda}   + \frac{1}{24}\, H_{\mu\nu\rho}{H}^{\mu\ \, \lambda}_{\ \,\sigma}{H}^{\nu\ \,\tau}_{\ \,\lambda}{H}^{\rho\ \,\sigma}_{\ \,\tau}
  \nonumber \\
  &\quad -\frac{1}{8}\,{H}^{2}_{\mu\nu}{H}^{2\,\mu\nu}
 +\frac{1}{16}\,\Tr{\nabla_{\mu}\S\nabla_{\nu}\S\nabla^{\mu}\S\nabla^{\nu}\S} - \frac{1}{32}\, \Tr{\nabla_{\mu}\S\nabla_{\nu}\S} \Tr{\nabla^{\mu}\S\nabla^{\nu}\S} \nonumber \\
  &\quad+\frac{1}{8}\, {\F_{\mu\nu}}^{M}{\S_{M}}^{N}\F_{\rho\sigma\,N}\F^{\mu\rho\,P}{\S_{P}}^{Q}{\F^{\nu\sigma}}_{Q} -\frac{1}{2}\,{\F_{\mu\nu}}^{M}{\S_{M}}^{N}{\F^{\mu\rho}}_{N}\F^{\nu\sigma\,P}{\S_{P}}^{Q}\F_{\rho\sigma\,Q} \nonumber \\
  &\quad+\frac{1}{8}\,{\F_{\mu\nu}}^{M}{\F_{\rho\sigma}}_{M}\F^{\mu\rho\,N}{\F^{\nu\sigma}}_{N}  -\frac{1}{2}\, R_{\mu\nu\rho\sigma} \F^{\mu\nu\,M}{\S_{M}}^{N}{\F^{\rho\sigma}}_{N} \nonumber \\
  &\quad+\frac{1}{8}\,H^{2}_{\mu\nu}\Tr{\nabla^{\mu}\S\nabla^{\nu}\S}-\frac{1}{2}\,H^{2}_{\mu\nu}{{\F^{\mu}}_{\rho}}^{M}{\S_{M}}^{N}{\F^{\nu\rho}}_{N} +\frac{1}{4}\, {H}^{\mu\nu\lambda}{H}^{\rho\sigma}_{\ \ \,\lambda}{\F_{\mu\rho}}^{M}{\S_{M}}^{N}\F_{\nu\sigma\,N} \nonumber\\
  &\quad-\frac{1}{2}\,{\F_{\mu\nu}}^{M}{\left(\S\nabla_{\rho}\S\nabla^{\nu}\S\right)_{M}}^{N} {\F^{\mu\rho}}_{N} + \frac{1}{4}\,\F^{\mu\rho\,M}{\S_{M}}^{N}{\F^{\nu}}_{\rho\,N}\Tr{\nabla_{\mu}\S\nabla_{\nu}\S}\nonumber \\
  &\quad-\frac{1}{2}\,H^{\mu\nu\rho}{\F_{\mu\sigma}}^{M}{\left(\S\nabla_{\nu}\S\right)_{M}}^{N}{{\F_{\rho}}^{\sigma}}_{N} \Big)\Bigg]\;.
  \end{align}

Let us finally note that one could have started equivalently from the ten-dimensional action formulated
in terms of the gravitational Chern-Simons form built from the Christoffel connection
\begin{equation}
  \hat{\Omega}^{(\hat\Gamma)}_{\hat\mu\hat\nu\hat\rho} = \hat\Gamma_{[\hat\mu\vert\hat\sigma}^{\hat\tau}\partial_{\vert\hat\nu}\,\hat\Gamma_{\hat\rho]\hat\tau}^{\hat\sigma}+\frac{2}{3}\,\hat\Gamma_{[\hat\mu\vert\hat\sigma}^{\hat\tau}\,\hat\Gamma_{\vert\hat\nu\vert\hat\eta}^{\hat\tau}\,\hat\Gamma_{\vert\hat\rho]\hat\sigma}^{\hat\eta}\;.
\end{equation}
This form is invariant under Lorentz transformations and 
related to Eq.~(\ref{eq:gravCS}) by~\cite{Hohm:2014eba}
\begin{equation}
 \hat{\Omega}^{(\hat\Gamma)}_{\hat\mu\hat\nu\hat\rho} = \hat{\Omega}^{(\hat\omega)}_{\hat\mu\hat\nu\hat\rho} + \partial_{[\hat\mu}\left(\partial_{\hat\nu\vert}\hat{e}_{\hat\sigma}{}^{\hat\alpha}\hat{e}_{\hat\beta}{}^{\hat\sigma}\hat{\omega}_{\vert\hat{\rho}]\hat\alpha}{}^{\hat\beta}\right) + \frac{1}{3}\,\Tr{\partial_{[\hat\mu}\hat{e}\hat{e}^{-1}\partial_{\hat\nu}\hat{e}\hat{e}^{-1}\partial_{\hat\rho]}\hat{e}\hat{e}^{-1}}\;,
 \label{CSCS}
\end{equation}
with the difference given by two closed terms that can be absorbed by a ten-dimensional field redefinition.
Dimensional reduction of the resulting ten-dimensional action then induces a lower-dimensional action
in which the ${\rm Tr}(\tilde{J}^3)$ term from Eq.~(\ref{eq:CSgravreduced}) is no longer present.
The field redefinition required in order to absorb the closed terms of Eq.~(\ref{CSCS}) precisely 
corresponds to the lower-dimensional field redefinition we have encountered in Eq.~(\ref{eq:delBCS}).

\section{Conclusions}

In this paper we have set up a systematic procedure for analyzing the 
higher-derivative corrections of the bosonic and the heterotic string upon toroidal compactification.
In particular, we have discussed how to control the ambiguities that arise due to 
non-linear field redefinitions and partial integration.
This establishes the basis for analyzing the realization of \Odd invariance
of the dimensionally reduced action.
At first order in $\alpha'$, we have presented the explicit reduction of the bosonic string 
and cast the result into a manifestly \Odd invariant form upon identification of the
necessary field redefinitions.
In particular, the analysis confirms that at order $\alpha'$, the \Odd invariance
of the dimensionally reduced action fixes all the couplings in higher dimensions
(up to an overall factor).
The analysis has revealed the need for a Green-Schwarz type mechanism by which the
lower-dimensional two-form (which is originally singlet under \Odd) acquires a
non-trivial transformation of order~$\alpha'$.
This is a genuine deformation which cannot be eliminated by further field redefinitions. 

We have also extended the analysis to the bosonic sector of the heterotic string
(in absence of the ten-dimensional vector fields).
In particular, we have given the complete set of non-linear field redefinitions 
(\ref{eq:fieldredef}), (\ref{eq:fieldredefhet})
which translate between the original ten-dimensional fields and the 
\Odd-covariant lower-dimensional fields.
This dictionary allows to exploit the \Odd symmetry as a solution
generating method for the heterotic string~\cite{Gasperini:1991qy,Ali:1992mj}
to first order in $\alpha'$. Examples of such solutions have been constructed in Refs.
\cite{Cano:2018qev,Cano:2018brq,Chimento:2018kop}.
It would be very interesting to extend the analysis to also include the ten-dimensional 
vector fields~\cite{Bergshoeff:1989de},
resulting in an ${\mathrm{O}(d,d+K,\mathbb{R})}$ extension of the present results
with the larger group broken down by the non-abelian gauge couplings~\cite{Hohm:2014sxa}.

In principle, the method we have outlined is fully systematic and could be applied to 
higher-order $\alpha'$-corrections. In practice, the number of terms quickly explodes
and calls for complementary techniques to be combined with the present approach.
As noted above, already at order $\alpha'{}^2$ the number of manifestly $\Odd$ invariant
terms in lower dimensions  amounts to 1817.
Nevertheless, it would be interesting to compare the resulting structures 
to related work in Ref.~\cite{Garousi:2019cdn,Garousi:2019mca}.
It would also be interesting to investigate the effect of $\alpha'$-corrections on the more general
Yang-Baxter type deformations recently explored in Ref.~\cite{Borsato:2020bqo}.

Finally, it will be interesting to further study the simplifications arising in the
resulting actions upon reduction to particularly low dimensions $D$.
For $D=1$, all terms other than the scalar couplings disappear 
from Eqs.~(\ref{eq:fullaction}) and (\ref{eq:fullactionhet}),
and we recover the lowest-order result of Refs.~\cite{Meissner:1996sa,Hohm:2015doa,Hohm:2019jgu}.
At $D=2$, the two-form couplings disappear and the vector fields may be integrated out.
Particularly interesting is the three-dimensional case. 
At $D=3$, the two-form may be integrated out. With a field equation of the type
\bea
\nabla^\mu (e^{-\Phi}\,H_{\mu\nu\rho}) &=& {\cal O}(\alpha')
\;, 
\eea
this introduces an integration constant 
which in particular turns the coupling (\ref{Omegafirst}) into  
a three-dimensional analogue of the WZW model, \textit{c.f.} Ref.~\cite{Ferretti:1992fg}.
Furthermore, in $D=3$, the (abelian) vector fields may be dualized into scalars.
While this dualization is still possible in the presence of $\alpha'$-corrections, 
the symmetry enhancement to ${\mathrm{O}(d+1,d+1,\mathbb{R})}$
encountered for the two-derivative action breaks down at order $\alpha'$
and is replaced by the appearance of the relevant automorphic forms
\cite{Lambert:2006he,Bao:2007er}.

\subsection*{Acknowledgements}
We thank J. Maharana and D. Marqu\'es for helpful discussions.
The work of O.H. is supported by the ERC Consolidator Grant ``Symmetries \& Cosmology''.


\begin{appendix}
\section*{Appendix}

\section{Basis at order \texorpdfstring{$\boldsymbol{\alpha'}$}{alpha'}}
\label{app:order1basis}

In this appendix, we explicitly spell out the \Odd invariant basis schematically given in Eq.~(\ref{basisalpha1}), 
whose existence we have
deduced in Sec.~\ref{subsec:basis} and which we have used in order to bring the reduced action into 
manifestly \Odd invariant form.
The basis is built from 61 terms which we list according to their different structures.

\paragraph*{$\boldsymbol{R^2}:$}
\begin{align}
  \Big\{ & R_{\mu\nu\rho\sigma}R^{\mu\nu\rho\sigma}\Big\}
\end{align}

\paragraph*{$\boldsymbol{H^4}:$}
\begin{align}
  \Big\{ & (H^{2})^{2},H^{2\,\mu\nu}H^{2}_{\mu\nu},H_{\mu\nu\rho}{{H^{\mu}}_{\alpha}}^{\beta}{{H^{\nu}}_{\beta}}^{\gamma}{{H^{\rho}}_{\gamma}}^{\alpha} \Big\}
\end{align}

\paragraph*{$\boldsymbol{(\nabla\Phi)^4}:$}
\begin{align}
  \Big\{ & \nabla_{\mu}\Phi\nabla^{\mu}\Phi\nabla_{\nu}\Phi\nabla^{\nu}\Phi \Big\}
\end{align}

\paragraph*{$\boldsymbol{(\nabla\S)^4}:$}
\begin{equation}
\begin{aligned}
  \Big\{ & \Tr{\nabla_{\mu}\S\nabla^{\mu}\S\nabla_{\nu}\S\nabla^{\nu}\S}, \Tr{\nabla_{\mu}\S\nabla_{\nu}\S\nabla^{\mu}\S\nabla^{\nu}\S}, \Tr{\S\nabla_{\mu}\S\nabla^{\mu}\S\nabla_{\nu}\S\nabla^{\nu}\S} \\
  & \Tr{\nabla_{\mu}\S\nabla^{\mu}\S}\,\Tr{\nabla_{\nu}\S\nabla^{\nu}\S},\Tr{\nabla_{\mu}\S\nabla_{\nu}\S}\,\Tr{\nabla^{\mu}\S\nabla^{\nu}\S}\Big\}
\end{aligned}
\end{equation}

\paragraph*{$\boldsymbol{\F^4}:$}
\begin{equation}
\begin{aligned}
 \Big\{ & {\F_{\mu\nu}}^{M}{\F^{\mu\nu}}_{M}\,{\F_{\rho\sigma}}^{N}{\F^{\rho\sigma}}_{N}, {\F_{\mu\nu}}^{M}{\S_{M}}^{N}{\F^{\mu\nu}}_{N}\,{\F_{\rho\sigma}}^{P}{\F^{\rho\sigma}}_{P}, {\F_{\mu\nu}}^{M}{\S_{M}}^{N}{\F^{\mu\nu}}_{N}\,{\F_{\rho\sigma}}^{P}{\S_{P}}^{Q}{\F^{\rho\sigma}}_{Q}, \\
 & {\F_{\mu\nu}}^{M}{\F_{\rho\sigma}}_{M}\,{\F^{\mu\nu}}^{N}{\F^{\rho\sigma}}_{N}, {\F_{\mu\nu}}^{M}{\S_{M}}^{N}{\F_{\rho\sigma}}_{N}\,{\F^{\mu\nu}}^{P}{\F^{\rho\sigma}}_{P}, {\F_{\mu\nu}}^{M}{\S_{M}}^{N}{\F_{\rho\sigma}}_{N}\,{\F^{\mu\nu}}^{P}{\S_{P}}^{Q}{\F^{\rho\sigma}}_{Q}, \\
 & {\F_{\mu\nu}}^{M}{\F_{\rho\sigma}}_{M}\,{\F^{\mu\rho}}^{N}{\F^{\nu\sigma}}_{N}, {\F_{\mu\nu}}^{M}{\S_{M}}^{N}{\F_{\rho\sigma}}_{N}\,{\F^{\mu\rho}}^{P}{\F^{\nu\sigma}}_{P}, {\F_{\mu\nu}}^{M}{\S_{M}}^{N}{\F_{\rho\sigma}}_{N}\,{\F^{\mu\rho}}^{P}{\S_{P}}^{Q}{\F^{\nu\sigma}}_{Q}, \\
 & {\F_{\mu\nu}}^{M}{\F^{\mu\rho}}_{M}\,{\F^{\nu\sigma}}^{N}{\F_{\rho\sigma}}_{N}, {\F_{\mu\nu}}^{M}{\S_{M}}^{N}{\F^{\mu\rho}}_{N}\,{\F^{\nu\sigma}}^{P}{\F_{\rho\sigma}}_{P}, {\F_{\mu\nu}}^{M}{\S_{M}}^{N}{\F^{\mu\rho}}_{N}\,{\F^{\nu\sigma}}^{P}{\S_{P}}^{Q}{\F_{\rho\sigma}}_{Q}\Big\}
\end{aligned}
\end{equation}

\paragraph*{$\boldsymbol{R\,H^2}:$}
\begin{align}
  \Big\{ & R_{\mu\nu\rho\sigma}H^{\mu\nu\lambda}{H^{\rho\sigma}}_{\lambda} \Big\} 
\end{align}

\paragraph*{$\boldsymbol{R\,\F^2}:$}
\begin{equation}
 \begin{aligned}
\Big\{&R_{\mu\nu\rho\sigma}\F^{\mu\nu\,M}{\F^{\rho\sigma}}_{M},\,R_{\mu\nu\rho\sigma}\F^{\mu\nu\,M}{\S_{M}}^{N}{\F^{\rho\sigma}}_{N}\Big\} 
\end{aligned}
\end{equation}

\paragraph*{$\boldsymbol{H^2\,(\nabla\Phi)^2}:$}
\begin{align}
  \Big\{ & H^{2}_{\mu\nu}\,\nabla^{\mu}\Phi\nabla^{\nu}\Phi, H^{2}\,\nabla_{\mu}\Phi\nabla^{\mu}\Phi \Big\}
\end{align}

\paragraph*{$\boldsymbol{H^2\,(\nabla\S)^2}:$}
\begin{align}
  \Big\{ & H^{2}_{\mu\nu}\,\,\Tr{\nabla^{\mu}\S\nabla^{\nu}\S}, H^{2}\,\Tr{\nabla_{\mu}\S\nabla^{\mu}\S} \Big\}
\end{align}

\paragraph*{$\boldsymbol{H^2\,\F^2}:$}
\begin{equation}
\begin{aligned}
  \Big\{ & H^{2}\,{\F_{\mu\nu}}^{M}{\F^{\mu\nu}}_{M},H^{2}\,{\F_{\mu\nu}}^{M}{\S_{M}}^{N}{\F^{\mu\nu}}_{N}, H^{2}_{\mu\nu}\,{\F^{\mu\rho}}^{M}{{\F_{\rho}}^{\nu}}_{M}, H^{2}_{\mu\nu}\,{\F^{\mu\rho}}^{M}{\S_{M}}^{N}{{\F_{\rho}}^{\nu}}_{N}, \\
  & {H_{\mu\nu}}^{\lambda}H_{\lambda\rho\sigma}\,{\F^{\mu\nu}}^{M}{\F^{\rho\sigma}}_{M}, {H_{\mu\nu}}^{\lambda}H_{\lambda\rho\sigma}\,{\F^{\mu\nu}}^{M}{\S_{M}}^{N}{\F^{\rho\sigma}}_{N},{H_{\mu\nu}}^{\lambda}H_{\lambda\rho\sigma}\,{\F^{\mu\sigma}}^{M}{\F^{\rho\nu}}_{M},  \\
  & {H_{\mu\nu}}^{\lambda}H_{\lambda\rho\sigma}\,{\F^{\mu\sigma}}^{M}{\S_{M}}^{N}{\F^{\rho\nu}}_{N}\Big\}
\end{aligned}
\end{equation}

\paragraph*{$\boldsymbol{(\nabla\Phi)^2\,(\nabla\S)^2}:$}
\begin{align}
  \Big\{ & \nabla_{\mu}\Phi\nabla_{\nu}\Phi\,\Tr{\nabla^{\mu}\S\nabla^{\nu}\S}, \nabla_{\mu}\Phi\nabla^{\mu}\Phi \,\Tr{\nabla_{\nu}\S\nabla^{\nu}\S} \Big\}
\end{align}

\paragraph*{$\boldsymbol{(\nabla\Phi)^2\,\F^2}:$}
\begin{equation}
\begin{aligned}
  \Big\{ & \nabla_{\rho}\Phi\nabla^{\rho}\Phi\,{\F_{\mu\nu}}^{M}{\F^{\mu\nu}}_{M},\nabla_{\rho}\Phi\nabla^{\rho}\Phi\,{\F_{\mu\nu}}^{M}{\S_{M}}^{N}{\F^{\mu\nu}}_{N}, \\
  & \nabla^{\mu}\Phi\nabla_{\nu}\Phi\,{\F_{\rho\mu}}^{M}{\F^{\rho\nu}}_{M}, \nabla^{\mu}\Phi\nabla_{\nu}\Phi\,{\F_{\rho\mu}}^{M}{\S_{M}}^{N}{\F^{\rho\nu}}_{N}\Big\}
\end{aligned}
\end{equation}

\paragraph*{$\boldsymbol{(\nabla\S)^2\,\F^2}:$}
\begin{equation}
\begin{aligned}
  \Big\{ & \Tr{\nabla_{\rho}\S\nabla^{\rho}\S}\,{\F_{\mu\nu}}^{M}{\F^{\mu\nu}}_{M},\Tr{\nabla_{\rho}\S\nabla^{\rho}\S}\,{\F_{\mu\nu}}^{M}{\S_{M}}^{N}{\F^{\mu\nu}}_{N}, \\
  & \Tr{\nabla^{\mu}\S\nabla^{\nu}\S}\,{\F_{\mu\rho}}^{M}{{\F_{\nu}}^{\rho}}_{M},\Tr{\nabla^{\mu}\S\nabla^{\nu}\S}\,{\F_{\mu\rho}}^{M}{\S_{M}}^{N}{{\F_{\nu}}^{\rho}}_{N}, \medskip \\
  & {\F_{\mu\nu}}^{M} \nabla_{\rho}{\S_{M}}^{N} \nabla^{\rho}{\S_{N}}^{P}{\F^{\mu\nu}}_{P},{\F_{\mu\nu}}^{M} \nabla_{\rho}{\S_{M}}^{N} \nabla^{\rho}{\S_{N}}^{P}{\S_{P}}^{Q}{\F^{\mu\nu}}_{Q}  \medskip\\
  & {\F_{\mu\nu}}^{M} \nabla^{\nu}{\S_{M}}^{N} \nabla_{\rho}{\S_{N}}^{P}{\F^{\mu\rho}}_{P},{\F_{\mu\nu}}^{M} \nabla^{\nu}{\S_{M}}^{N} \nabla_{\rho}{\S_{N}}^{P}{\S_{P}}^{Q}{\F^{\mu\rho}}_{Q},  \\
   & {\F_{\mu\nu}}^{M} \nabla_{\rho}{\S_{M}}^{N} \nabla^{\nu}{\S_{N}}^{P}{\F^{\mu\rho}}_{P},{\F_{\mu\nu}}^{M} \nabla_{\rho}{\S_{M}}^{N} \nabla^{\nu}{\S_{N}}^{P}{\S_{P}}^{Q}{\F^{\mu\rho}}_{Q}\Big\}
\end{aligned}
\end{equation}

\paragraph*{$\boldsymbol{H\,\nabla\Phi\,\F^2}:$}
\begin{align}
  \Big\{ & H^{\mu\nu\rho}\,\nabla^{\sigma}\Phi\,{\F_{\mu\nu}}^{M}{\F_{\rho\sigma}}_{M},H^{\mu\nu\rho}\,\nabla^{\sigma}\Phi\,{\F_{\mu\nu}}^{M}{\S_{M}}^{N}{\F_{\rho\sigma}}_{N}\Big\}
\end{align}

\paragraph*{$\boldsymbol{H\,\nabla\S\,\F^2}:$}
\begin{align}
 \Big\{H^{\mu\nu\rho}\,{\F_{\mu\sigma}}^{M}\nabla_{\nu}{\S_{M}}^{N}{\S_{N}}^{P}\,{{\F_{\rho}}^{\sigma}}_{P},H^{\mu\nu\rho}\,{\F_{\mu\nu}}^{M}\nabla^{\sigma}{\S_{M}}^{N}\,{\F_{\rho\sigma}}_{N},H^{\mu\nu\rho}\,{\F_{\mu\nu}}^{M}\nabla^{\sigma}{\S_{M}}^{N}{\S_{N}}^{P}\,{\F_{\rho\sigma}}_{P}\Big\}
\end{align}

\paragraph*{$\boldsymbol{\nabla\Phi\,\nabla\S\,\F^2}:$}
\begin{align}
  \Big\{ \nabla^{\rho}\Phi\,{\F_{\mu\nu}}^{M}\nabla_{\rho}{\S_{M}}^{N}{\F^{\mu\nu}}_{N},\nabla^{\mu}\Phi\,{\F_{\mu\rho}}^{M}\nabla_{\nu}{\S_{M}}^{N}{\F^{\nu\rho}}_{N},\nabla^{\mu}\Phi\,{\F_{\mu\rho}}^{M}\nabla_{\nu}{\S_{M}}^{N}{\S_{N}}^{P}{\F^{\nu\rho}}_{P}\Big\}
\end{align}

\section{Partial integration and explicit field redefinitions}
\label{app:IPP}
In this appendix, we give some details about the computations of the dimensionally reduced actions
presented in Sec.~\ref{sec;reducRR},
and Sec.~\ref{sec:het} respectively. We show explicitly how to eliminate all second order derivatives 
by partial integration 
up to terms appearing in the first column of Tab.~\ref{tab:fieldredef},
amenable to subsequent elimination by field redefinitions.

\subsection[\texorpdfstring{$\hat{R}_{\hat\mu\hat\nu\hat\rho\hat\sigma}\hat{R}^{\hat\mu\hat\nu\hat\rho\hat\sigma}$}{RR}]{\texorpdfstring{$\boldsymbol{\hat{R}_{\hat\mu\hat\nu\hat\rho\hat\sigma}\hat{R}^{\hat\mu\hat\nu\hat\rho\hat\sigma}}$}{RR}}

Let us begin with the terms appearing in the reduction of $\hat{R}_{\hat\mu\hat\nu\hat\rho\hat\sigma}\hat{R}^{\hat\mu\hat\nu\hat\rho\hat\sigma}$, as presented in Sec.~\ref{sec;reducRR}. 
We give the explicit expression of 
the five last terms in Eq.~\eqref{eq:RRreducedorder2} after integration by parts
(and use of Bianchi identities).
Up to boundary terms (which we ignore), the first two terms can be rewritten as
\begin{align}
\frac{\alpha'}{4} \int \d^{D}x& \,\sqrt{-g}\,e^{-\Phi}
\Tr{\nabla_{\mu}\nabla_{\nu}G^{-1}G\nabla^{\mu}\nabla^{\nu}G^{-1}G} \nonumber\\
=~ & \frac{\alpha'}{4} \int \d^{D}x\,\sqrt{-g}\,e^{-\Phi}\,\Big[{\rm Tr}\,\Big(\left(\Box G^{-1}-\nabla_{\mu}\Phi\nabla^{\mu}G^{-1}\right)G\left(\Box G^{-1}-\nabla_{\nu}\Phi\nabla^{\nu}G^{-1}\right)G\Big) \nonumber\\
&+\!2\,{\rm Tr}\Big(\!\!\left(\Box G^{-1}\!-\!\nabla_{\mu}\Phi\nabla^{\mu}G^{-1}\right)G\nabla_{\nu}G^{-1}\nabla^{\nu}G\Big)\!-\!{\rm Tr}\Big(\!\!\left(\Box G\!-\!\nabla_{\mu}\Phi\nabla^{\mu}G\right)G^{-1}\nabla_{\nu}G\nabla^{\nu}G^{-1}\Big) \nonumber\\
& +\Tr{\nabla_{\mu}G^{-1}\nabla_{\nu}G\nabla^{\mu}G^{-1}\nabla^{\nu}G} +\left(R_{\mu\nu}+\nabla_{\mu}\nabla_{\nu}\Phi\right)\Tr{\nabla^{\mu}G^{-1}\nabla^{\nu}G} \Big]\;,
\label{eq:term51}
\end{align}
and
\begin{align}
\frac{3\,\alpha'}{4} \int \d^{D}x\,\sqrt{-g}\,e^{-\Phi}\,&\Tr{\nabla_{\mu}\nabla_{\nu}G^{-1}\nabla^{\mu}G\nabla^{\nu}G^{-1}G} \nonumber\\
=~ &\int \d^{D}x\,\sqrt{-g}\,e^{-\Phi}\,\frac{\alpha'}{4}\,\Bigg[\frac{3}{2}\,{\rm Tr}\,\Big(\left(\Box G-\nabla_{\mu}\Phi\nabla^{\mu}G\right)G^{-1}\nabla_{\nu}G\nabla^{\nu}G^{-1}\Big) \nonumber\\
&
\qquad\qquad\qquad\qquad\qquad
-\frac{3}{2}\,\Tr{\nabla_{\mu}G^{-1}\nabla_{\nu}G\nabla^{\mu}G^{-1}\nabla^{\nu}G} \Bigg]\;,
\end{align}
respectively. The last three terms can be manipulated similarly 
and their sum takes the following form
\begin{align}
\frac{\alpha'}{4} \int \d^{D}x&\,\sqrt{-g}\,e^{-\Phi}\,\Big[F_{\mu\nu}^{(1)\,m}\left(G\nabla^{\mu}\nabla_{\rho}G^{-1}G\right)_{mn} F^{(1)\rho\nu\,n}-2\,\nabla_{\rho}F_{\mu\nu}^{(1)\,m}G_{mn}\nabla^{\mu}F^{(1)\nu\rho\,n}\nonumber\\
&\qquad\qquad\qquad-6\,\nabla_{\rho}F_{\mu\nu}^{(1)\,m}\nabla^{\mu}G_{mn}F^{(1)\nu\rho\,n}\Big]\nonumber \\[1ex]
= &\int \d^{D}x\,\sqrt{-g}\,e^{-\Phi}\,\frac{\alpha'}{4}\,\Big[ 2\big(\nabla^{\mu}F_{\mu\nu}^{(1)\,m}-\nabla^{\mu}\Phi \,F_{\mu\nu}^{(1)\,m}\big)G_{mn}\left(\nabla_{\rho}F^{(1)\rho\nu\,n}-\nabla_{\rho}\Phi F^{(1)\rho\nu\,n}\right)\nonumber \\
& -2\,\left(R_{\mu\nu}+\nabla_{\mu}\nabla_{\nu}\Phi\right)\,F_ {\mu\rho}^{(1)\,m}G_{mn}F_{\ \nu}^{(1)\,\rho\,n}-\frac{5}{4}\,F_{\mu\nu}^{(1)\,m}\left(\Box G-\nabla_{\rho}\Phi\nabla^{\rho}G\right)_{mn}F^{(1)\mu\nu\,n}  \nonumber \\
&+F_{\mu\nu}^{(1)\,m}\left(\nabla_{\rho}G\nabla^{\mu}G^{-1}G\right)_{mn}F^{(1)\nu\rho\,n} +F_{\mu\nu}^{(1)\,m}\left(\nabla^{\mu}G\nabla_{\rho}G^{-1}G\right)_{mn}F^{(1)\nu\rho\,n}  \nonumber \\
&+3\,F_{\mu\nu}^{(1)\,m}\nabla^{\mu}G_{mn}\left(\nabla_{\rho}F^{(1)\rho\nu\,n}-\nabla_{\rho}\Phi \,F^{(1)\rho\nu\,n}\right) + R^{\mu\nu\rho\sigma} F_{\mu\nu}^{(1)\,m}G_{mn}F_{\rho\sigma}^{(1)\,n}\Big]\;,
\label{eq:term53}
\end{align}
again up to boundary contributions. In the form (\ref{eq:term51})--(\ref{eq:term53}), all 
the remaining second order derivatives are of the form 
appearing in the first column of Tab.~\ref{tab:fieldredef}.
They can thus be reabsorbed into field redefinitions as discussed in Sec.~\ref{sec:GLdredef}.
Explicitly, this induces the order $\alpha'$ field redefinitions
\begin{align}
\ \delta \Phi&= \dfrac{1}{8}\,\Big[-2\,F_{\mu\nu}^{(1)\, m}G_{mn}F^{(1)\mu\nu\,n}+\Tr{\nabla_{\mu}G^{-1}\nabla^{\mu}G}\Big]
    \;,\nonumber\\
    \ \delta g_{\mu\nu}&= \dfrac{1}{4}\,\Big[2\,F_{\mu\rho}^{(1)\, m}G_{mn}F_{\ \nu}^{(1)\rho\,n}-\Tr{\nabla_{(\mu}G^{-1}\nabla_{\nu)}G}\Big]    \;,\nonumber\\
    \delta B_{\mu\nu}&=\dfrac{1}{8}\,\Big(-2\,\nabla^{\rho}F_{\rho\mu}^{(1)\,m}+2\,\nabla^{\rho}\Phi F_{\rho\mu}^{(1)\,m} +H_{\mu\rho\sigma}{H^{\rho\sigma}}_{p}G^{pm} +F_{\mu\rho}^{(1)\,p}{\left(\nabla^{\rho}GG^{-1}\right)_{p}}^{m}    \nonumber\\
    &{}\qquad+2\,H_{\mu\rho p}\left(G^{-1}\nabla^{\rho}BG^{-1}\right)^{pm}\Big)\Big(A_{\nu\,m}^{(2)}-B_{mn}A_{\nu}^{(1)n}\Big) - \Big(\mu\leftrightarrow\nu\Big)    \;,\nonumber\\
    \delta G^{mn}&=\dfrac{1}{4}\,\Big[-2\,\Box G^{mn}+2\,\nabla_{\mu}\Phi\nabla^{\mu}G^{mn}-G^{mp}H_{\mu\nu p}G^{nq}{H^{\mu\nu}}_{q} - \dfrac{3}{2}\,F_{\mu\nu}^{(1)\,m}F^{(1)\mu\nu\,n}    \nonumber\\
    &{}\qquad-\left(G^{-1}\nabla_{\mu}G\nabla^{\mu}G^{-1}\right)^{mn}+2\,\left(G^{-1}\nabla_{\mu}BG^{-1}\nabla^{\mu}BG^{-1}\right)^{mn}\Big]    \;,\nonumber\\
    \delta A_{\mu}^{(1)m}&=\dfrac{1}{4}\,\Big[-2\,\nabla^{\nu}F_{\nu\mu}^{(1)\,m}+2\,\nabla^{\nu}\Phi F_{\nu\mu}^{(1)\,m} +H_{\mu\nu\rho}{H^{\nu\rho}}_{n}G^{nm}    \nonumber\\
    &{}\qquad+F_{\mu\nu}^{(1)\,n}{\left(\nabla^{\nu}GG^{-1}\right)_{n}}^{m}+2\,H_{\mu\nu n}\left(G^{-1}\nabla^{\nu}BG^{-1}\right)^{nm}\Big]     \;,\nonumber\\
    \delta A_{\mu\,m}^{(2)}&=\dfrac{1}{4}\,\Big[2\,\nabla^{\nu}F_{\nu\mu}^{(1)\,n}B_{nm}-2\,\nabla^{\nu}\Phi F_{\nu\mu}^{(1)\,n}B_{nm} -H_{\mu\nu\rho}{H^{\nu\rho}}_{n}{\left(G^{-1}B\right)^{n}}_{m}    \nonumber\\
    &{}\qquad-F_{\mu\nu}^{(1)\,n}\left(\nabla^{\nu}GG^{-1}B\right)_{nm}-2\,H_{\mu\nu n}{\left(G^{-1}\nabla^{\nu}BG^{-1}B\right)^{n}}_{m}\Big]
    \;.   
\label{eq:fieldrefRR}
\end{align}

\subsection[\texorpdfstring{$\hat{R}_{\hat\mu\hat\nu\hat\rho\hat\sigma}\hat{H}^{\hat\mu\hat\nu\hat\lambda}\hat{H}^{\hat\rho\hat\sigma}_{\quad\hat\lambda}$}{RHH}]{\texorpdfstring{$\boldsymbol{\hat{R}_{\hat\mu\hat\nu\hat\rho\hat\sigma}\hat{H}^{\hat\mu\hat\nu\hat\lambda}\hat{H}^{\hat\rho\hat\sigma}_{\quad\hat\lambda}}$}{RHH}}
Here, we consider the four last terms in the reduction~\eqref{eq:RHHreducedorder2} of $RHH$. 
After partial integration, they can be brought into the form
\begin{align}
\frac{\alpha'}{4} & \int\d^{D}x\,\sqrt{-g}\,e^{-\Phi}\,\Tr{\nabla_{\mu}\nabla_{\nu}G^{-1}\nabla^{\mu}BG^{-1}\nabla^{\nu}B} \nonumber \\[1ex]
=~&\frac{\alpha'}{4} \int\d^{D}x\,\sqrt{-g}\,e^{-\Phi}\,\Big[ -\Tr{\nabla_{\mu}B\nabla^{\mu}G^{-1}\nabla_{\nu}B\nabla^{\nu}G^{-1}} +\frac{1}{2}\, \Tr{\nabla_{\mu}B\nabla_{\nu}G^{-1}\nabla^{\mu}B\nabla^{\nu}G^{-1}}\nonumber \\
& -{\rm Tr}\,\Big(\left(\Box B-\nabla_{\mu}\Phi\nabla^{\mu}B\right)G^{-1}\nabla_{\nu}B\nabla^{\nu}G^{-1}\Big)+\frac{1}{2}\,{\rm Tr}\,\Big(\left(\Box G^{-1}-\nabla_{\mu}\Phi\nabla^{\mu}G^{-1}\right)\nabla_{\nu}BG^{-1}\nabla^{\nu}B\Big) \Big]\;,
\end{align}
\begin{align}
-\frac{\alpha'}{4}& \int\d^{D}x\,\sqrt{-g}\,e^{-\Phi}\,{H^{\mu\rho}}_{m}\nabla_{\mu}\nabla_{\nu}G^{mn}{H^{\nu}}_{\rho n}\nonumber \\[1ex]
=~&\frac{\alpha'}{4} \int\d^{D}x\,\sqrt{-g}\,e^{-\Phi}\,\Big[ -\left(\nabla_{\mu}{H^{\mu\nu}}_{m}-\nabla_{\mu}\Phi{H^{\mu\nu}}_{m}\right)\nabla^{\rho}G^{mn}H_{\nu\rho n}-\frac{1}{2}\,F_{\mu\nu}^{(1)\,m}{\left(\nabla_{\rho}B\nabla^{\rho}G^{-1}\right)_{m}}^{n}{H^{\mu\nu}}_{n}\nonumber\\
& -\frac{1}{4}\,{H^{\mu\nu}}_{m}\left(\Box G^{-1}-\nabla_{\rho}\Phi\nabla^{\rho}G^{-1}\right)^{mn}H_{\mu\nu n} -F_{\nu\rho}^{(1)\,m}{\left(\nabla_{\mu}B\nabla^{\rho}G^{-1}\right)_{m}}^{n}{H^{\mu\nu}}_{n}\Big]\;,
\end{align}
\begin{align}
-\frac{\alpha'}{2}\, \int\d^{D}x& \,\sqrt{-g}\,e^{-\Phi}\,\nabla_{\mu}F_{\nu\rho}^{(1)\, m}{\left(\nabla^{\rho}BG^{-1}\right)_{m}}^{n}{H^{\mu\nu}}_{n}  \nonumber \\[1ex]
=~&\frac{\alpha'}{4}\int\d^{D}x\,\sqrt{-g}\,e^{-\Phi}\,\Big[ -\left(\nabla_{\mu}{H^{\mu\nu}}_{m}-\nabla_{\mu}\Phi{H^{\mu\nu}}_{m}\right){\left(G^{-1}\nabla^{\rho}B\right)^{m}}_{n}F_{\nu\rho}^{(1)\,n}\nonumber \\
&-\left(\nabla_{\mu}F^{(1)\mu\nu\,m}-\nabla_{\mu}\Phi F^{(1)\mu\nu\,m}\right){\left(\nabla^{\rho}BG^{-1}\right)_{m}}^{n}H_{\nu\rho n}+\frac{1}{2}\,F_{\mu\nu}^{(1)\,m}\left(\nabla_{\rho}BG^{-1}\nabla^{\rho}B\right)_{mn}F^{(1)\,\mu\nu\,n}\nonumber\\
& -\frac{1}{2}\,F_{\mu\nu}^{(1)\,m}{\left(\nabla_{\rho}B\nabla^{\rho}G^{-1}\right)_{m}}^{n}{H^{\mu\nu}}_{n}-\frac{1}{2}\,F_{\mu\nu}^{(1)\,m}{\big(\left(\Box B-\nabla_{\rho}\Phi\nabla^{\rho}B\right)G^{-1}\big)_{m}}^{n}{H^{\mu\nu}}_{n}\nonumber\\
&+ F_{\nu\rho}^{(1)\,m}{\left(\nabla^{\rho}B\nabla_{\mu}G^{-1}\right)_{m}}^{n}{H^{\mu\nu}}_{n} -F_{\nu\rho}^{(1)\,m}{\left(\nabla_{\mu}B\nabla^{\rho}G^{-1}\right)_{m}}^{n}{H^{\mu\nu}}_{n}\nonumber\\
&+F_{\mu\nu}^{(1)\,m}\left(\nabla_{\rho}BG^{-1}\nabla^{\nu}B\right)_{mn}F^{(1)\,\rho\mu\,n}\Big]\;,
\end{align}
\begin{align}
\frac{\alpha'}{4} &\int\d^{D}x\,\sqrt{-g}\,e^{-\Phi}\,2\,H^{\mu\nu\lambda}\nabla_{\mu}F_{\nu\rho}^{(1)\,m}{H^{\rho}}_{\lambda m} \nonumber \\[1ex]
=~&\frac{\alpha'}{4} \int\d^{D}x\,\sqrt{-g}\,e^{-\Phi}\,\Big[ \frac{1}{2}\,\left(\nabla_{\mu}{H^{\mu\nu}}_{m}-\nabla_{\mu}\Phi{H^{\mu\nu}}_{m}\right)H_{\nu\rho\sigma}F^{(1)\rho\sigma\,m}-H^{\mu\nu\rho}F_{\nu\sigma}^{(1)\,m}\nabla_{\mu}B_{mn}F_{\ \rho}^{(1)\,\sigma\,n} \nonumber \\
& -\left(\nabla_{\mu}H^{\mu\nu\rho}-\nabla_{\mu}\Phi H^{\mu\nu\rho}\right)F_{\nu\sigma}^{(1)\,m}{H^{\sigma}}_{\rho\, m}+\frac{1}{2}\, \left(\nabla_{\mu}F^{(1)\mu\nu\,m}-\nabla_{\mu}\Phi F^{(1)\mu\nu\,m}\right) H_{\nu\rho\sigma}{H^{\rho\sigma}}_{m}\nonumber\\
& +\frac{1}{2}\,H^{\mu\nu\rho}F_{\nu\sigma}^{(1)\,m}\nabla^{\sigma}B_{mn}F_{\rho\mu}^{(1)\,n} +F_{\mu\nu}^{(1)\,m}H_{\rho\sigma\,m}F^{(1)\,\rho\nu\,n}H^{\nu\sigma}{}_{n}-\frac{1}{4}\,F_{\mu\nu}^{(1)\,m}H_{\rho\sigma\,m}F^{(1)\,\mu\nu\,n}H^{\rho\sigma}{}_{n} \nonumber \\
&-\frac{1}{4}\,F_{\mu\nu}^{(1)\,m}H_{\rho\sigma\,m}F^{(1)\,\rho\sigma\,n}H^{\mu\nu}{}_{n}  \Big]\;,
\end{align}
respectively. Again, all left-over terms 
carrying second-order derivatives can be converted to products of first order derivatives 
by means of the rules of Tab.~\ref{tab:fieldredef}. This induces the explicit field redefinitions
\begin{align}
   \delta B_{\mu\nu}&=
    \dfrac{1}{8}\,\Big[-A_{\mu}^{(1)m}{\left(G\nabla_{\rho}G^{-1}\right)_{m}}^{n}H_{\nu\rho n} -A_{\mu}^{(1)m} \nabla^{\rho}B_{mn}F^{(1)n}_{\nu\rho} +2\,F_{\mu\rho}^{(1)m}{H^{\rho}}_{\nu m}\nonumber\\
    &{}\qquad+\dfrac{1}{2}\,A_{\mu}^{(1)m}H_{\nu\rho\sigma}G_{mn}F^{(1)\rho\sigma\,n} - \dfrac{1}{2}\,H_{\mu\rho\sigma}{H^{\rho\sigma}}_{m}G^{mn}\Big(A_{\nu\,n}^{(2)}-B_{np}A_{\nu}^{(1)p}\Big)\nonumber\\
    &{} \qquad-H_{\mu\rho m}\left(G^{-1}\nabla^{\rho}BG^{-1}\right)^{mn}\Big(A_{\nu\,n}^{(2)}-B_{np}A_{\nu}^{(1)p}\Big)\Big] - \Big(\mu\leftrightarrow\nu\Big)\;,\nonumber\\
    \delta G^{mn}&=\dfrac{1}{4}\,\Big[\dfrac{1}{2}\,G^{mp}H_{\mu\nu p}G^{nq}{H^{\mu\nu}}_{q}-\left(G^{-1}\nabla_{\mu}BG^{-1}\nabla^{\mu}BG^{-1}\right)^{mn}\Big]\;,\nonumber\\
    \delta B_{mn}&=\dfrac{1}{4}\,\Big[\left(\nabla_{\mu}B\nabla^{\mu}G^{-1}G\right)_{mn}+\left(G\nabla_{\mu}G^{-1}\nabla^{\mu}B\right)_{mn} \nonumber\\
    &{}\qquad- \dfrac{1}{2}\, H_{\mu\nu m}F^{(1)\mu\nu\,p}G_{pn} + \dfrac{1}{2}\,G_{mp}F^{(1)\mu\nu\,p}H_{\mu\nu n}\Big]\;,\nonumber\\
    \delta A_{\mu}^{(1)m}&=\dfrac{1}{4}\,\Big[-H_{\mu\nu n}\left(G^{-1}\nabla^{\nu}BG^{-1}\right)^{nm}-\dfrac{1}{2}\,H_{\mu\nu\rho}{H^{\nu\rho}}_{n}G^{nm}\Big]\;,\nonumber\\
    \delta A_{\mu\,m}^{(2)}&=\dfrac{1}{4}\,\Big[H_{\mu\nu n}{\left(\nabla^{\nu}G^{-1}G\right)^{n}}_{m}-F^{(1)n}_{\mu\nu}\nabla^{\nu}B_{nm}+H_{\mu\nu n}{\left(G^{-1}\nabla^{\nu}BG^{-1}B\right)^{n}}_{m} \nonumber\\
    &{}\qquad-\dfrac{1}{2}\,H_{\mu\nu\rho}F^{(1)\nu\rho\,n}G_{nm}+\dfrac{1}{2}\,H_{\mu\nu\rho}{H^{\nu\rho}}_{n}{\left(G^{-1}B\right)^{n}}_{m}\Big]\;.
\label{eq:fieldrefRHH}
\end{align}

\subsection[\texorpdfstring{$\hat{H}^{\hat\mu\hat\nu\hat\rho}\, \hat{\Omega}^{(\hat\omega)}_{\hat\mu\hat\nu\hat\rho}$}{RHH}]{\texorpdfstring{$\boldsymbol{\hat{H}^{\hat\mu\hat\nu\hat\rho}\, \hat{\Omega}^{(\hat\omega)}_{\hat\mu\hat\nu\hat\rho}}$}{RHH}}
\label{app:heterotic}
Finally, we give the result for the six last terms in Eq.~\eqref{eq:hetsecondorder}. After partial integration they are rewritten as
\begin{align}
\frac{\alpha'}{4} &\int\d^{D}x\,\sqrt{-g}\,e^{-\Phi}\left(- \frac{1}{2}\right)\,H^{\mu\nu}{}_{m}G^{mn}\nabla_{\mu}\left(\nabla_{\nu}e_{\sigma}{}^{\alpha}e_{\alpha}{}^{\rho} F_{\rho}^{(1)\,\sigma\,p}G_{np}\right) \nonumber \\
=~&\frac{\alpha'}{4}\int\d^{D}x\,\sqrt{-g}\,e^{-\Phi}\,\Big[\frac{1}{2}\left(\nabla_{\mu}H^{\mu\nu}{}_{p}-\nabla_{\mu}\Phi H^{\mu\nu}{}_{p}+H^{\mu\nu}{}_{m}\nabla_{\mu}G^{mn}G_{np}\right)\nabla_{\nu}e_{\sigma}{}^{\alpha}e_{\alpha}{}^{\rho}F_{\rho}^{(1)\,\sigma\,p}\Big]
\;,
\end{align}
\begin{align}
\frac{\alpha'}{4} &
\int\d^{D}x\,\sqrt{-g}\,e^{-\Phi}\,\frac{1}{2}\,\Tr{\nabla_{\mu}\nabla_{\nu}G\nabla^{\mu}G^{-1}\nabla^{\nu}BG^{-1}} \nonumber \\
=~&\frac{\alpha'}{4}\int\d^{D}x\,\sqrt{-g}\,e^{-\Phi}\,\Big[-\frac{1}{2}\, {\rm Tr}\Big(\left(\Box B-\nabla_{\mu}\Phi\nabla^{\mu}B\right)G^{-1}\nabla_{\nu}G\nabla^{\nu}G^{-1}\Big) \nonumber \\
&+\frac{1}{2}\, {\rm Tr}\Big(\left(\Box G-\nabla_{\mu}\Phi\nabla^{\mu}G\right)\nabla_{\nu}G^{-1}\nabla^{\nu}BG^{-1}\Big) -\Tr{\nabla_{\mu}B\nabla^{\mu}G^{-1}\nabla_{\nu}G\nabla^{\nu}G^{-1}}\Big]\;,
\end{align}
\begin{align}
\frac{\alpha'}{4} & \int\d^{D}x\,\sqrt{-g}\,e^{-\Phi}\,\left[- \frac{1}{2}\,H_{\mu\nu\,m}\left(G^{-1}\nabla_{\rho}G\right)^{m}{}_{n}\nabla^{\mu}F^{(1)\nu\rho\,n}+ \frac{1}{2}\,H_{\mu\nu\,m}\left(G^{-1}\nabla^{\mu}\nabla_{\rho}G\right)^{m}{}_{n}F^{(1)\nu\rho\,n}\right] \nonumber \\
=~&\frac{\alpha'}{4} \int\d^{D}x\,\sqrt{-g}\,e^{-\Phi}\,\Big[-\frac{1}{2}\left(\nabla_{\mu}F^{(1)\mu\nu\,m}-\nabla_{\mu}\Phi F^{(1)\mu\nu\,m}\right)\left(\nabla^{\rho}GG^{-1}\right)_{m}{}^{n}H_{\nu\rho\,n} \nonumber \\
&-\frac{1}{4}\,H_{\mu\nu\,m}\Big(G^{-1}\left(\Box G-\nabla_{\rho}\Phi\nabla^{\rho}G+\nabla_{\rho}G\nabla^{\rho}G^{-1}G\right)\Big)^{m}{}_{n}F^{(1)\mu\nu\,n} -\frac{1}{2}\,H_{\nu\rho\,m}\left(\nabla_{\mu}G^{-1}\nabla^{\rho}G\right)^{m}{}_{n}F^{(1)\mu\nu\,n}\nonumber \\
& - \frac{1}{2}\,F_{\nu\rho}^{(1)\,m}\left(\nabla_{\mu}BG^{-1}\nabla^{\rho}G\right)_{mn}F^{(1)\mu\nu\,n} -\frac{1}{4}\,F_{\mu\nu}^{(1)\,m}\left(\nabla_{\rho}BG^{-1}\nabla^{\rho}G\right)_{mn}F^{(1)\mu\nu\,n}\Big]\;,
\end{align}
\begin{align}
\frac{\alpha'}{4} &\int\d^{D}x\,\sqrt{-g}\,e^{-\Phi}\,\frac{1}{2}\,H^{\mu\nu\rho}F_{\mu\sigma}^{(1)\,m}G_{mn}\nabla_{\nu}^{}F_{\rho}^{(1)\,\sigma\,n} \nonumber \\
=~&\frac{\alpha'}{4} \int\d^{D}x\,\sqrt{-g}\,e^{-\Phi}\,\Big[-\frac{1}{4}\,H_{\mu\nu\rho}\left(\nabla_{\sigma}F^{(1)\sigma\mu\,m}-\nabla_{\sigma}\Phi F^{(1)\sigma\mu\,m}+F^{(1)\sigma\mu\,p}\left(\nabla_{\sigma}GG^{-1}\right)_{p}{}^{m}\right)G_{mn}F^{(1)\,n}_{\nu\rho} \nonumber \\
&-\frac{1}{4}\,F^{(1)\,m}_{\mu\nu}H_{\rho\sigma\,m}\,F^{(1)\mu\rho\,p}G_{pq}F^{(1)\nu\sigma\,q}+\frac{1}{8}\,F^{(1)\,m}_{\mu\nu}H_{\rho\sigma\,m}\,F^{(1)\mu\nu\,p}G_{pq}F^{(1)\rho\sigma\,q}\Big]\;,
\end{align}
\begin{align}
\frac{\alpha'}{4} &\int\d^{D}x\,\sqrt{-g}\,e^{-\Phi}\,\left(- \frac{1}{4}\right)\,F_{\mu\nu}^{(1)\,m}\nabla_{\rho}B_{mn}\nabla^{\rho}F^{(1)\mu\nu\,n} \nonumber \\
=~&\frac{\alpha'}{4} \int\d^{D}x\,\sqrt{-g}\,e^{-\Phi}\,\Big[-\frac{1}{2}\left(\nabla_{\mu}F^{(1)\mu\nu\,m}-\nabla_{\mu}\Phi F^{(1)\mu\nu\,m}\right)\nabla^{\rho}B_{mn}F^{(1)\,n}_{\nu\rho}\Big]\;.
\end{align}
Again, the remaining terms carrying second-order derivatives can be eliminated by
field redefinitions as discussed in Sec.~\ref{sec:GLdredef}. 
The explicit form of the induced field redefinitions
has been given in Eq.~\eqref{eq:fieldredefhet} in the main text.

\section{\texorpdfstring{$\boldsymbol{\GL(d)}$ expressions of some $\boldsymbol{\Odd}$ terms}{GL(d) expressions od some O(d,d) terms}}
\label{app:OddtoGLd}
In this appendix, we present the $\GL(d)$ decomposition of some of the
\Odd invariant terms, that are relevant for the identifications made in 
Secs.~\ref{sec:alpha'action} and \ref{sec:het}.
\begin{equation}
\Tr{\nabla_{\mu}\S\nabla_{\nu}\S} = 2\,\Tr{\nabla_{(\mu}G\nabla_{\nu)}G^{-1}}+ 2\,\Tr{\nabla_{\mu}BG^{-1}\nabla_{\nu}BG^{-1}}.
\end{equation}
%
\begin{align}
\Tr{\nabla_{\mu}\S\nabla_{\nu}\S\nabla^{\mu}\S\nabla^{\nu}\S} = ~&2\,\Tr{\nabla_{\mu}G^{-1}\nabla_{\nu}G\nabla^{\mu}G^{-1}\nabla^{\nu}G} +4\,\Tr{\nabla_{\mu}B\nabla_{\nu}G^{-1}\nabla^{\mu}B\nabla^{\nu}G^{-1}}\nonumber\\
& +8\,\Tr{G^{-1}\nabla_{\mu}BG^{-1}\nabla_{\nu}G\nabla^{\mu}G^{-1}\nabla^{\nu}B} \\
& +2\,\Tr{G^{-1}\nabla_{\mu}BG^{-1}\nabla_{\nu}BG^{-1}\nabla^{\mu}BG^{-1}\nabla^{\nu}B}, \nonumber
\end{align}
%
\begin{align}
\Tr{\nabla_{\mu}\S\nabla^{\mu}\S\nabla_{\nu}\S\nabla^{\nu}\S} =~& 2\,\Tr{\nabla_{\mu}G^{-1}\nabla^{\mu}G\nabla_ {\nu}G^{-1}\nabla^{\nu}G}+ 4\,\Tr{G^{-1}\nabla_{\mu}BG^{-1}\nabla_{\nu}G\nabla^{\nu}G^{-1}\nabla^{\mu}B} \nonumber \\
&+4\,\Tr{\nabla_{\mu}G^{-1}\nabla^{\mu}B\nabla_{\nu}G^{-1}\nabla^{\nu}B}+4\,\Tr{G^{-1}\nabla_{\mu}B\nabla^{\mu}G^{-1}\nabla_{\nu}GG^{-1}\nabla^{\nu}B} \nonumber \\
&+ 2\,\Tr{G^{-1}\nabla_{\mu}BG^{-1}\nabla^{\mu}BG^{-1}\nabla_{\nu}BG^{-1}\nabla^{\nu}B},
\end{align}
 \begin{align}
  &\Tr{\S\nabla_{\mu}\S\nabla_{\nu}\S\nabla_{\rho}\S\nabla_{\sigma}\S} = 2\Big[\Tr{\nabla_{\mu}G^{-1}\nabla_{\nu}G\nabla_{\rho}G^{-1}\nabla_{\sigma}B}-\Tr{\nabla_{\sigma}G^{-1}\nabla_{\mu}G\nabla_{\nu}G^{-1}\nabla_{\rho}B}  \nonumber\\
  &\qquad\qquad\qquad
  +\Tr{\nabla_{\rho}G^{-1}\nabla_{\sigma}G\nabla_{\mu}G^{-1}\nabla_{\nu}B}-\Tr{\nabla_{\nu}G^{-1}\nabla_{\rho}G\nabla_{\sigma}G^{-1}\nabla_{\mu}B} \nonumber \\
  &\qquad\qquad\qquad
  -\Tr{\nabla_{\mu}BG^{-1}\nabla_{\nu}BG^{-1}\nabla_{\rho}B\nabla_{\sigma}G^{-1}}+\Tr{\nabla_{\sigma}BG^{-1}\nabla_{\mu}BG^{-1}\nabla_{\nu}B\nabla_{\rho}G^{-1}} \nonumber \\
  &\qquad\qquad\qquad
  -\Tr{\nabla_{\rho}BG^{-1}\nabla_{\sigma}BG^{-1}\nabla_{\mu}B\nabla_{\nu}G^{-1}}+\Tr{\nabla_{\nu}BG^{-1}\nabla_{\rho}BG^{-1}\nabla_{\sigma}B\nabla_{\mu}G^{-1}} \Big]
  \;,
  \end{align}
\begin{align}
{\F_{\mu\nu}}^{M}{\S_{M}}^{N}\F_{\rho\sigma\,N} &= F_{\mu\nu}^{(1)\,m}G_{mn}F_{\rho\sigma}^{(1)\,n}+H_{\mu\nu\,m}G^{mn}H_{\rho\sigma\, n}
\;,
\\[2ex]
{\F_{\mu\nu}}^{M}\F_{\rho\sigma\,M} &= F_{\mu\nu}^{(1)\,m}F_{\rho\sigma\,m}^{(2)}+F_{\rho\sigma}^{(1)\,m}F_{\mu\nu\,m}^{(2)}
\;,
\\[2ex]
{\F_{\mu\nu}}^{M}{\S_{M}}^{N}\nabla_{\rho}{\S_{N}}^{P}\F_{\sigma\lambda\,
P} &= F_{\mu\nu}^{(1)\,m}{\left(G\nabla_{\rho}G^{-1}\right)_{m}}^{n}H_{\rho\lambda\,n} - F_{\mu\nu}^{(1)\,m}\nabla_{\rho}B_{mn}F_{\sigma\lambda}^{(1)\,n}  \nonumber\\
&{}
+H_{\mu\nu\,m}\left(G^{-1}\nabla_{\rho}BG^{-1}\right)^{mn}H_{\sigma\lambda\, n} +H_{\mu\nu\, m}{\left(G^{-1}\nabla_{\rho}G\right)^{m}}_{n}F_{\sigma\lambda}^{(1)\,n}\;,\\[2ex]
  {\F_{\mu\nu}}^{M}\nabla_{\rho}{\S_{M}}^{N}\F_{\sigma\lambda\,N} &= F_{\mu\nu}^{(1)\,m}\nabla_{\rho}G_{mn}F_{\sigma\lambda}^{(1)\,n}+F_{\mu\nu}^{(1)\,m}{\left(\nabla_{\rho}BG^{-1}\right)_{m}}^{n}H_{\rho\lambda\,n}  \nonumber\\
  &{}-H_{\mu\nu\, m}{\left(G^{-1}\nabla_{\rho}B\right)^{m}}_{n}F_{\sigma\lambda}^{(1)\,n}+H_{\mu\nu\,m}\nabla_{\rho}G^{mn}H_{\sigma\lambda\, n}\;,
\end{align}

\bea
&&{}
{\F_{\mu\nu}}^{M}{\S_{M}}^{N}\nabla_{\rho}{\S_{N}}^{P}\nabla_{\sigma}{\S_{P}}^{Q}\F_{\lambda\tau\,Q} ~=
\nonumber\\[.5ex]
&&{} \hspace*{15mm}
F_{\mu\nu}^{(1)\,m}\left(G\nabla_{\rho}G^{-1}\nabla_{\sigma}G\right)_{mn}F_{\lambda\tau}^{(1)\,n} +F_{\mu\nu}^{(1)\,m}{\left(G\nabla_{\rho}G^{-1}\nabla_{\sigma}BG^{-1}\right)_{m}}^{n}H_{\lambda\tau\,n}  \nonumber\\
&&{} \hspace*{15mm}
-F_{\mu\nu}^{(1)\,m}{\left(\nabla_{\rho}B\nabla_{\sigma}G^{-1}\right)_{m}}^{n}H_{\lambda\tau\,n} + H_{\mu\nu m}{\left(\nabla_{\rho}G^{-1}\nabla_{\sigma}B\right)^{m}}_{n}F_{\lambda\tau}^{(1)\,n}  \nonumber\\
&&{} \hspace*{15mm}
+H_{\mu\nu m}{\left(G^{-1}\nabla_{\rho}BG^{-1}\nabla_{\sigma}G\right)^{m}}_{n}F_{\lambda\tau}^{(1)\,n} + F_{\mu\nu}^{(1)\,m}\left(\nabla_{\rho}BG^{-1}\nabla_{\sigma}B\right)_{mn}F_{\lambda\tau}^{(1)\,n} \nonumber\\
&&{} \hspace*{15mm}
+H_{\mu\nu\,m}\left(G^{-1}\nabla_{\rho}G\nabla_{\sigma}G^{-1}\right)^{mn}H_{\lambda\tau\, n}+H_{\mu\nu\,m}\left(G^{-1}\nabla_{\rho}BG^{-1}\nabla_{\sigma}BG^{-1}\right)^{mn}H_{\lambda\tau\, n}\;.\quad
\eea

\bea
 && {\F_{\mu\nu}}^{M} \nabla_{\rho}{\S_{M}}^{N}\nabla_{\sigma}{\S_{N}}^{P}\F_{\lambda\tau\,P}~=
  \nonumber\\[.5ex]
&&{} \hspace*{20mm}  F_{\mu\nu}^{(1)\,m}\left(\nabla_{\rho}BG^{-1}\nabla_{\sigma}G\right)_{mn}F_{\lambda\tau}^{(1)\,n} -F_{\mu\nu}^{(1)\,m}\left(\nabla_{\rho}GG^{-1}\nabla_{\sigma}B\right)_{mn}F_{\lambda\tau}^{(1)\,n} \nonumber\\
 &&{} \hspace*{20mm} +F_{\mu\nu}^{(1)\,m}{\left(\nabla_{\rho}G\nabla_{\sigma}G^{-1}\right)_{m}}^{n}H_{\lambda\tau\,n} + F_{\mu\nu}^{(1)\,m}{\left(\nabla_{\rho}BG^{-1}\nabla_{\sigma}BG^{-1}\right)_{m}}^{n}H_{\lambda\tau\,n}  \nonumber\\
 &&{} \hspace*{20mm} +H_{\mu\nu m}{\left(\nabla_{\rho}G^{-1}\nabla_{\sigma}G\right)^{m}}_{n}F_{\lambda\tau}^{(1)\,n} +H_{\mu\nu m}{\left(G^{-1}\nabla_{\rho}BG^{-1}\nabla_{\sigma}B\right)^{m}}_{n}F_{\lambda\tau}^{(1)\,n} \nonumber\\
 &&{} \hspace*{20mm} +H_{\mu\nu\,m}\left(\nabla_{\rho}G^{-1}\nabla_{\sigma}BG^{-1}\right)^{mn}H_{\lambda\tau\, n}-H_{\mu\nu\,m}\left(G^{-1}\nabla_{\rho}B\nabla_{\sigma}G^{-1}\right)^{mn}H_{\lambda\tau\, n}
 \;.\quad
 \eea
 
\end{appendix}


\providecommand{\href}[2]{#2}\begingroup\raggedright\endgroup

\end{document}